\UseRawInputEncoding
\pdfoutput=1
\documentclass{article}
\usepackage[T1]{fontenc} 
\usepackage{framed}
\usepackage{multirow}
\usepackage[percent]{overpic}
\usepackage{amsfonts,amssymb,graphics,epsfig,verbatim,bm,latexsym,amsmath,url,amsbsy}
\usepackage[capposition=top]{floatrow}
\usepackage{float}
\usepackage{graphicx}
\usepackage{booktabs}
\usepackage{caption}
\usepackage{tabularx}
\usepackage{multirow}
\usepackage[flushleft]{threeparttable}
\usepackage{enumitem}
\usepackage{mathtools}
\usepackage{doi}
\usepackage{accents}

\newcommand{\ubar}[1]{\underaccent{\bar}{#1}}
\newcommand*{\at}[2][]{#1|_{#2}} 

\newtheorem{theorem}{Theorem}
\newtheorem{assumption}{Assumption}
\newtheorem{proposition}{Proposition}
\newtheorem{corollary}{Corollary}
\newtheorem{lemma}{Lemma}

\newtheorem{remark}{Remark}
\newcommand{\eps}{\varepsilon}
\newcommand{\hlambda}{\hat{\lambda}_{T}}
\newcommand{\hnu}{\hat{\nu}_{T}}
\newcommand{\htheta}{\hat{\theta}_{T,S}}
\newcommand{\hthetaq}{\hat{\theta}_{T,S,q}}
\newcommand{\ttheta}{\tilde{\theta}}

\newcommand{\hpsi}{\hat{\psi}_{T,S,i,j}}

\newcommand{\hpsiq}{\hat{\psi}_{T,S,q}}
\newcommand{\hTpsiq}{\hat{\psi}_{T,q}}
\newcommand{\hpsik}{\hat{\psi}_{T,S,i,j,k}}
\newcommand{\hTpsi}{\hat{\psi}_{T,i,j}}
\newcommand{\hTpsik}{\hat{\psi}_{T,i,j,k}}
\newcommand{\epsi}{\epsilon}
\newcommand{\Ex}{\textsf{E}}
\newcommand{\eq}{\,=\,}
\newcommand{\var}{\textsf{var}}
\newcommand{\cov}{\textsf{cov}}

\newcommand{\st}{{\raisebox{.1pt}{\textnormal{\footnotesize$*$}}}}
\newcommand{\St}{{\raisebox{1pt}{\textnormal{\footnotesize$*$}}}}

\newcommand{\sumt}{\sum_{t \eq 1}^T}
\newcommand{\sums}{\sum_{s \eq 1}^S}
\newcommand{\inn}{\,\in\,}
\newcommand{\dd}{\textsf{d}}
\newcommand{\DD}{\textsf{D}}
\newcommand{\FF}{\textsf{F}}
\newcommand{\hF}{\hat{\textsf{F}}}
\newcommand{\ff}{\textsf{f}}
\newcommand{\HH}{\textsf{H}}
\newcommand{\GG}{\textsf{G}}
\renewcommand{\gg}{\textsf{g}}
\newcommand{\hG}{\hat{\textsf{G}}}
\newcommand{\PP}{\textsf{P}}
\newcommand{\CC}{\textsf{C}}
\newcommand{\tCC}{\tilde{\textsf{C}}}
\newcommand{\hCTS}{\hat{\CC\,}_{T,S,i,j}}

\newcommand{\hGTSi}{\hat{\GG\,}_{T,S,i}}
\newcommand{\hGTSj}{\hat{\GG\,}_{T,S,j}}
\newcommand{\hCT}{\hat{\CC\,}_{T,i,j}}

\newcommand{\hFTi}{\hat{\FF\,}_{T,i}}
\newcommand{\hFTj}{\hat{\FF\,}_{T,j}}

\newcommand{\UU}{\textsf{U}}

\newcommand{\nn}{\nonumber}

\renewcommand{\theequation}{\arabic{section}.\arabic{equation}}

\renewcommand{\theassumption}{\arabic{section}.\arabic{assumption}}

\newcommand{\ii}{i\textnormal{,}\hspace*{0.025cm}j}
\newcommand{\ijn}{1 \leq i < j \leq n}
\newcommand{\lleq}{\,\leq\,}
\newcommand{\nala}{\nabla_{\hspace*{-.05cm}\lambda}}
\newcommand{\nanu}{\nabla_{\hspace*{-.05cm}\nu}}
\usepackage{harpoon}
\usepackage{chngcntr}
\usepackage{apptools}
\AtAppendix{\counterwithin{theorem}{section}}
\AtAppendix{\counterwithin{lemma}{section}}
\AtAppendix{\counterwithin{remark}{section}}
\usepackage{vmargin}
\setmarginsrb{3cm}{0cm}{3cm}{2cm}{1cm}{2cm}{2cm}{2cm}
\DeclarePairedDelimiter\floor{\lfloor}{\rfloor}

\newcommand\norm[1]{\lVert#1\rVert}
\RequirePackage[authoryear]{natbib}
\usepackage{accents}
\newcommand\munderbar[1]{%
  \underaccent{\bar}{#1}}
\usepackage{chngcntr}
\counterwithin{proposition}{section}
\counterwithin{corollary}{section}
\counterwithin{lemma}{section}
\counterwithin{theorem}{section}
\counterwithin{remark}{section}
\counterwithin{figure}{section}
\counterwithin{table}{section}

\renewcommand{\theassumption}{\arabic{assumption}}

\DeclareMathOperator*{\suplim}{\overline{\textsf{lim}}}
\DeclareMathOperator*{\ssup}{\textsf{sup}}
\DeclareMathOperator*{\iinf}{\textsf{inf}}
\DeclareMathOperator*{\mmin}{\textsf{min}}
\DeclareMathOperator*{\mmax}{\textsf{max}}
\usepackage{framed}
\allowdisplaybreaks
\RequirePackage[authoryear]{natbib}
\usepackage[affil-it]{authblk}

 \title{Estimation and Inference in Factor Copula Models with Exogenous Covariates\thanks{Financial support by Deutsche Forschungsgemeinschaft (DFG grant `Strukturbrüche und Zeitvariation in hochdimensionalen Abhängigkeitsstrukturen') is gratefully acknowledged. We would like to thank two anonymous referees for many helpful
comments and suggestions which improved the quality of the paper. Moreover, we are grateful for helpful discussions with Hans Manner, Sven Otto, and Florian Stark as well as for computational support from Sebastian Valet. Suggestions and comments made by participants of research seminars at the universities of Bonn, Cologne, and the Balearic Islands, as well as by participants of the following workshops are highly appreciated:  the 7th RCEA Time Series Workshop (virtual), the Symposium of the Hausdorff Center for Mathematics in Bonn 2021, the Workshop on High Dimensional Data Analysis at the University Carlos III de Madrid 2021 (virtual), the European Winter Meeting of the Econometric Society 2021 (virtual), the CFE 2021 (virtual), and the DAGStat 2022.}\vspace*{-.5cm}}
\author[1,a]{Alexander Mayer}
\author[2]{Dominik Wied\vspace*{-.5cm}}
\affil[1]{%
Department of Economics, Universit\`{a} Ca' Foscari Venezia}
\affil[2]{%
   Institute of Econometrics and Statistics, University of Cologne} 
\affil[a]{\small \textnormal{Corresponding author, email: \href{mailto:alexandersimon.mayer@unive.it}{alexandersimon.mayer@unive.it}}}
	\date{\today}

\begin{document}

\maketitle
 \vspace*{-1cm}
    \begin{abstract} 
 A factor copula model is proposed in which factors are either simulable or estimable from exogenous information. Point estimation and inference are based on a simulated methods of moments (SMM) approach with non-overlapping simulation draws. Consistency and limiting normality of the estimator is established and the validity of bootstrap standard errors is shown. Doing so, previous results from the literature are verified under low-level conditions imposed on the individual components of the factor structure. Monte Carlo evidence confirms the accuracy of the asymptotic theory in finite samples and an empirical application illustrates the usefulness of the model to explain the cross-sectional dependence between stock returns.
		\end{abstract}
  \normalsize
	\newpage

\section{Introduction}

Factor copula models have been successfully introduced as a means to cope with data of high cross-sectional dimensionality; see, e.g., \cite{kj13}, \cite{crts15}, and \cite{ohpa17}. The use of a latent factor structure offers an economically intuitive yet flexible way to multivariate modeling that parsimoniously handles commonly encountered characteristics of financial time series like, for example, the tail asymmetry and tail dependence described by \cite{hansen94}.  Recently, some research effort has been devoted to incorporate time variation and exogenous information to factor copula models; see, e.g., \cite{crts15}, \cite{ohpa18}, \cite{opetal20}, and \cite{kj20}. For example, \cite{ohpa18} and \cite{opetal20}, by utilizing the generalized autoregressive score (GAS) framework of  \cite{cretal13}, consider specifications with latent factors and time-varying loadings that may depend on exogenous information. We contribute to this literature by introducing a class of factor copula models with exogenous, (partly) observable \textit{factors}, an idea reminiscent of \cite{bertal05}, \cite{boetal09}, and \cite{stwa05}. Contrary to the above cited factor copula models, we take a step back and treat the, possibly group-specific, loadings as time-invariant constants and the SMM estimator employed here is build on the \textit{un}conditional copula---a concession in the name of tractability that frees us from the necessity of specifying parametric marginals [e.g., \cite{ohpa18}] or a closed form likelihood of the copula [e.g., \cite{opetal20}] and thereby allows for a large variety of `covariate-augmented' factor copulas, nesting the model \cite{ohpa17} as a special case.  

Since the copula likelihood is rarely available in closed form for the model class considered here, an SMM framework for estimation and inference is proposed which uses the general principles outlined by \cite{ohpa13}. Our main contribution is a novel distinction between simulable factors and factors that are estimable from exogenous information. Following the seminal SMM literature of \cite{mcf89}, \cite{papo89}, and \cite{lee92}, we exploit the benefits from non-overlapping simulation draws.  The incorporation of exogenous covariates considerably complicates the development of an asymptotic theory as many arguments made by \cite{ohpa13} do not apply. Nevertheless, we show that all technical hurdles can be overcome by combining recent developments from copula empirical process theory [see, e.g., \cite{buvo13}, \cite{betal17}, and \cite{neumetal19}] with a seminal result for extremum estimation with nonsmooth objective function due to \cite{nemc94}. In consequence, consistency, limiting normality, and validity of bootstrap standard errors are established. In doing so, we derive the stochastic equicontinuity of the objective function from primitive conditions on the distributional characteristics of the factor structure using the functional central limit theorem (FCLT) of \cite{anpo94} for \(\alpha\)-mixing triangular arrays.  The theory developed here verifies earlier equicontinuity results from the literature that made use of high-level conditions; see, e.g., \cite{ohpa13}, \cite{manetal19}, and \cite{manetal20}. Since stochastic equicontinuity is an essential ingredient of the asymptotic theory that links pointwise and uniform properties, more primitive conditions are of utmost interest. An application to dependence modeling of a cross-section of stock returns of eleven financial companies illustrates the theoretical results and highlights how the incorporation of estimable factors can help to achieve improvements in model performance.

The remainder of this paper is organized as follows. Section \ref{copula_model} introduces the model. The main results for SMM estimation and inference are contained in  Section \ref{copula_estimation}. A small Monte Carlo exercise is conducted in Section \ref{copula_mc} and an empirical application can be found in Section \ref{copula_empapp}.  Section \ref{copula_con} briefly summarizes and concludes the paper.

\section{Model}\label{copula_model}

Our aim is to capture the dependence structure among the cross-sectional entities of the \(n \times 1\) vector of financial assets \(Y_{t} \coloneqq (Y_{1,t},\dots,Y_{n,t})'\) in time-period \(t \in \{1,\dots,T\}\),  conditional on the available information \(\mathcal{F}_{t} \coloneqq \sigma(\{Y_{j}^\St,Y_{j-1}: j \leq t\})\), where \(Y_{t}^\St\) represents a vector of exogenous regressors. The number of financial assets \(n\) might be large but is assumed finite. If the marginal conditional distributions \(Y_{i,t} \mid \mathcal{F}_{t} \sim \HH_{i,t}\) are continuous, we can follow \cite{pat06} and uniquely decompose the joint conditional distribution \(Y_t \mid \mathcal{F}_{t} \sim \HH_t\) into its \(n\) margins and a copula function \(\CC_t: [0,1]^n \mapsto [0,1]\), where \(\CC_t(\cdot)\) completely describes the dependence conditionally on \(\mathcal{F}_{t}\); i.e., \(\HH_t(x_1,\dots,x_n) = \CC_t\{\HH_{1,t}(x_1),\dots,\HH_{n,t}(x_n)\}\), \(x_i \in \mathbb{R},\) \(i \in \{1,\dots,n\}\).
Following, among others, \cite{chef06}, \cite{ohpa13}, and \citet[Section 2.1]{fapa14}, we assume for the \(n\) assets parametric location-scale specifications
of the form
\begin{equation}\label{modelY}
Y_{i,t} =  \mu_{1,i}(\mathcal{F}_t,\lambda_0) + \mu_{2,i}(\mathcal{F}_t,\lambda_0)\eta_{i,t},
\end{equation}
where, for each \(i \in \{1,\dots,n\}\), \(\{\eta_{i,t}: t \geq 1\}\) are \textit{i.i.d.} innovations independent of \(\mathcal{F}_{t}\), while \(\mu_{1,i}\) and \(\mu_{2,i}\) are \(\mathcal{F}_{t}\)-measurable parametric specifications of the conditional mean \(\mu_{1,i}(\mathcal{F}_t,\lambda) = \Ex[Y_{i,t} \mid \mathcal{F}_t]\) and the conditional standard deviation \(\mu_{2,i}(\mathcal{F}_t,\lambda) = \sqrt{\var[Y_{i,t} \mid \mathcal{F}_t]}\) that are known up to the true \(r \times 1\) parameter vector \(\lambda = \lambda_{0} \in \Lambda_{0} \subset \mathbb{R}^r\). In particular, we assume that
\begin{equation}\label{modelYa}
\mu_{j,i}(\mathcal{F}_t,\lambda) = \mu_{j,i}(R_t(\lambda),\lambda),\;\; j \in \{1,2\},\,i \in \{1,\dots,n\},
\end{equation}
where, for any \(\lambda \in \Lambda\), the \(p_{\textsf{R}} \times 1\) random vector \(R_t(\lambda)\) is \(\mathcal{F}_t\)-measurable with components that may parametrically depend on the entire history \(\mathcal{F}_t\) through \(\lambda\). For example, this model class includes nonlinear autoregressions (AR) with nonlinear autoregressive conditional heteroskedasticity (ARCH), in which case  the dependence of \(R_{t}(\lambda)\) on \(\lambda\) is superfluous, as well as nonlinear generalized autoregressive conditional heteroskedasticity (GARCH) as illustrated below for the GJR model of \cite{glostetal93}.

\noindent \textbf{Example:} \textit{Consider the \textnormal{AR(1)} model \(Y_{i,t} =  \gamma_i Y_{i,t-1} + \mu_{2,i,t}\eta_{i,t}\), with conditional heteroskedasticity
\(
\mu_{2,i,t}^2 =  \omega_i + (\beta_i + \eta_i1\{\eps_{i,t-1}(\gamma_i) < 0\})\eps_{i,t-1}^2(\gamma_i) + \alpha_i  \mu_{2,i,t-1}^2
\)
for \(\eps_{i,t}(\gamma_{i}) \coloneqq Y_{i,t} -  \gamma_{i}Y_{i,t-1}\). In accordance with our notation, let \(\lambda = (\gamma_1,\dots,\gamma_n,\alpha_1,\dots,\alpha_n,\omega_1,\dots,\omega_n,\beta_1,\dots,\beta_n,\eta_1,\dots,\eta_n)'\) be the \(r\times 1\) parameter vector collecting all unknowns for \(r = 5n\).  Following the discussion in \textnormal{\cite{yang96}}, by suitably restricting \(\Lambda_{0} \subset \mathbb{R}^{r}\), this specification can be cast in form of \eqref{modelY} and \eqref{modelYa} if we set \(\mu_{1,i}(\mathcal{F}_t,\lambda) = \gamma_iY_{i,t-1}\),  \(R_t(\lambda) = (Y_{t-1}',\tilde{R}_{t}(\lambda)')'\), where \(\tilde{R}_{t}(\lambda) = (\tilde{R}_{1,t}(\lambda),\dots,\tilde{R}_{n,t}(\lambda))'\), with \(\tilde{R}_{i,t}(\lambda) = \sum_{j \eq 1}^\infty \alpha_{i}^{j-1}g_i\{Y_{i,t-j}- \gamma_iY_{i,t-1},\eta_{i}\}\), \(g(x,a) = x^2+ a1\{x<0\}x^2\), and \(\mu_{2,i}(R_t(\lambda),\lambda) = \beta_i \tilde{R}_{i,t}(\lambda) + \omega_i/(1-\alpha_i)\).}

\noindent Since we assume that individually \(\eta_{i,t}\) are independent of \(\mathcal{F}_t\) and, by Eq. \eqref{modelY}, \(\eta_{i,t} = \{Y_{i,t}-\mu_{1,i}(\mathcal{F}_t,\lambda_0)\}/\mu_{2,i}(\mathcal{F}_t,\lambda_0)\), we can --assuming continuous margins \(\FF_i(\eta_{i,t} \leq x_i) \coloneqq \PP(\eta_{i,t} \leq x_i)\), \(i \in \{1,\dots,n\}\)-- rephrase the conditional joint distribution of \(Y_t\) in terms of the innovation ranks \(V_{i,t} \coloneqq  \FF_{i}(\eta_{i,t})\) as
\begin{equation}\label{Ycop_cond}
\begin{split}
\HH_t(x_1,\dots,x_n) = \,& \CC_t\{\FF_{1}(x_1),\dots,\FF_{n}(x_n)\}, \\
\CC_t(u_1,\dots,u_n) = \,& \PP(V_{1,t} \leq u_1,\dots,V_{n,t} \leq u_n \mid \mathcal{F}_t),
\end{split}
\end{equation}
\(u_i \in [0,1]\), \(i \in \{1,\dots,n\}\). While the effect of \(\mathcal{F}_{t}\) on the margins has been completely removed by the use of the location-scale model \eqref{modelY}, the \textit{joint} cross-sectional distribution of \(\eta_{t} \coloneqq (\eta_{1,t},\dots,\eta_{n,t})'\) is allowed to depend on \(\mathcal{F}_{t}\) through some `exogenous' vector \(Z_t\). That is, we assume that
\begin{equation}
\CC_t(u_1,\dots,u_n) = \PP(V_{1,t} \leq u_1,\dots,V_{n,t} \leq u_n \mid \mathcal{F}_t) = \PP(V_{1,t} \leq u_1,\dots,V_{n,t} \leq u_n \mid Z_t).
\end{equation}
We do not model the conditional copula directly but we assume that the unconditional copula
\begin{equation}\label{copula:filtered}
\CC(u_1,\dots,u_n) = \PP(V_{1,t} \leq u_1,\dots,V_{n,t} \leq u_n),
\end{equation}
can be generated from an auxiliary factor model \textit{via} 
\begin{equation}\label{Gfactora}
\GG(x_1,\dots,x_n) = \CC\{G_{1}(x_1),\dots,G_{n}(x_n)\},\; x_i \in \mathbb{R},
\end{equation}
where \(G_{i}(x_i) \coloneqq \PP(X_{i,t} \leq x_i)\) represents the \(i\)-th margin of  
\begin{equation}
X_{i,t} =   a_{0,i}'F_t + b_{0,i}'Z_t + \eps_{i,t}\label{Xfactora}
\end{equation}
and \(\GG(x) \coloneqq \PP(X_{1,t}\leq x_1,\dots,X_{n,t} \leq x_n)\), \(x \coloneqq (x_1,\dots,x_n)' \in \mathbb{R}^n\), is the corresponding joint distribution. As defined below, \(F_t\) and \(\eps_{i,t}\) denote latent factors and the idiosyncratic component, respectively. It is crucial to stress that the margins \(G_i\), \(i \in \{1,\dots,n\}\), can differ from the univariate distributions of the observed data and are not of interest here. Rather, Eq. \eqref{Xfactora} serves as a means to generate the copula \(\CC\) that determines the joint distribution. Since any effect of \(\mathcal{F}_t\) on the margins has been filtered out while the conditional distribution is only affected by \(\mathcal{F}_t\) through the regressor, the unconditional copula obtains directly by taking the expectation with respect to \(Z_t\), i.e. \(\CC(\cdot) = \Ex[\CC_t(\cdot)]\).

\cite{ohpa18} or \cite{opetal20} consider a related specification. Contrary to our approach, however, they model the conditional copula directly, i.e. they consider a conditional copula \(\CC_t = \CC(\theta_{0,t})\) indexed by a time-varying copula parameter \(\theta_{0,t} \coloneqq \theta_0(Z_t)\) that is driven by GAS-dynamics so that \(Z_t = (\eta_{1}',\dots,\eta_{t-1}')'\). We, on the other hand, assume that all components of the factor model including \(Z_t\) are \(i.i.d.\). To make these notions precise, Assumption \ref{asu-B} formalizes the characteristics of the factor model. It constitutes a naturally extension of \cite{ohpa17}.  
\renewcommand{\theassumption}{A}
\begin{assumption} \textcolor[rgb]{1,1,1}{.}\label{asu-B} \normalfont 
\begin{enumerate}[label= (A\arabic*)]
\item \(\{Z_t: t \geq 1\}\), \(\{F_t: t \geq 1\}\) and \(\{\eps_t: t \geq 1\}\) are mutually independent \(i.i.d.\) sequences with \(\eps_{i,t} \perp \eps_{j,t}\), \(i,j \in \{1,\dots,n\}\) and \(F_{t,i} \perp F_{t,j}\), \(i,j \in \{1,\dots,p_\alpha\}\);
\item \(\{X_t,\eta_t: t\geq 1\}\) is an \textit{i.i.d.} sequence.
\end{enumerate}
\end{assumption} 

The peculiar feature of Eq. \eqref{Xfactora}, and the main contribution of this paper, is the distinction between \textit{simulable} and \textit{observable} factors: while \(F_t \coloneqq (F_{t,1},\dots,F_{t,p_\alpha})'\) is a \(p_\alpha \times 1\) vector of latent random variables with known parametric distribution, the \(p_\beta \times 1\) vector \(Z_t \coloneqq (Z_{t,1},\dots,Z_{t,p_\beta})'\)  can be recovered from observed data based on econometric tools. Therefore, \(Z_t\) is also referred to as the estimable factor.  More specifically, both $F_t$ and \(\eps_{i,t}\) are \textit{i.i.d.} with parametric distributions 
\[
\DD_\eps(x;\,\delta_0) \coloneqq    \PP(\eps_{i,t} \leq x),\;\;\;\DD_{F,j}(x;\,\gamma_{0,j}) \coloneqq   \PP(F_{t,j} \leq x),\,j \in \{1,\dots, p_\alpha\},
\]
which are partially known up to the \(p_\delta \times 1\) vector \(\delta_0 \coloneqq (\delta_{0,1},\dots,\delta_{0,p_\delta})'\) and the \(p_\alpha p_\gamma \times 1\) vector \(\gamma_0 \coloneqq (\gamma_{0,1}',\dots,\gamma_{0,p_\alpha}')'\), with  \(\gamma_{0,j} \coloneqq (\gamma_{0,j,1},\dots,\gamma_{0,j,{p_\gamma}})'\), respectively. On the other hand, the distribution of the vector \(Z_{t}\)  is unknown but we assume that its components \(Z_{t,j}\) can be represented as \textit{i.i.d.} innovations of an observable \(\mathcal{F}_t\)-measurable processes \(W_{t,j}\) given by
\begin{equation}
Z_{t,j} = W_{t,j} - \sigma_j(\mathcal{W}_t,\nu_{0}),\;\;j \in \{1,\dots,p_{\beta}\}, \label{model-WS1} 
\end{equation}
where the measurable functions \(\sigma_j(\cdot,\nu)\) are known up to the \(m \times 1\) vector \(\nu = \nu_{0} \in \mathcal{V}_{0} \subset \mathbb{R}^m\); the \textit{i.i.d.} innovations \(Z_{t,j}\) are independent of the history \(\mathcal{W}_t \subset \mathcal{F}_{t-1}\). Similar to the specification of Eq. \eqref{modelYa}, we assume that \(\sigma_j(\mathcal{W}_t,\nu) = \sigma_j(M_t(\nu),\nu)\), where the \(p_{\textsf{M}} \times 1\) vector \(M_{t}(\nu)\) comprises short-range dependent covariates, possibly including lagged dependent variables, whose components may be parametrically generated from the complete history \(\mathcal{W}_t\). We cannot allow for both time-varying conditional means \textit{and} variances to ensure that the limiting distribution of the SMM estimator is unaffected by the first step estimation of \(\nu_0\). Note, however, that several models with time-varying conditional variance that obey \(\tilde{W}_t = \tilde{\sigma}(M_t(\nu),\nu)\tilde{Z}_t\), \(\ssup_{x,y}\tilde{\sigma}(x,y) > 0\), can be cast in form of \eqref{model-WS1} by setting \(W_t \coloneqq \textsf{log} |\tilde{W}_t|\), \(\sigma_t(M_t(\nu),\nu) \coloneqq \textsf{log}\,\tilde{\sigma}_t(M_t(\nu),\nu)\), and \(Z_t \coloneqq \textsf{log} |\tilde{Z}_t|\).\footnote{The argument is inspired by \citet{genestetal07}, who propose this transformation in their Example 1 to ensure a nuisance-free distribution of the BDS-type test studied there; see also \cite{carpetal07} for a study of the BDS test based on the logarithm of absolute GARCH(1,1)-residuals.}

As in \citet[Section 4.2]{ohpa17} and \citet[Section 2.1.1]{opetal20}, it is assumed that the factor loadings \(a_{0,i} \coloneqq (a_{0,i,1},\dots,a_{0,i,p_\alpha})'\) and \(b_{0,i}  \coloneqq (b_{0,i,1},\dots,b_{0,i,p_\beta})'\) can be grouped into a small number of \(Q\) group-specific coefficients \(\alpha_{0,q} \coloneqq (\alpha_{0,q,1},\dots,\alpha_{0,q,p_\alpha})'\) and \(\beta_{0,q} \coloneqq (\beta_{0,q,1},\dots,\beta_{0,q,p_\beta})'\), \(q \in \{ 1,\dots,Q\}\). Put differently, there exists a finite collection of disjoint sets  \(\{\mathcal{G}_1,\dots,\mathcal{G}_Q\}\) partitioning the cross-sectional index set  \(\{1,\dots,n\}\) such that \(a_{0,i} = a_{0,j} = \alpha_{0,q}\) and \(b_{0,i} = b_{0,j} = \beta_{0,q}\) for any  \(i,j \in \mathcal{G}_q\), \(q \in \{ 1,\dots,Q\}\). Importantly, this `block-equidependent' factor structure implies that the number of latent marginals needed to be specified  reduces from \(n\) to \(Q\)  distinct distributions \(\GG_1,\dots,\GG_Q\), say, so that \(G_{i}(x) = G_{j}(x) \eqqcolon \GG_q(x)\) for any  \(i,j \in \mathcal{G}_q\), \(q \in \{ 1,\dots,Q\}\). Throughout, the group assignment is assumed to be known.

\section{SMM-based Estimation}\label{copula_estimation}

The object of interest is the \(p \times 1\) vector \(\theta_0 \coloneqq (\alpha_{0,1}',\beta_{0,1}',\dots,\alpha_{0,Q}',\beta_{0,Q}',\gamma_0',\delta_0')' \in \Theta \subseteq \mathbb{R}^p\), which, in view of the latent factor structure \eqref{Gfactora}, collects all \(p \coloneqq Q(p_\alpha + p_\beta) +p_\alpha p_\gamma + p_\delta\) unknown copula parameters. A different parameter vector \(\theta \coloneqq (\alpha_{1}',\beta_{1}',\dots,\alpha_{Q}',\beta_{Q}',\gamma',\delta')' \in \Theta\) gives rise to an alternative factor structure
\begin{equation}\label{Xfactorb}
X_{i,t}(d_i) \coloneqq  a_i'F_t(\gamma) + b_i'Z_t + \eps_{i,t}(\delta),
\end{equation}
say, where the notational conventions \(X_{i,t}(d_i)\), \(F_t(\gamma) \coloneqq (F_{t,1}(\gamma_1),\dots,F_{t,p_\alpha}(\gamma_{p_\alpha}))'\), and \(\eps_{i,t}(\delta)\) are used to make the dependence of the various quantities on the \((p_\alpha + p_\beta + p_\alpha p_\gamma + p_\delta) \times 1\) vector \(d_i \coloneqq (a_i',b_i',\gamma',\delta')'\) explicit; e.g., \(\eps_{i,t}(\delta) \sim \DD_\eps(\delta)\) and \(F_{t,j}(\gamma_j) \sim \DD_{F,j}(\gamma_j)\), \(j \in \{1, \dots, p_\alpha\}\). The block-equidependent design ensures that 
\begin{equation}\label{loading_block}
d_i = d_j = \theta_q, \quad \forall\,i,j \in \mathcal{G}_q,\; q \in \{1,\dots,Q\},
\end{equation}
where the \((p_\alpha + p_\beta + p_\alpha p_\gamma + p_\delta) \times 1\) vector \(\theta_q \coloneqq (\alpha_q',\beta_q',\delta',\gamma')'\)  contains the parameters specific to the \(q\)-th group. Thus, with a slight abuse of notation, \({\displaystyle\theta =\cup_{q \eq 1}^Q \theta_q}\). For each \(\theta \in \Theta\), Eq. \eqref{Xfactorb} generates a differently parametrized copula 
\begin{equation}\label{cop_X}
\begin{split}
\CC(u_1,\dots,u_n;\theta) \coloneqq \,& \PP\{U_{1,t}(d_1) \leq u_1, \dots, U_{n,t}(d_n)\leq u_n\},\\
 U_{i,t}(d_i) \coloneqq \,&  G_i(X_{i,t}(d_i);d_i),\\
 G_i(x_i;d_i) \coloneqq \,& \PP\{X_{i,t}(d_i) \leq x_i\},\,u_i \in [0,1],\,x_i \in \mathbb{R}. 
\end{split}
\end{equation}
 Importantly, due to the block-equidependent design and Eq. \eqref{loading_block}, we have for any two cross-sectional indices belonging to the same group 
\[
G_{i}(x;d_i) = G_{j}(x;d_j) \eqqcolon \GG_q(x;\theta_q), \quad i,j \in \mathcal{G}_q,\, q \in \{1,\dots,Q\},
\]
say, so that \(G_{i}(X_{i,t}(d_i);d_i) = U_{i,t}(d_i) = U_{i,t}(\theta_q) = \GG_q(X_{i,t}(\theta_q); \theta_q)\). 

The simulation-based estimation uses Assumption \ref{asu-A} below to estimate the true value of \(\theta \in \Theta\) given by \(\theta_0 = \cup_{q \eq 1}^Q \theta_{0,q}\), \(\theta_{0,q} = (\alpha_{0,q}',\beta_{0,q}',\gamma_0',\delta_0')'\).

\renewcommand{\theassumption}{B}
\begin{assumption}\label{asu-A}  \textcolor[rgb]{1,1,1}{.} \normalfont
\begin{enumerate}[label= (B\arabic*)]
\item \label{asuA-2} \(\CC(u_1,\dots,u_n) = \CC(u_1,\dots,u_n;\theta)\) uniformly in \((u_1,\dots,u_n)' \in [0,1]^n\) if and only if \(\theta = \theta_0\), where \(\theta_0 \in \Theta \subset \mathbb{R}^p\). The parameter space \(\Theta\) is compact.
\item \label{asu-A1} 
\begin{enumerate}[label=($\roman*$)]
\item The joint distribution \(\FF(x_1,\dots,x_n) = \PP(\eta_{1,t} \leq x_1,\dots,\eta_{n,t} \leq x_n)\) is continuous with continuous marginal distributions \(\FF_i(x_i)\), \(x_i \in \mathbb{R}\), \(i \in \{1,\dots,n\}\).
\item The joint distribution \(\GG(x_1,\dots,x_n;\theta)= \PP(X_{1,t}(d_1) \leq x_1,\dots,X_{n,t}(d_n) \leq x_n)\) is continuous with continuous marginal distributions \(G_i(x_i;d_i)\), \(x_i \in \mathbb{R}\), \(i \in \{1,\dots,n\}\), uniformly in \(\theta \in \Theta\).
\end{enumerate}
\end{enumerate}
\end{assumption}
Assumption \ref{asu-A} formalizes the introductory notion of a factor copula; i.e., the unknown copula \(\CC(u_1,\dots,u_n)\)  can be generated by the latent factor structure for a suitable choice of \(\theta = \theta_0 \in \Theta\).

\subsection{Independent Simulations}

Akin to the `independent simulation' scheme known from classical SMM estimation [see, e.g. \cite{mcf89}, \cite{papo89}, and \cite{lee92}],  we generate for a given candidate value \(\theta \in \Theta \subset \mathbb{R}^p\)   a random sample \(\{(F_{t,s}(\gamma)',\eps_{i,t,s}(\delta))': i=1,\dots,n,\,t = 1,\dots,T,\,s = 1,\dots,S\}\) to obtain a version of the auxiliary factor model
\begin{equation}
X_{i,t,s}(d_i)  \coloneqq  a_{i}'F_{t,s}(\gamma) + b_i'Z_t + \eps_{i,t,s}(\delta),\label{simXa}
\end{equation}
where \(F_{t,s}(\gamma) \coloneqq (F_{t,s,1}(\gamma_1),\dots,F_{t,s,p_\alpha}(\gamma_{p_\alpha}))'\). Hence, we sample for each time period \(t\) a new batch of \(S\) random variables from \(\DD_\eps\)  and   \(\DD_{F,j}\), \(j \in \{1,\dots,p_\alpha\}\).  More specifically, let \(\DD_\eps^{-1}\) and \(\DD_{F,j}^{-1}\) denote the inverse distribution functions of \(\eps_t\) and \(F_{t,j}\), \(j \in \{1,\dots,p_\alpha\}\), respectively. We may then write \(\eps_{i,t,s}(\delta) \coloneqq \DD_\eps^{-1}(\eps_{i,t,s}^\st;\delta)\) and \(F_{t,s}(\gamma) \coloneqq  \DD_F^{-1}(F_{t,s}^\St;\gamma)\), with \(\DD_F^{-1}(F_{t,s}^\St;\gamma) \coloneqq (\DD_{F,1}^{-1}(F_{t,s,1}^\St;\gamma_1),\dots,\DD_{F,p_\alpha}^{-1}(F_{t,s,p_\alpha}^\St;\gamma_{p_\alpha}))'\),  where  \(F_{t,s}^\St \coloneqq (F_{t,s,1}^\St, \dots, F_{t,s,p_\alpha}^\St)'\) and \(\eps_{i,t,s}^\st\) denote independent draws of \textit{i.i.d.} standard uniform random variates which are drawn once. Note that for our estimation procedure we hold the underlying random draws \(F_{t,s}^\St\) and \(\eps_{i,t,s}^\st\) fixed while \(\theta\) is allowed to vary over the compact set \(\Theta\). This is important to ensure uniform convergence of simulated moments and to facilitate convergence of the numerical optimization routine used to find \(\theta_0\); see \citet[p. 29]{gofo97} and \citet[p. 1048]{papo89} for further remarks.  Since \(Z_{t}\) might be unobservable, we will replace the unknown innovation with \(\hat{Z}_{t}(\hnu) \coloneqq (\hat{Z}_{t,1}(\hat{\nu}_{T}),\dots,\hat{Z}_{t,p_\beta}(\hat{\nu}_{T}))'\),  where \(\hat{Z}_{t,j}(\nu) \coloneqq W_{t,j}-\sigma_j(M_t(\nu),\nu)\), \(j \in \{1,\dots,p_\beta\}\),  represents  the generalized residual. The estimator \(\hnu\) is assumed to be \(\sqrt{T}\)-consistent for the \(m \times 1\) vector \(\nu_{0}\) satisfying certain mild regularity conditions outlined below; for example, in the empirical application maximum likelihood estimation is used. Therefore, a feasible counterpart of Eq. \eqref{simXa} is obtained from
\begin{equation}
\hat{X}_{i,t,s}(d_i,\hnu) \coloneqq   a_{i}'F_{t,s}(\gamma) + b_{i}'\hat{Z}_t(\hnu)+ \eps_{i,t,s}(\delta). \label{sim-S1} 
\end{equation}

\subsection{The Estimator}

Throughout, the cross-sectional dimension \(n\) might be large but is considered fixed, while asymptotics are carried out as \(T \rightarrow \infty\); the number of simulation draws \(S\) can either be fixed or a function of \(T\) such that \(S \coloneqq S(T) \rightarrow \infty\) as \(T \rightarrow \infty\). For the sake of brevity we report only results for the latter case.

Similar to \cite{ohpa13}, estimation aims at minimizing the difference between empirical and simulated rank dependence measures that only depend on the unknown bivariate marginal copulae. Importantly, Assumption \ref{asu-A}  implies also the equivalence at \(\theta = \theta_0\) between each of the \(n(n-1)/2\) bivariate marginals copulae of the joint copula from Eq. \eqref{copula:filtered}, given by  \(\CC_{i,j}(u_i,u_j) \coloneqq \PP(V_{i,t} \leq u_i, V_{j,t} \leq u_j)\), and the bivariate marginals of the joint factor copula from Eq. \eqref{cop_X}, given by \(\CC_{i,j}(u_i,u_j;d_i,d_j) \coloneqq \PP(U_{i,t}(d_i) \leq u_i, U_{j,t}(d_j) \leq u_j)\), \(\ijn\). By block-equidependence, 
\[
\CC_{i,j}(u_i,u_j;d_i,d_j) \eqqcolon \CC_{q}(u_i,u_j;\theta_{q}), \quad i,j \in \mathcal{G}_q,\; q \in \{1,\dots,Q\}.
\]
Put differently, the number of distinct marginal copulae reduces from \(n(n-1)/2\) to \(Q\) block-specific copulae \(\CC_1(\theta_{1}),\dots,\CC_Q(\theta_{Q})\) for which \(\CC_{i,j}(u_i,u_j) =  \CC_{q}(u_i,u_j;\theta_{q})\) if \(i,j \in \mathcal{G}_q\), \(q \in \{1,\dots,Q\}\), and \(\theta_q = \theta_{0,q}\). To illustrate the main idea behind the following SMM estimator, suppose \(i,j \in \mathcal{G}_q\) for some \(q \in \{1,\dots,Q\}\) and introduce, for some estimator \(\sqrt{T}\)-consistent estimator \(\hlambda\) of \(\lambda_0\) satisfying some regularity conditions outlined below, the following two \(\ell \times 1\) vectors  
\begin{equation}
\begin{split}
\hTpsi(\hlambda) \coloneqq  \,& (\hat{\psi}_{T,i,j,1}(\hlambda),\dots,\hat{\psi}_{T,i,j,\ell}(\hlambda))' \\
\hpsi(\theta_q,\hnu) \coloneqq \,& (\hat{\psi}_{T,S,i,j,1}(\theta_q,\hnu),\dots,\hat{\psi}_{T,S,i,j,\ell}(\theta_q,\hnu))',
\end{split}
\end{equation}
which collect bivariate dependence measures like, for example, Spearman's \(\rho\) , Blomqvist's \(\beta\),  Gini's \(\gamma\), or the measures of quantile dependence used by \citet[p. 691]{ohpa13}. Formally, these statistics can be expressed with the help of a suitable collection of bivariate functions \(\{\varphi_k: [0,1]^2 \mapsto \mathbb{R}, 1\leq k \leq \ell\}\) as follows 
\begin{equation}\label{rankstat}
\begin{split}
\hTpsik(\hlambda) \coloneqq  \;&\frac{1}{T}\sum_{t \eq 1}^T\varphi_k(\hat{V}_{i,t}(\hlambda),\hat{V}_{j,t}(\hlambda)) \\
\hpsik(\theta_q,\hnu) \coloneqq \;&\frac{1}{TS}\sumt\sums\varphi_k(\hat{U}_{i,t,s}(\theta_q,\hnu),\hat{U}_{j,t,s}(\theta_q,\hnu)), 
\end{split}
\end{equation}
where $\hat{V}_{i,t}(\hlambda)$ and $\hat{U}_{i,t,s}(\theta_q,\hnu)$ represent the rank of $\hat{\eta}_{i,t}(\hlambda) \coloneqq \{Y_{i,t}-\mu_{1,i}(\mathcal{F}_t,\hlambda)\}/\mu_{2,i}(\mathcal{F}_t,\hlambda)$ among $\{\hat{\eta}_{i,t}(\hlambda): t = 1,\dots,T\}$ and the rank of $\hat{X}_{i,t,s}(\theta_q,\hnu)$ among $\{\hat{X}_{i,t,s}(\theta_q,\hnu): t = 1,\dots,T;s=1,\dots,S\}$, respectively.

The \(\ell\) different bivariate dependence measures are then aggregated according to the group-specific factor structure. To provide some intuition, note that \(\hTpsik(\hlambda)\) and \(\hpsik(\theta_q,\hnu)\), \(k \in \{1,\dots,\ell\}\), can be viewed as sample estimates of the population statistics \(\Ex[\varphi_k(V_{i,t},V_{j,t})]\) and \(\Ex[\varphi_k(U_{i,t,s}(\theta_q),U_{j,t,s}(\theta_q))]\).
These statistics depend only on the bivariate copulae \(\CC_{q}(u_i,u_j)\) and  \(\CC_{q}(u_i,u_j;\theta_q)\), which, due to the block-equidependence, exhibit within-group homogeneity; i.e., each of the \(\ell\) statistics depends on the cross-sectional index set \(\{1,2,\dots,n\}\) only \textit{via} the group-identifier \(q \in \{1,\dots,Q\}\):
\begin{equation} \label{pop_gq}
\begin{split}
\psi_{q,k}(\theta) \coloneqq \,&\Ex[\varphi_k(U_{i,t,s}(\theta_q),U_{j,t,s}(\theta_q))] = \int_{[0,1]^2} \varphi_k(u_i,u_j) \textsf{d} \CC_{q}(u_i,u_j;\theta_q)\\
\psi_{q,k}  \coloneqq \,& \Ex[\varphi_k(V_{i,t},V_{j,t})] = \int_{[0,1]^2} \varphi_k(u_i,u_j) \textsf{d} \CC_{q}(u_i,u_j),\;\;k \in \{1,\dots,\ell\},\; \forall\, i,j \in \mathcal{G}_q.
\end{split}
\end{equation}
Therefore, the following aggregation scheme of bivariate dependence measures is justified
\begin{equation}\label{aggPsi}
\begin{split}  
\hpsiq(\theta,\hnu) \coloneqq  \,& \frac{1}{\displaystyle{|\mathcal{G}_q| \choose 2}}\mathop{\sum\sum}\limits_{\substack{\ijn\\i,j \inn \mathcal{G}_q}}\hpsi(\theta_q,\hnu)\\
\hTpsiq(\hlambda) \coloneqq  \,& \frac{1}{\displaystyle{|\mathcal{G}_q| \choose 2}}\mathop{\sum\sum}\limits_{\substack{\ijn\\i,j \inn \mathcal{G}_q}}\hTpsi(\hlambda),\; q \in \{1,\dots,Q\}.
\end{split} 
\end{equation}
We thus obtain the  \(\bar{\ell} \times 1\), \(\bar{\ell} \coloneqq Q\ell\), vector of empirical  dependence measures 
\[
\hat{\psi}_{T}(\hlambda) \coloneqq (\hat{\psi}_{T,1}(\hlambda)',\dots,\hat{\psi}_{T,Q}(\hlambda)')'
\] 
and the \(\bar{\ell} \times 1\) vector of simulated dependence measures
\[
\hat{\psi}_{T,S}(\theta,\hnu) \coloneqq (\hat{\psi}_{T,S,1}(\theta_1,\hnu)',\dots,\hat{\psi}_{T,S,Q}(\theta_Q,\hnu)')',
\] respectively. Following the literature on extremum estimators [see, e.g., \citet[Section 7]{nemc94}], we define, for some stochastically bounded and positive-definite weight matrix \(\hat{L}_{T,S}\), an SMM estimator \(\hat{\theta}_{T,S}\) as an estimator that  minimizes the objective function 
\[\hat{A}_{T,S}(\theta,\hlambda,\hnu) \coloneqq \hat{\Psi}_{T,S}(\theta,\hlambda,\hnu)'\hat{L}_{T,S}\hat{\Psi}_{T,S}(\theta,\hlambda,\hnu),\]
with \(\hat{\Psi}_{T,S}(\theta,\hlambda,\hnu) \coloneqq \hat{\psi}_{T}(\hlambda) - \hat{\psi}_{T,S}(\theta,\hnu)\), in the sense that
\begin{equation}\label{SMMest}
\hat{A}_{T,S}(\hat{\theta}_{T,S},\hlambda,\hnu) \leq  \inf\limits_{\theta \inn \Theta} \hat{A}_{T,S}(\theta,\hlambda,\hnu) + o_p(1/T).
\end{equation}

\subsection{Limiting Normality}

As pointed out by \cite{ohpa13},  the objective function is non-differentiable and, in general, does not posses a population counterpart in known closed form. Thus, some care is required in deriving the asymptotic distribution of \(\htheta\). Due to the mutual dependence on the covariate, \(\hat{\psi}_{T}(\hlambda)\) and \(\hat{\psi}_{T,S}(\theta,\hnu)\) are not independent, which considerably complicates the analysis and precludes a direct application of the arguments developed by the aforementioned authors. To shed some light, let us recall  from \cite{nemc94} that two crucial high-level conditions for asymptotic normality of \(\sqrt{T}(\htheta-\theta_0)\) are (1) limiting normality of the normalized sample `moments' \(\sqrt{T}\hat{\Psi}_{T,S}(\theta,\hlambda,\hnu)\) evaluated at \(\theta = \theta_0\)  and (2) the stochastic equicontinuity of the map \(\theta \mapsto \sqrt{T}\hat{\Psi}_{T,S}(\theta,\hlambda,\hnu)\). Both conditions are shown to be closely tied to the limiting behaviour of the triangular-array empirical process
\begin{equation}\label{Bemp}
\begin{split}
\hat{\mathbb{B}}_{T,S,i,j}(&u_i,u_j;\theta_q,\hlambda,\hnu)\\
 \coloneqq  \,&\frac{1}{\sqrt{T}}\sumt \big[1\{\hat{\eta}_{i,t}(\hlambda) \leq \hF_{T,i}^{-}(u_i;\hlambda),\hat{\eta}_{j,t}(\hlambda) \leq \hF_{T,j}^{-}(u_j;\hlambda)\}\\
\,& \hspace*{.25cm}- \frac{1}{S}\sums 1\{\hat{X}_{i,t,s}(\theta_q,\hnu) \leq \hG_{T,S,i}^{-}(u_i;\theta_q,\hnu),\hat{X}_{j,t,s}(\theta,\hnu) \leq \hG_{T,S,j}^{-}(u_j;\theta_q,\hnu)\}\big],
\end{split}
\end{equation}
where  $\HH^-(p) \coloneqq \iinf\{x \in \mathbb{R}: \HH(x) \geq p,\,p \in (0,1]\}$ denotes the left-continuous generalized inverse function of a distribution function $\HH$, and
\[
\hF_{T,k}(x;\hlambda) \coloneqq  \frac{1}{T}\sum_{t \eq 1}^T1\{\hat{\eta}_{k,t}(\hlambda) \leq x\},\;\; \hat{\GG}_{T,S,k}(x;\theta_q,\hnu) \coloneqq   \frac{1}{TS}\sum_{t \eq 1}^T\sum_{s \eq 1}^S1\{\hat{X}_{k,t,s}(\theta_q,\hnu) \leq x\}, 
\]
for $x \in \mathbb{R}$, $k \in \{i,j\}$, with \(i,j \in \mathcal{G}_q\), \(q \in \{1,\dots,Q\}\).  More specifically, taking \citet[Theorem 6]{fetal04} and \citet[Lemma 7.2]{buse13} into account, we can express the  $k$-th entry of $\hat{\Psi}_{T,S,i,j}(\theta_q,\hlambda,\hnu) \coloneqq \hTpsi(\hlambda)-\hpsi(\theta_q,\hnu)$ as a Lebesgue-Stieltjes integral
\begin{equation}\label{Psik}
\sqrt{T}\hat{\Psi}_{T,S,i,j,k}(\theta_q,\hlambda,\hnu) = \int_{[0,1]^2}\hat{\mathbb{B}}_{T,S,i,j}(u_i,u_j;\theta_q,\hlambda,\hnu)\,\dd\varphi_k(u_i,u_j) + o_p(1)
\end{equation}
for \(k \in \{1,\dots,\ell\}\). Hence, the limiting distribution of $\sqrt{T}\hat{\Psi}_{T,S}(\theta_0,\hlambda,\hnu)$ can be deduced from the weak convergence of the process $\{\hat{\mathbb{B}}_{T,S,i,j}(u_i,u_j;\theta_{0,q},\hlambda,\hnu): u_i,u_j \in [0,1]\}$, which is readily recognized as the difference between two empirical copula processes. Since filtered data is used, an invariance result with respect to the corresponding statistics based on the unknown counterparts is desirable. If empirical and simulated rank statistics are independent, then it suffices to show that the empirical copula processes based on $\hat{V}_{i,t}(\hlambda)$ and $\hat{U}_{i,t,s}(\theta_{0,q},\hnu)$ share the same weak limit; a proof strategy employed by \cite{ohpa13} who argue along the lines of \cite{rem17}. Here, we require the somewhat stronger notion of uniform asymptotic negligibility; i.e.,  we show, under sufficient regularity of the data, that 
\begin{equation}  
\ssup\limits_{\substack{u_i,u_j \inn [0,1]}}\ssup\limits_{\theta_q \inn \Theta} |\hat{\mathbb{B}}_{T,S,i,j}(u_i,u_j;\theta_q,\hlambda,\hnu)-\hat{\mathbb{B}}_{T,S,i,j}(u_i,u_j;\theta_q,\lambda_{0},\nu_0)| = o_p(1), \label{asy-neg}
\end{equation}
for each \(i,j \in \mathcal{G}_q\), \(q \in \{1,\dots,Q\}\). There exist already similar results in the literature for $\beta$-mixing processes [see \cite{neumetal19} and \cite{chetal20} who rely on \cite{detal09} and \citet[Lemma 1]{avk01}]; the underlying stochastic equicontinuity result of \cite{dotal95} is, however, not directly applicable to the triangular-array case considered here. In order to overcome this difficulty, we resort to the FCLT of \citet[Theorem 2.2]{anpo94}. The following regularity conditions are assumed to hold:

\renewcommand{\theassumption}{C}
\begin{assumption}\label{asu-C}  \normalfont 
For each $i \in \{1,\dots,n\}$, the $i$-th first-order partial derivative $\partial_i \CC(u_1,\dots,u_n;\theta)$ exists and is continuous on the set $\{(u_1,\dots,u_n)' \in [0, 1]^n: 0< u_i < 1\}$ uniformly in $\theta \in \Theta$. The same holds for each bivariate copula $\{\CC_{q}(u_i,u_j;\theta_q): q \in \{1,\dots,Q\}\}$. 
\end{assumption}

\renewcommand{\theassumption}{D}
\begin{assumption}\label{asu-D}  \textcolor[rgb]{1,1,1}{.} \normalfont
\begin{enumerate}[label= (D\arabic*)]
\item\label{asu-D1}
\begin{enumerate}[label=($\roman*$)] 
\item For each $i \in \{1,\dots,n\}$, \(\FF_i\) has density function \(\ff_i\) satisfying  $\ssup\limits_{x \inn \mathbb{R}} \ff_i(x) < \infty$, $\ssup\limits_{x \inn \mathbb{R}} |x\ff_i(x)| < \infty$, $\ff_i\{\FF_i^{-1}(x)\}(1+\FF_i^{-1}(x)) = o(1)$ as $x \rightarrow 0$ or $x \rightarrow 1$
\item For each $1 \leq i < j \leq n$, the bivariate distribution functions $\FF_{i,j}(x_i,x_j) \coloneqq \PP(\eta_{i,t} \leq x_i,\eta_{j,t} \leq x_j)$ satisfy $\mmax\limits_{k,l \inn \{\ii\}}\ssup\limits_{x_i,x_j \inn \mathbb{R}}  |\partial_l\partial_k \FF_{i,j}(x_i,x_j)(1+x_k)(1+x_l)| < \infty$.
\end{enumerate}
\item\label{asu-D2} For each $q \in \{1,\dots,Q\}$, $\GG_q$ density function $\gg_q$ satisfying $\ssup\limits_{\theta_q \inn \Theta}\ssup\limits_{x \inn \mathbb{R}} |\gg_q(x;\theta_q)| < \infty$ and $\ssup\limits_{\theta_q \inn \Theta}\gg_q\{\GG_q^{-1}(x;\theta_q);\theta_q\} = o(1)$ as $x \rightarrow 0$ or $x \rightarrow 1$.
\item\label{asu-D3} $\DD_\eps(x;\delta)$ and $\DD_{\textsf{F},j}(x;\gamma_j)$, \(j \in \{1,\dots,p_\alpha\}\), are continuous and strictly increasing distribution functions, which are known up to finite dimensional parameter $\delta$ and \( \gamma \coloneqq (\gamma_1',\dots,\gamma_{p_\alpha}')'\), respectively.
\begin{enumerate}[label=($\roman*$)] 
\item\label{A-bracket1} $\ssup\limits_{\delta \inn \Theta}\ssup\limits_{x \inn \mathbb{R}} \dd_\eps(x;\delta) < \infty$, where $\dd_\eps$ denotes the marginal density of $\DD_\eps$.
\item\label{A-bracket2} There exists an integrable function $\dot{Q}_{F}: [0,1] \mapsto \mathbb{R}_+$ bounding $\ssup\limits_{\gamma_j \inn \Theta} |\DD_{F,j}^{-1}(u,\gamma_j)|$ and  $\ssup\limits_{\gamma_j \inn \Theta}\norm{\nabla_{\hspace*{-.03cm}\gamma_j} \DD_{F,j}^{-1}(u,\gamma_j)}$ from above for any \(j \in \{1,\dots,p_\alpha\}\) and \(u \in [0,1]\).
\end{enumerate}
\end{enumerate}
\end{assumption}

\renewcommand{\theassumption}{E}
\begin{assumption}\label{asu-E} \textcolor[rgb]{1,1,1}{.} \normalfont
\begin{enumerate}[label= (E\arabic*)]
\item \label{asu-E1} 
\begin{enumerate}[label=($\roman*$)] 
\item  \(\sqrt{T}\norm{\hat{\lambda}_T-\lambda_{0}} = O_p(1)\) and the true parameter \(\lambda_0\) is element of the compact set $\Lambda_{0} \subset \mathbb{R}^{r}$.
\item  \(\sqrt{T}\norm{\hat{\nu}_{T}-\nu_{0}} = O_p(1)\) and the true parameter \(\nu_0\) is element of the compact set $\mathcal{V}_{0} \subset \mathbb{R}^{m}$.
\end{enumerate}
\item \label{asu-E2} 
\begin{enumerate}[label=($\roman*$)]
\item Let \(\mathcal{U}(\lambda_0) \subset \Lambda_{0}\) be a neighborhood around \(\lambda_0\) and set \(R_t \coloneqq R_t(\lambda_0)\).
There exists a measurable function \(\dot{\mu}(R_t)\) with \(\Ex[\dot{\mu}(R_t)^4] < \infty\) such that \(a.s.\)
\[
\ssup\limits_{\lambda \inn \mathcal{U}(\lambda_0) } \mmax\{\norm{\nala   \mu_{j,i}(\mathcal{F}_t,\lambda)/\mu_{2,i}(R_t,\lambda_{0})}, \norm{\nala^2   \mu_{j,i}(\mathcal{F}_t,\lambda)/\mu_{2,i}(R_t,\lambda_{0})}\} \leq \dot{\mu}(R_t)
\] 
for each \(i \in \{1,\dots,n\}\), \(t \in \{1,\dots,T\}\), and any \(j \in \{1,2\}\). Moreover, there exists some constant \(\munderbar{b} \in (0,\infty)\) such that  \(\iinf\limits_{\lambda \inn \mathcal{U}(\lambda_0) }\iinf\limits_{1 \,\leq\, t \,\leq\, T}\mu_{2,i}(\mathcal{F}_t,\lambda) \geq 1/\munderbar{b}\) \(a.s.\) for each \(i \in \{1,\dots,n\}\).
 \item  Let \(\mathcal{U}(\nu_0) \subset \nu_{0}\) be a neighborhood around \(\nu_0\) and set \(M_t \coloneqq M_t(\nu_0)\). There exists a measurable function \(\dot{\sigma}(M_t)\) with \(\Ex[\dot{\sigma}(M_t)^2] < \infty\) such that \(a.s.\)
\[
\ssup\limits_{\nu \inn \mathcal{U}(\nu_0)}\mmax\{\norm{\nanu   \sigma_j(\mathcal{W}_t,\nu)},\norm{\nanu^2    \sigma_j(\mathcal{W}_t,\nu)}\} \leq \dot{\sigma}(M_t),
\]
for each \(j \in \{1,\dots,p_\beta\}\), \(t \in \{1,\dots,T\}\).
\end{enumerate}
\item \label{asu-E0} 
\begin{enumerate}[label=($\roman*$)]
\item The process \(\{R_t: t \geq 1\}\) is strictly stationary and \(\alpha\)-mixing with mixing number $\alpha_{\textsf{R}}(j) = O(j^{-c})$, $j \in \mathbb{N}$, where $c > (c_1-1)(c_1+c_2)/c_2$ for $c_1 \coloneqq \mmin\{i \in \mathbb{N}:  i > 4(1+r)(2+c_2)\}$ and some $c_2 > 0$. 
\item The process \(\{M_{t}: t\geq 1\}\) is strictly stationary and \(\alpha\)-mixing with mixing number $\alpha_{\textsf{M}}(j) = O(j^{-k})$, $j \in \mathbb{N}$, where $k > (k_1-1)(k_1+k_2)/k_2$ for $k_1 \coloneqq \mmin\{i \in \mathbb{N}: i > 2(2+p_\alpha(1+ p_\gamma) + p_\beta(1+ m)+ p_\delta)(2+k_2)\}$ and some $k_2 > 0$.
\end{enumerate}
\end{enumerate}
\end{assumption}

\renewcommand{\theassumption}{F}
\begin{assumption}\label{asu-F} \textcolor[rgb]{1,1,1}{.}	\newline \normalfont  Define \(\CC_{t}(u_1,\dots,u_n) \coloneqq \PP(V_{1,t} \leq u_1, \dots, V_{n,t} \leq u_n \mid Z_t)\) and \(\mathcal{C}_{t}(u_1,\dots,u_n) \coloneqq \PP(U_{1,t,s} \leq u_1, \dots, U_{n,t,s} \leq u_n \mid Z_t)\). Then, for \(w_k = (u_{1,k},\dots,u_{n,k})' \in [0,1]^n\), \(k \in \{1,\dots,m\}\), such that \(\CC_{t}(w_1),\dots,\CC_{t}(w_m)\) are all distinct \(a.s.\) and \(\mathcal{C}_{t}(w_1),\dots,\mathcal{C}_{t}(w_m)\) are all distinct \(a.s.\) with values in \((0,1)\) \(a.s.\) the matrix 
\[
\Big(\Ex[\CC_{t}(w_k \wedge w_l) -\CC_{t}(w_k)\CC_{t}(w_l)]\Big)_{1\leq k,l \leq m}
\]
is positive definite and the matrices 
\[
\Big(\Ex[\mathcal{C}_{t}(w_k \wedge w_l) - \mathcal{C}_{t}(w_k)\mathcal{C}_{t}(w_l)]\Big)_{1\leq k,l \leq m},\;\Big(\Ex[(\CC_{t}(w_k) - \mathcal{C}_{t}(w_k))(\CC_{t}(w_l) - \mathcal{C}_{t}(w_l))]\Big)_{1\leq k,l \leq m},
\]
are positive semi-definite for any \(m \in \mathbb{N}_+\). 
\end{assumption}

\renewcommand{\theassumption}{G}
\begin{assumption}\normalfont\label{asu-G} $\{\varphi_k: 1 \leq k \leq \ell\}$ are  of bounded Hardy-Krause variation; see, e.g., \citet[Definition 2]{owen05}.
\end{assumption}

\noindent The smoothness condition \ref{asu-C}---due to \cite{seg12}---is needed to apply the functional delta method; see also \cite{buvo13}. Assumptions \ref{asu-D1} and \ref{asu-D2} are similar to regularity conditions imposed by \citet[p. 141]{neumetal19}; as discussed in \cite{cotetal19} and \cite{ometal20}, this assumption can be relaxed at the expense of additional technicalities. Assumption \ref{asu-D3} summarizes conditions which, in conjunction with the remaining assumptions, ensure the asymptotic equicontinuity of \(\theta_q \mapsto \hat{\mathbb{B}}_{T,S,i,j}(\theta_q)\).  When compared to similar conditions used by \citet[Assumption 5]{manetal20}, Assumption \ref{asu-D3} is relatively primitive. The assumption is, for example, satisfied if factors and idiosyncratic errors are Gaussian. To give a less trivial example, suppose a scalar factor \(F_t\) follows a (standardized) Student's \(t\)-distribution with degrees of freedom parameter \(2 < \ubar{\gamma} \leq \gamma_0 \leq \bar{\gamma} < \infty\). Then, Assumption \ref{asu-D3} ($ii$) is satisfied by setting \(\dot{Q}_F(u) \coloneqq  \sqrt{\bar{\gamma} \vee 1/(\ubar{\gamma}-2)}|\tilde{D}^{-1}(u,\ubar{\gamma})|\), where \(\tilde{D}^{-1}(u,\ubar{\gamma})\) is the inverse of the non-standardized  Student's \(t\)-distribution so that
\[
\frac{\ubar{\gamma}-1}{\ubar{\gamma}}\int_{[0,1]}\dot{Q}_F(u)\, \dd u \leq \sqrt{\frac{2}{\pi}\left(\bar{\gamma} \vee \frac{1}{\ubar{\gamma}-2}\right)} < \infty;
\]
see the online supplement for details. Assumption \ref{asu-E} concerns the marginal time-series models: part \ref{asu-E1} is high-level and can be verified for many estimators of AR-GARCH type-models [see \cite{fraz04} for a more primitive underpinning]; part \ref{asu-E2} is similar to \citet[Assumption N]{chef06} and means that the gradient vectors of the location and scale functions are locally dominated. The \(\alpha\)-mixing sizes in part \ref{asu-E0} are chosen as to match the conditions of the FCLT in \citet[Theorem 2.2]{anpo94}. Assumption \ref{asu-E} would, for instance, be satisfied by many stationary AR-GARCH processes with geometric mixing rate; see, e.g., \cite{cachen02}, \cite{fryrao11}, or \cite{liuyang16}. Assumption \ref{asu-F} is a regularity condition needed to establish the weak convergence of the `finite-dimensional distributions' of \eqref{Bemp} based on an argument borrowed from \cite{boist17}. The assumption does not seem overly restrictive as similar results exist for univariate unconditional distributions; see, e.g., \citet[Lemma 9.5]{boist17}. Bounded variation in the sense of Hardy-Krause, imposed by Assumption \ref{asu-G} on the functions \(\varphi_k: [0,1]^2 \mapsto \mathbb{R}\), ensures an integration by parts formula for bivariate integrals; see \cite{fetal04} and \cite{radetal17}. Since the identity map or indicators of axis-parallel boxes are of bounded Hardy-Krause variation [see, e.g., \cite{owen05}], dependence measures used here and in \cite{ohpa13} like Spearman's \(\rho\) and quantile dependence can be expressed in terms of Eq. \eqref{rankstat} using functions that satisfy this assumption. Assumption \ref{asu-G} implies Riemann-integrability and thus boundedness [see \citet[Lemma 1]{owenrud21}], a strong assumption, admittedly, but one which completely suffices here and that could in principle be relaxed as pointed out by \citet[Section 3.1]{betal17}.

Proposition \ref{prop-1} distills the main ingredients needed to derive the asymptotic properties of the SMM estimator.

\begin{proposition}\label{prop-1}\normalfont  Suppose assumptions  \ref{asu-B}, \ref{asu-A}, \ref{asu-C}, \ref{asu-D}, \ref{asu-E}, \ref{asu-F}, and \ref{asu-G} hold true. 
\begin{enumerate}[label= ($\alph*$)]  
\item\label{prop-1-a} For each \(\theta \in \Theta\), \(\hat{\Psi}_{T,S}(\theta,\hlambda,\hnu) \stackrel{p}{\longrightarrow} \Psi(\theta) \coloneqq \psi-\psi(\theta)\), where \(\psi \coloneqq (\psi_1',\dots,\psi_Q')'\) and \(\psi(\theta)\coloneqq (\psi_1(\theta_1)',\dots,\psi_Q(\theta_Q)')'\) are \(\bar{\ell} \times 1\) vectors; typical elements of the \(\ell \times 1\) vectors \(\psi_{q} \coloneqq (\psi_{q,1},\dots,\psi_{q,\ell})'\) and \(\psi_{q}(\theta) \coloneqq (\psi_{q,1}(\theta),\dots,\psi_{q,\ell}(\theta))'\) are given in Eq. \eqref{pop_gq}.

\item\label{prop-1-b} \(\sqrt{T}\hat{\Psi}_{T,S}(\theta_0,\hlambda,\hnu) \stackrel{d}{\longrightarrow} \mathcal{N}(0_{\bar{\ell}},\Sigma_{0})\), where \(\Sigma_0\) is a positive-definite variance-covariance matrix.

\item\label{prop-1-c} For any $\epsi,\eta>0$, there exists some $\delta > 0$ such that
\[
\suplim\limits_{T \rightarrow \infty} \PP\bigg[\ssup\limits_{\theta,\ttheta\inn\Theta:\norm{\theta-\ttheta} \lleq \delta} \sqrt{T}\norm{\hat{\psi}_{T,S}(\theta,\hat{\nu}_T)-\psi(\theta)-\hat{\psi}_{T,S}(\ttheta,\hat{\nu}_T)+\psi(\ttheta)} > \eta \bigg] < \epsi.
\]
\end{enumerate}
\end{proposition}

\begin{remark}
It is instructive to take a look at the limiting distribution of part \text{\normalfont($b$)}. 
As shown in Appendix \text{\ref{app:A1}}, \(\Sigma_{0} \coloneqq (\Sigma_{0}(g,q))_{1 \leq g,q\leq Q}\) is a block-symmetric matrix whose \((g,q)\)-th block is given by the \(\ell \times \ell\) matrix \(\Sigma_{0}(g,q) \coloneqq  (\sigma_{0}(g,q \mid k,l))_{1 \leq k,l\leq \ell}\)
with typical element
\[\normalfont
\sigma_{0}(g,q \mid k,l) \coloneqq \int_{[0,1]^2}\int_{[0,1]^2} \Ex[\mathbb{C}_{g}(u_{1},v_{1})\mathbb{C}_{q}(u_{2},v_{2})] \textsf{d}\varphi_k(u_{1},v_{1})\textsf{d}\varphi_l(u_{2},v_{2}),\]
with
\[\normalfont
\mathbb{C}_{q}(u,v) \coloneqq \mathbb{B}_{q}(u,v) - \partial_u\CC_{q}(u,v) \mathbb{B}_{q}(u,1)- \partial_v\CC_{q}(u,v) \mathbb{B}_{q}(1,v),
\] 
where \(\mathbb{B}_{q}\) is a mean-zero Gaussian process concentrated on \(\mathbb{D}_0 \coloneqq \{\alpha \in C[0,1]^2: \alpha(1,1) = \alpha(u,0) = \alpha(0,u) = 0,\,u \in (0,1)\}\) such that 
\begin{align}\normalfont
\cov[\mathbb{B}_{q}(u_{1},v_{1}),\mathbb{B}_{q}(u_{2},v_{2})] = \,& \normalfont\Ex[\CC_{q,t}(u_1 \wedge u_2, v_1 \wedge v_2) -\CC_{q,t}(u_1,v_1)\CC_{q,t}(u_2,v_2)] \nn\\
 \,& + \normalfont\Ex[(\CC_{q,t}(u_1,v_1)-\mathcal{C}_{q,t}(u_1,v_1))(\CC_{q,t}(u_2,v_2)-\mathcal{C}_{q,t}(u_2,v_2))], \nn
\end{align}
where  \(\normalfont\CC_{q,t}(u,v) = \PP(V_{i,t} \leq u, V_{j,t} \leq v \mid Z_t)\) and \(\normalfont\mathcal{C}_{q,t}(u,v) = \PP(U_{i,t,s} \leq u, U_{j,t,s} \leq v \mid Z_t)\), \(i,j \in \mathcal{G}_q\), are the bivariate counterparts of \(\CC_{t}\) and \(\mathcal{C}_{t}\) defined in Assumption \text{\normalfont\ref{asu-F}}. Thus, the limiting distribution of $\sqrt{T}\hat{\Psi}_{T,S}(\theta_0,\hlambda,\hnu)$  is unaffected by the first-step estimation error, a finding which, in view of Eq. \eqref{asy-neg}, was to be expected.  A closer inspection of the limiting variance-covariance matrix reveals that the limiting distribution depends on the covariate \(Z_t\) and the partial derivatives of the copula through the covariance kernel of the Gaussian processes \(\mathbb{C}_{q}\), \(q \in \{1,\dots,Q\}\).\end{remark} 

  \noindent The stochastic equicontinuity result of Proposition \ref{prop-1} provides the link between the pointwise properties of $\theta \mapsto \hat{\Psi}_{T,S}(\theta,\hlambda,\hnu)$ and the asymptotic behavior of $\htheta$. To make the argument precise, an additional set of regularity conditions is imposed.

\renewcommand{\theassumption}{H}
\begin{assumption}\label{asu-H}\textcolor[rgb]{1,1,1}{.} \normalfont
\begin{enumerate}[label= (H\arabic*)]
\item\label{asuE-0} \(\hat{L}_{T,S} = L_{0} + o_p(1)\), \(L_{0}\) is a deterministic \(\bar{\ell} \times \bar{\ell}\) positive definite matrix;
\item\label{asuE-1} \(\Psi(\theta) \neq 0_{\bar{\ell}}\) for \(\theta \neq \theta_0\); 
\item\label{asuE-2} \(\theta_0\) is an interior point of the compact set \(\Theta \subset \mathbb{R}^p\), with \(p \leq \bar{\ell} < \infty\) 
\item\label{asuE-3} \(\psi(\theta)\) is differentiable at \(\theta_0\) with \(\bar{\ell} \times p\) dimensional Jacobian matrix \(\nabla_\theta\psi(\theta)\) such that \(\dot{\psi}_0'L_{0} \dot{\psi}_0\) is nonsingular for \(\dot{\psi}_0 \coloneqq \nabla_\theta\psi(\theta_0)\) 
\item \label{asuE-4} \(\hat{A}_{T,S}(\htheta,\hlambda,\hnu) \leq \iinf\limits_{\theta \inn \Theta}\hat{A}_{T,S}(\theta,\hlambda,\hnu) + o_p(1/T)\). 
\end{enumerate}
\end{assumption}

\noindent Assumption \ref{asu-H} is common for extremum estimators with non-smooth objective function; see, e.g., \citet[Section 7]{nemc94}.   Analogously to \citet[Proposition 2]{ohpa13}, we make use of \citet[Theorem 7.2]{nemc94} to derive the asymptotic normality of the SMM estimator.

\begin{proposition}\label{prop-3}\normalfont Suppose assumptions \ref{asu-B}, \ref{asu-A}, \ref{asu-C}, \ref{asu-D}, \ref{asu-E}, \ref{asu-F}, \ref{asu-G}, and  \ref{asu-H} hold. Then,
\[
\sqrt{T}(\htheta-\theta_0) \stackrel{d}{\longrightarrow} \mathcal{N}(0_{\bar{\ell}}, \Omega_{0}),\;\Omega_{0} \coloneqq (\dot{\psi}_0'L_{0} \dot{\psi}_0)^{-1}\dot{\psi}_0'L_{0}\Sigma_{0} L_{0}\dot{\psi}_0(\dot{\psi}_0'L_{0} \dot{\psi}_0)^{-1},
\]
where \(\Sigma_{0}\) is the variance-covariance matrix given in Proposition \ref{prop-1}.
\end{proposition}

\subsection{Standard Errors and Inference}
We follow \cite{ohpa13} by resorting to numerical derivatives and the bootstrap to estimate \(\dot{\psi}_0\) and the limiting variance-covariance matrix \(\Sigma_{0}\), respectively. The former estimator is almost completely analogous to the one used by \citet[p. 692]{ohpa13}; i.e., for a step size \(\pi_T \rightarrow 0^+\) define \(\hat{\dot{\psi}}_{T,S}\), whose \(k\)-th column is given by
\[
\hat{\dot{\psi}}_{T,S,k} \coloneqq \frac{\hat{\psi}_{T,S}(\htheta+e_k\pi_{T},\hnu)-\hat{\psi}_{T,S}(\htheta-e_k\pi_{T},\hnu)}{2\pi_{T}},\;\; k \in \{1,\dots,p\},
\]
where \(e_k\) denotes the \(p\)-dimensional vector of zeros with one at \(k\)-th position. 

Contrary to the aforementioned authors, however, the estimator of \(\Sigma_{0}\) needs to account for the dependence structure induced by the exogenous regressor. Therefore, we propose standard errors based on bootstrap replications of both empirical \textit{and} simulated statistics. More specifically, draw for each \(i,j \in \mathcal{G}_q\), \(q \in \{1,\dots,Q\}\), with replacement \(B\) bootstrap samples \(\big\{\mathcal{Z}_{S,t,i,j}^{(b)}(\hthetaq,\hlambda,\hnu)\big\}_{t = 1}^T\), \(b \in \{1,\dots,B\}\), from   \(\big\{\mathcal{Z}_{S,t,i,j}(\hthetaq,\hlambda,\hnu)\big\}_{t = 1}^T\), where 
\begin{equation}\nn
\begin{split}
\mathcal{Z}_{S,t,i,j}(\hthetaq,\hlambda,\hnu) \coloneqq   (\hat{\eta}_{i,t}(\hlambda),\hat{\eta}_{j,t}(\hlambda),\;&\hat{X}_{i,t,1}(\hthetaq,\hnu),\hat{X}_{j,t,1}(\hthetaq,\hnu),\dots\\
\,& \dots,\hat{X}_{i,t,S}(\hthetaq,\hnu),\hat{X}_{j,t,S}(\hthetaq,\hnu))',
\end{split}
\end{equation}
with \(\hthetaq\) representing the \((p_\alpha + p_\beta + p_\alpha p_\gamma + p_\delta) \times 1\) sub-vector of \(\htheta\) that contains the SMM estimates pertaining to group \(\mathcal{G}_q\). Next, denote the corresponding ranks by \(\hat{V}_{k,t}^{(b)}(\hlambda)\), \(\hat{U}_{k,t,s}^{(b)}(\htheta,\hnu)\), \(k \in \{i,j\}\). In view of Eq. \eqref{rankstat}, introduce the bootstrap rank-based dependence measures
\begin{equation}\label{rankstatb}
\begin{split}
\hat{\psi}_{T,i,j,k}^{(b)}(\hlambda) \coloneqq \,&  \frac{1}{T}\sum_{t \eq 1}^T\varphi_k(\hat{V}_{i,t}^{(b)}(\hlambda),\hat{V}_{j,t}^{(b)}(\hlambda)), \\
 \hat{\psi}_{T,S,i,j,k}^{(b)}(\hthetaq,\hnu)\coloneqq \,& \frac{1}{ST}\sum_{s \eq 1}^S\sum_{t \eq 1}^T\varphi_k(\hat{U}_{i,t,s}^{(b)}(\hthetaq,\hnu),\hat{U}_{j,t,s}^{(b)}(\hthetaq,\hnu)).
\end{split}
\end{equation}
Akin to the discussion surrounding \eqref{aggPsi}, let \(\hat{\Psi}_{T,S}^{(b)}(\htheta,\hlambda,\hnu)\) denote the \(\bar{\ell} \times 1\) vector of group-averages of \(\hat{\Psi}_{T,S,i,j}^{(b)}(\hthetaq,\lambda,\lambda,\hnu)\coloneqq \hat{\psi}_{T,i,j}^{(b)}(\hnu)-\hat{\psi}_{T,S,i,j}^{(b)}(\hthetaq,\hlambda)\). Following the arguments made by \citet[Theorem 5]{fetal04}, we can then show that the conditional distribution of \(\sqrt{T}(\hat{\Psi}_{T,S}^{(b)}(\htheta,\hlambda,\hnu)-\hat{\Psi}_{T,S}(\htheta,\hlambda,\hnu))\) consistently estimates the limiting distribution of \(\sqrt{T}\hat{\Psi}_{T,S}(\theta_0,\hlambda,\hnu)\). Importantly, since the first-step estimation of the location-scale parameters does not contaminate the limiting distribution of \(\sqrt{T}\hat{\Psi}_{T,S}(\theta_0,\hlambda,\hnu)\), there is no need to adjust for this source of uncertainty as is, for example, done in \cite{gonetal19}. Hence, the limiting variance-covariance \(\Sigma_{0,S}\) is consistently estimable by the bootstrap second moment\footnote{Throughout, `$\st$' indicates that the given probability/moment has been computed under the bootstrap distribution conditional on the original sample.} \(\cov^\St[\sqrt{T}\hat{\Psi}_{T,S}^{(b)}(\htheta,\hlambda,\hnu)]\), provided \(\sqrt{T}(\hat{\Psi}_{T,S}^{(b)}(\htheta,\hlambda,\hnu)-\hat{\Psi}_{T,S}(\htheta,\hlambda,\hnu))\) is uniformly square integrable; see, e.g., \citet[Theorem 7]{brwe02} or \citet[Lemma 1]{cheng15}. Since an estimator of \(\cov^\St[\sqrt{T}\hat{\Psi}_{T,S}^{(b)}(\htheta,\hlambda,\hnu)]\) is given by
\[
\hat{\Sigma}_{T,S,B} \coloneqq \frac{T}{B}\sum_{b \eq 1}^B (\hat{\Psi}_{T,S}^{(b)}(\htheta,\hlambda,\hnu)-\hat{\Psi}_{T,S}(\htheta,\hlambda,\hnu))(\hat{\Psi}_{T,S}^{(b)}(\htheta,\hlambda,\hnu)-\hat{\Psi}_{T,S}(\htheta,\hlambda,\hnu))', 
\]
we can introduce the following consistent estimator of \(\Omega_{0,S}\) 
\begin{equation}\label{bootSE}
\hat{\Omega}_{T,S,B} \coloneqq  (\hat{\dot{\psi}}_{T,S}'\hat{L}_{T,S} \hat{\dot{\psi}}_{T,S})^{-1}\hat{\dot{\psi}}_{T,S}'\hat{L}_{T,S}\hat{\Sigma}_{T,S,B}\hat{L}_{T,S} \hat{\dot{\psi}}_{T,S}(\hat{\dot{\psi}}_{T,S}'\hat{L}_{T,S} \hat{\dot{\psi}}_{T,S})^{-1}.
\end{equation}

\begin{corollary}\label{cor-1}\normalfont Suppose that \(\Ex^\St[\norm{\sqrt{T}(\hat{\Psi}_{T,S}^{(b)}(\htheta,\hlambda,\hnu)-\hat{\Psi}_{T,S}(\htheta,\hlambda,\hnu))}^{2+\delta}] < \infty\) \(a.s.\) for some \(\delta > 0\) and \(\sqrt{T}\pi_{T} \rightarrow \infty\). Then,  \(\hat{\Omega}_{T,S,B}  \stackrel{p}{\longrightarrow} \Omega_{0}\), as \(B,T \rightarrow \infty\).
\end{corollary}

\noindent Corollary \ref{cor-1} allows to conduct inference about \(\theta_0\) and to obtain the two-step SMM estimator with optimal weight matrix \(\hat{L}_{T,S} = \hat{\Sigma}_{T,S,B}^{-1}\). More primitive conditions under which the the uniform square integrability holds are not readily available. However, as discussed by \cite{hahnliao21}, if this  assumption fails, bootstrap standard errors based on \(\hat{\Sigma}_{T,S,B}\) are likely to yield conservative tests.  Provided \(\bar{\ell} > p\), the preceding result can also be used to ascertain overidentifying restrictions based on the Sargan-Hansen type \(J\)-statistic
\begin{equation}\label{overJ}
J_{T,S} \coloneqq T \hat{A}_{T,S}(\htheta,\hlambda,\hnu) \stackrel{d}{\longrightarrow} u'\mathcal{A}_{0}'\mathcal{A}_{0}'u, \;\;u \sim \mathcal{N}(0_{\bar{\ell}},I_{\bar{\ell} \times \bar{\ell}}),
\end{equation}
where  \(\mathcal{A}_{0} \coloneqq L_{0}^{1/2}\Sigma_{0}^{1/2}R_{0}\), with \(R_{0} \coloneqq I_{\bar{\ell} \times \bar{\ell}} -\Sigma_{0}^{-1/2}\dot{\psi}_0(\dot{\psi}_0'L_{0} \dot{\psi}_0)^{-1}\dot{\psi}_0'L_{0}\Sigma_{0}^{1/2}.\) Critical values for \(J_{T,S}\) need to be simulated (using estimators of \(\Sigma_{0}\) and  \(\dot{\psi}_0\)) unless the optimal weight matrix is used, in which case the common result \(J_{T,S} \stackrel{d}{\longrightarrow} \chi^2(\bar{\ell}-p)\) obtains; see also \citet[Proposition 4]{ohpa13}.

\section{Monte Carlo Experiment}\label{copula_mc}
The Monte Carlo experiment uses a data generating process similar to that in \citet[p. 695]{ohpa13}; that is, we consider an AR(1)-GARCH(1,1) process to describe the evolution of each of \(n\) assets over time:
\begin{equation}\label{y_margins_mc}
Y_{i,t} = 0.01 + 0.05 Y_{i,t-1} + \sigma_{i,t}\eta_{i,t}, \quad \sigma_{i,t}^2 =  0.05 + 0.85 \sigma_{i,t-1}^2 + 0.1\sigma_{i,t-1}^2\eta_{i,t-1}^2,  
\end{equation}
where \(\eta_t \coloneqq  (\eta_{1,t},\dots,\eta_{n,t}) \sim \CC(\Phi_1,\dots,\Phi_n),\) with \(\Phi_i\), \(i\in \{1,\dots,n\}\), denoting the marginal (Gaussian) distribution function of \(\eta_{i,t}\). The copula \(\CC\) is generated by the following `block-equidependent' factor model
\begin{equation}\label{mc:model}
X_{i,t} = \alpha_{0,j}F_t + \beta_{0,j}Z_{j,t} + \eps_{i,t},\quad i  \in \mathcal{G}_q,\, q \in \{1,2,3\}.
\end{equation}
We consider three groups \(\mathcal{G}_1, \mathcal{G}_2,\) and \(\mathcal{G}_3\), of equal size partitioning the cross-sectional index set; i.e., \(\{1,2,\dots,n\} =  \mathcal{G}_1 \cup \mathcal{G}_2 \cup \mathcal{G}_3\), with \(|\mathcal{G}_q| = n/3, q \in \{1,2,3\}\) for \(n \in \{15,\,30\}\). The factor \(F_t\) and the idiosyncratic component \(\eps_{i,t}\) are latent but simulable. It is assumed that \(F_t \stackrel{\textsf{IID}}{\sim} t(\zeta_0,\xi_0)\), with \(\zeta_0 = 1/4,\,\xi_0 = -1/2\), where \(t(\zeta_0,\xi_0)\) denotes Hansen's standardized skewed \textit{t}-distribution with tail thickness parameter \(2 < 1/\zeta < \infty\) and skewness parameter \(-1 < \xi < 1\). For the idiosyncratic component, we either consider \(\eps_{i,t} \stackrel{\textsf{IID}}{\sim} t(\zeta_0,0)\) (called the \textsf{skew}-\(t\)/\(t\) specification) or \(\eps_{i,t} \stackrel{\textsf{IID}}{\sim} \mathcal{N}(0,1)\) (called the \textsf{skew}-\(t\)/\textsf{normal} specification). Turning to the loadings and the estimable factors, we consider two `block-equidependent' specifications similar to the multilevel model of \citet[Proposition 3]{bw15}; see the summary of Table \ref{tab:factor structure}
. 
\setlength{\tabcolsep}{1pt}
\renewcommand{\arraystretch}{0.6}
\begin{table}[!htbp]
\centering
\small
\vspace*{1cm}
\caption[Table: Factor Structure]{Factor Structure}
\label{tab:factor structure}
\vspace*{-1.3cm}
\centering 
\begin{tabular}{l@{\hskip .25cm}cc@{\hskip .25cm}clc@{\hskip .25cm}cccl}
\toprule
&\multicolumn{4}{l}{\textsf{design 1}}&&\multicolumn{4}{l	}{\textsf{design 2}}\vspace*{0.085cm}\\
\cmidrule{2-5} \cmidrule{7-10}
&\multicolumn{2}{c}{\textsf{loadings}}&&\multicolumn{2}{c}{\textsf{estimable factors}}&\multicolumn{2}{c}{\textsf{loadings}}&\multicolumn{2}{c}{\textsf{estimable factors}}\\
\cmidrule{2-10}
$i\in \mathcal{G}_1$&$\alpha_1= 1\textcolor[rgb]{1,1,1}{/2}$&$\beta = 1/2$&&&&$\alpha= 2$&$\beta_1 = 1/2$&&$Z_{i,t} = \tilde{Z}_t$\hspace*{2cm}\textsf{AR}(1)\\
&&&&&&&&&$\tilde{W}_t =  0.65\tilde{W}_{t-1}+ \tilde{Z}_t$ \\
&&&&\multirow{2}{*}{$\forall i,\;Z_{i,t} = Z_t$ \hspace*{1.25cm}\textsf{AR}(1)}&&&&&$\tilde{Z}_t \sim \mathcal{N}(0,1)$\vspace*{0.085cm}\\
\cmidrule{8-10}
$i\in \mathcal{G}_2$&$\alpha_2= 3/2$&$\beta = 1/2$&&$W_t =  0.65W_{t-1}+ Z_t$&&$\alpha = 2$&$\beta_2 = 1\textcolor[rgb]{1,1,1}{/2}$&&$Z_{i,t} = \textsf{log}|Z^\St_t|$\hspace*{.5cm}\textsf{GARCH}(1,1)\\
&&&&$Z_t \sim \mathcal{N}(0,1)$&&&&&$W^\St_t = \sigma_tZ^\St_t$\\
&&&&&&&&&$\sigma^2_t = 0.1+0.1\sigma^2_{t-1}+0.5W^\St_{t-1}$\\
&&&&&&&&&$Z^\St_t \sim \mathcal{N}(0,1)$\vspace*{0.085cm}\\
\cmidrule{8-10}
$i\in \mathcal{G}_3$&$\alpha_3 = 2\textcolor[rgb]{1,1,1}{/2}$&$\beta = 1/2$&&&&$\alpha= 2$&$\beta_3 = 3/2$&&$Z_{i,t} = Z_t$\hspace*{2.25cm}\textsf{WN}\\
&&&&&&&&&$Z_t \sim \mathcal{N}(0,1)$\vspace*{0.085cm}\\
	\toprule
	\end{tabular}
\end{table}
\renewcommand\arraystretch{1}\normalsize

Under \textsf{design 1}, loadings on \(F_t\) are group specific while the loading on the single estimable factor is common; the latter factor has to be estimated from an AR(1) model.  Under \textsf{design 2}, the loading on \(F_t\) is common while three estimable factors have group-specific loadings; the estimable factors have to be estimated from an AR(1) model, from a GARCH(1,1) model, or follow observable white noise.  Hence, \textsf{design 1} and \textsf{design 2} imply 6 unknown copula parameters \(\theta_0 = (\alpha_{0,1},\alpha_{0,2},\alpha_{0,3},\beta_0,\zeta_0,\xi_0)'\) and  \(\theta_0 = (\alpha_{0},\beta_{0,1},\beta_{0,2},\beta_{0,3},\zeta_0,\xi_0)'\), respectively. Estimation of \(\theta_0\) is in both cases based on Spearman's rank correlation 
\[
\hat{\varphi}_{T,i,j,1} \coloneqq \frac{12}{T}\sumt \hat{V}_{i,t}(\hlambda)\hat{V}_{j,t}(\hlambda)-3,\; \hat{\varphi}_{T,S,i,j,1} \coloneqq \frac{12}{ST}\sumt\sums \hat{U}_{i,t,s}(\theta_q,\hnu)\hat{U}_{j,t,s}(\theta_q,\hnu)-3 
\] 
and quantile dependence
\begin{equation}
\begin{split}
\hat{\varphi}_{T,i,j,2}^{(\tau)} \coloneqq \,& \begin{cases} \displaystyle\frac{1}{T\tau}\sumt 1\{\hat{V}_{i,t}(\hlambda) \leq \tau,\hat{V}_{j,t}(\hlambda)\leq \tau\}&\mbox{if } \tau  \in (0,1/2], \\
\displaystyle\frac{1}{T(1-\tau)}\sumt 1\{\hat{V}_{i,t}(\hlambda) > \tau,\hat{V}_{j,t}(\hlambda)> \tau\}&\mbox{if } \tau  \in (1/2,1)\end{cases} \\
\hat{\varphi}_{T,S,i,j,2}^{(\tau)} \coloneqq \,&\begin{cases} \displaystyle\frac{1}{TS\tau}\sumt\sums 1\{\hat{U}_{i,t,s}(\theta_q,\hnu) \leq \tau,\hat{U}_{j,t,s}(\theta_q,\hnu)\leq \tau\}&\mbox{if } \tau  \in (0,1/2], \\
\displaystyle\frac{1}{TS(1-\tau)}\sumt\sums 1\{\hat{U}_{i,t,s}(\theta_q,\hnu) > \tau,\hat{U}_{j,t,s}(\theta_q,\hnu)> \tau\}&\mbox{if } \tau  \in (1/2,1).\end{cases} 
\end{split}
\end{equation}
\noindent Throughout, we set \(T \in \{\)500,\, 1,000,\, 2,000\(\}\), \(S = 25\), and use the identity weight matrix \(L_{T,S} = I_{\bar{\ell}}\). We make use of quantile dependence for \(\tau \in \{0.05,0.10,0.90,0.95\}\) alongside Spearman's rank correlation, which yields \(\bar{\ell} =  3 \times 5 = 15\) rank-based dependence measures for estimation. Numerical optimization employs a derivative-free simplex search based on \texttt{MATLAB's} (2019a) \texttt{fminsearchbnd} routine; see \cite{derr21}. The starting values are obtained from a first-step surrogate minimization using \texttt{MATLAB}'s (2019a) \texttt{surrogateopt} optimization for time-consuming objective functions and the individual time-series models are estimated using maximum likelihood. 

\setlength{\tabcolsep}{1.65pt}
\renewcommand{\arraystretch}{0.6}
\begin{table}[!htbp]
\centering
\footnotesize
\vspace*{1cm}
\caption[Table: Simulation results]{Simulation results (\textsf{design 1}: \textsf{skew}-\(t\)/\textsf{normal})}
\label{tab1a}
\vspace*{-1.3cm}
\centering 
\begin{tabular}{lrrcccccccccccccccc}
\toprule
&&&&\multicolumn{6}{c}{\textsf{feasible}}&&\multicolumn{6}{c}{\textsf{unfeasible}}\vspace*{0.085cm}\\
&&&&$\zeta_0$&$\xi_0 $&$\beta_0$&$\alpha_{0,1}$&$\alpha_{0,2}$&$\alpha_{0,3}$&&$\zeta_0$&$\xi_0 $&$\beta_0$&$\alpha_{0,1}$&$\alpha_{0,2}$&$\alpha_{0,3}$\\
\cmidrule{4-10} \cmidrule{12-17}\vspace*{-.3cm}\\
$n$&$T$&&&$0.25$&$-0.5$&$0.5$&$1.0$&$1.5$&$2.0$&\hspace*{.1cm}&$0.25$&$-0.5$&$0.5$&$1.0$&$1.5$&$2.0$\vspace*{-.05cm}\\
\cmidrule{1-17}	
15	&	500	&	\textsf{mean}	&&	0.275	&	-0.535	&	0.574	&	0.994	&	1.547	&	2.083	&&	0.276	&	-0.531	&	0.572	&	0.996	&	1.548	&	2.087	\\
	&		&	\textsf{median}	&&	0.287	&	-0.515	&	0.586	&	0.971	&	1.507	&	2.020	&&	0.282	&	-0.520	&	0.582	&	0.975	&	1.510	&	2.021	\\
	&		&	\textsf{var}	&&	0.008	&	0.017	&	0.028	&	0.035	&	0.073	&	0.133	&&	0.008	&	0.016	&	0.030	&	0.036	&	0.066	&	0.133	\\
	&		&	\textsf{rmse}	&&	0.092	&	0.136	&	0.184	&	0.187	&	0.275	&	0.374	&&	0.092	&	0.129	&	0.189	&	0.190	&	0.261	&	0.375	\\
	&		&	\(t\)	&&	7.60	&	1.20	&	14.80	&	4.20	&	6.00	&	7.40	&&	7.60	&	1.80	&	16.00	&	5.80	&	7.00	&	8.00	\\
	&		&	\(J\)	&&			\multicolumn{6}{c}{4.20}									&&			\multicolumn{6}{c}{4.60}									\\
\cmidrule{2-17}																													
	&	1,000	&	\textsf{mean}	&&	0.264	&	-0.524	&	0.536	&	1.006	&	1.533	&	2.059	&&	0.262	&	-0.524	&	0.536	&	1.006	&	1.529	&	2.053	\\
	&		&	\textsf{median}	&&	0.257	&	-0.517	&	0.550	&	0.993	&	1.512	&	2.010	&&	0.254	&	-0.518	&	0.544	&	0.995	&	1.504	&	2.007	\\
	&		&	\textsf{var}	&&	0.005	&	0.007	&	0.017	&	0.016	&	0.033	&	0.069	&&	0.005	&	0.007	&	0.015	&	0.015	&	0.031	&	0.064	\\
	&		&	\textsf{rmse}	&&	0.071	&	0.087	&	0.135	&	0.125	&	0.186	&	0.269	&&	0.069	&	0.087	&	0.129	&	0.123	&	0.180	&	0.258	\\
	&		&	\(t\)	&&	6.20	&	1.20	&	12.00	&	2.60	&	4.60	&	6.00	&&	4.80	&	1.40	&	10.20	&	3.60	&	5.60	&	6.00	\\
	&		&	\(J\)	&&			\multicolumn{6}{c}{3.60}									&&			\multicolumn{6}{c}{4.20}									\\
\cmidrule{2-17}																													
	&	2,000	&	\textsf{mean}	&&	0.254	&	-0.510	&	0.514	&	1.000	&	1.511	&	2.020	&&	0.253	&	-0.509	&	0.510	&	1.002	&	1.514	&	2.023	\\
	&		&	\textsf{median}	&&	0.247	&	-0.508	&	0.523	&	0.992	&	1.498	&	1.997	&&	0.245	&	-0.507	&	0.521	&	0.994	&	1.499	&	1.993	\\
	&		&	\textsf{var}	&&	0.002	&	0.003	&	0.009	&	0.006	&	0.011	&	0.023	&&	0.002	&	0.003	&	0.009	&	0.006	&	0.014	&	0.034	\\
	&		&	\textsf{rmse}	&&	0.048	&	0.056	&	0.097	&	0.076	&	0.108	&	0.153	&&	0.047	&	0.054	&	0.096	&	0.079	&	0.120	&	0.185	\\
	&		&	\(t\)	&&	6.20	&	2.40	&	8.20	&	3.20	&	4.80	&	3.60	&&	5.80	&	3.40	&	6.80	&	3.20	&	4.80	&	4.40	\\
	&		&	\(J\)	&&			\multicolumn{6}{c}{4.80}									&&			\multicolumn{6}{c}{2.60}									\\
\cmidrule{1-17}																													
30	&	500	&	\textsf{mean}	&&	0.240	&	-0.510	&	0.505	&	0.998	&	1.498	&	1.999	&&	0.240	&	-0.509	&	0.503	&	1.002	&	1.506	&	2.009	\\
	&		&	\textsf{median}	&&	0.240	&	-0.497	&	0.512	&	0.982	&	1.467	&	1.956	&&	0.240	&	-0.495	&	0.518	&	0.985	&	1.470	&	1.959	\\
	&		&	\textsf{var}	&&	0.005	&	0.011	&	0.017	&	0.019	&	0.037	&	0.073	&&	0.005	&	0.011	&	0.018	&	0.024	&	0.054	&	0.109	\\
	&		&	\textsf{rmse}	&&	0.068	&	0.106	&	0.131	&	0.136	&	0.192	&	0.269	&&	0.069	&	0.103	&	0.133	&	0.154	&	0.232	&	0.330	\\
	&		&	\(t\)	&&	4.20	&	1.40	&	3.80	&	3.40	&	7.80	&	9.00	&&	4.20	&	1.20	&	4.20	&	4.40	&	8.40	&	9.40	\\
	&		&	\(J\)	&&			\multicolumn{6}{c}{2.00}									&&			\multicolumn{6}{c}{2.00}									\\
\cmidrule{2-17}																													
	&	1,000	&	\textsf{mean}	&&	0.243	&	-0.508	&	0.488	&	1.006	&	1.505	&	2.001	&&	0.243	&	-0.508	&	0.488	&	1.006	&	1.506	&	2.001	\\
	&		&	\textsf{median}	&&	0.240	&	-0.499	&	0.498	&	0.995	&	1.494	&	1.982	&&	0.241	&	-0.498	&	0.499	&	0.994	&	1.493	&	1.980	\\
	&		&	\textsf{var}	&&	0.002	&	0.005	&	0.010	&	0.008	&	0.016	&	0.028	&&	0.002	&	0.005	&	0.009	&	0.008	&	0.016	&	0.028	\\
	&		&	\textsf{rmse}	&&	0.050	&	0.069	&	0.100	&	0.092	&	0.126	&	0.167	&&	0.050	&	0.068	&	0.097	&	0.091	&	0.125	&	0.168	\\
	&		&	\(t\)	&&	2.60	&	3.40	&	3.00	&	4.40	&	5.80	&	6.60	&&	2.40	&	3.20	&	3.80	&	3.60	&	6.00	&	6.80	\\
	&		&	\(J\)	&&			\multicolumn{6}{c}{2.40}									&&			\multicolumn{6}{c}{1.80}									\\
\cmidrule{2-17}																													
	&	2,000	&	\textsf{mean}	&&	0.248	&	-0.506	&	0.497	&	1.003	&	1.506	&	2.007	&&	0.248	&	-0.504	&	0.496	&	1.003	&	1.506	&	2.006	\\
	&		&	\textsf{median}	&&	0.247	&	-0.501	&	0.501	&	1.000	&	1.497	&	1.984	&&	0.245	&	-0.501	&	0.501	&	0.998	&	1.498	&	1.983	\\
	&		&	\textsf{var}	&&	0.002	&	0.003	&	0.006	&	0.005	&	0.010	&	0.019	&&	0.001	&	0.002	&	0.005	&	0.004	&	0.008	&	0.015	\\
	&		&	\textsf{rmse}	&&	0.039	&	0.051	&	0.077	&	0.067	&	0.099	&	0.137	&&	0.039	&	0.046	&	0.074	&	0.065	&	0.091	&	0.124	\\
	&		&	\(t\)	&&	4.20	&	4.60	&	4.40	&	3.40	&	4.80	&	5.40	&&	4.20	&	4.60	&	4.60	&	2.80	&	5.00	&	5.00	\\
	&		&	\(J\)	&&			\multicolumn{6}{c}{3.40}									&&			\multicolumn{6}{c}{2.80}									\\
	\toprule
	\end{tabular}
\end{table}
\renewcommand\arraystretch{1}\normalsize
\newpage

\setlength{\tabcolsep}{1.65pt}
\renewcommand{\arraystretch}{0.6}
\begin{table}[!htbp]
\centering
\footnotesize
\vspace*{1cm}
\caption[Table: Simulation results]{Simulation results (\textsf{design 1}: \textsf{skew}-\(t\)/\(t\))}
\label{tab1b}
\vspace*{-1.3cm}
\centering 
\begin{tabular}{lrrcccccccccccccccc}
\toprule
&&&&\multicolumn{6}{c}{\textsf{feasible}}&&\multicolumn{6}{c}{\textsf{unfeasible}}\vspace*{0.085cm}\\
&&&&$\zeta_0$&$\xi_0 $&$\beta_0$&$\alpha_{0,1}$&$\alpha_{0,2}$&$\alpha_{0,3}$&&$\zeta_0$&$\xi_0 $&$\beta_0$&$\alpha_{0,1}$&$\alpha_{0,2}$&$\alpha_{0,3}$\\
\cmidrule{4-10} \cmidrule{12-17}\vspace*{-.3cm}\\
$n$&$T$&&&$0.25$&$-0.5$&$0.5$&$1.0$&$1.5$&$2.0$&\hspace*{.1cm}&$0.25$&$-0.5$&$0.5$&$1.0$&$1.5$&$2.0$\vspace*{-.05cm}\\
\cmidrule{1-17}	
15	&	500	&	\textsf{mean}	&	&	0.264	&	-0.514	&	0.488	&	0.969	&	1.471	&	1.971	&	&	0.267	&	-0.512	&	0.486	&	0.966	&	1.47	&	1.967	\\
	&		&	\textsf{median}	&	&	0.268	&	-0.500	&	0.492	&	0.974	&	1.466	&	1.971	&	&	0.273	&	-0.497	&	0.497	&	0.963	&	1.469	&	1.963	\\
	&		&	\textsf{var}	&	&	0.009	&	0.015	&	0.023	&	0.021	&	0.024	&	0.038	&	&	0.009	&	0.015	&	0.022	&	0.021	&	0.025	&	0.039	\\
	&		&	\textsf{rmse}	&	&	0.098	&	0.124	&	0.151	&	0.147	&	0.158	&	0.197	&	&	0.098	&	0.123	&	0.149	&	0.150	&	0.161	&	0.199	\\
	&		&	\(t\)	&	&	6.80	&	2.40	&	4.80	&	5.00	&	6.60	&	7.00	&	&	5.80	&	1.80	&	3.20	&	5.20	&	6.60	&	7.40	\\
	&		&	\(J\)	&	&			\multicolumn{6}{c}{1.20}									&	&			\multicolumn{6}{c}{1.00}									\\
\cmidrule{2-17}																															
	&	1,000	&	\textsf{mean}	&	&	0.27	&	-0.515	&	0.486	&	0.979	&	1.487	&	1.994	&	&	0.269	&	-0.513	&	0.485	&	0.980	&	1.487	&	1.993	\\
	&		&	\textsf{median}	&	&	0.268	&	-0.505	&	0.496	&	0.982	&	1.484	&	1.990	&	&	0.266	&	-0.504	&	0.493	&	0.989	&	1.481	&	1.990	\\
	&		&	\textsf{var}	&	&	0.008	&	0.008	&	0.013	&	0.010	&	0.012	&	0.021	&	&	0.008	&	0.006	&	0.013	&	0.010	&	0.012	&	0.020	\\
	&		&	\textsf{rmse}	&	&	0.090	&	0.089	&	0.116	&	0.101	&	0.108	&	0.144	&	&	0.089	&	0.080	&	0.113	&	0.102	&	0.108	&	0.142	\\
	&		&	\(t\)	&	&	6.80	&	2.40	&	3.60	&	6.40	&	5.00	&	5.60	&	&	6.80	&	3.20	&	3.40	&	5.40	&	5.20	&	5.20	\\
	&		&	\(J\)	&	&			\multicolumn{6}{c}{3.20}									&	&			\multicolumn{6}{c}{2.60}									\\
\cmidrule{2-17}																															
	&	2,000	&	\textsf{mean}	&	&	0.272	&	-0.504	&	0.485	&	0.983	&	1.493	&	1.998	&	&	0.267	&	-0.505	&	0.486	&	0.986	&	1.495	&	1.998	\\
	&		&	\textsf{median}		&	&	0.262	&	-0.500	&	0.496	&	0.984	&	1.491	&	1.994	&	&	0.260	&	-0.502	&	0.492	&	0.993	&	1.495	&	1.994	\\
	&		&	\textsf{var}			&	&	0.006	&	0.003	&	0.009	&	0.006	&	0.007	&	0.011	&	&	0.005	&	0.003	&	0.008	&	0.005	&	0.006	&	0.010	\\
	&		&	\textsf{rmse}			&	&	0.079	&	0.053	&	0.094	&	0.078	&	0.082	&	0.105	&	&	0.075	&	0.051	&	0.092	&	0.075	&	0.08	&	0.102	\\
	&		&	\(t\)							&	&	7.40	&	2.00	&	5.40	&	8.20	&	6.60	&	5.20	&	&	6.20	&	2.60	&	4.40	&	6.40	&	5.80	&	5.00	\\
	&		&	\(J\)							&	&			\multicolumn{6}{c}{5.80}									&	&			\multicolumn{6}{c}{3.60}			\\
\cmidrule{1-17}				
30	&	500	&	\textsf{mean}	& &	0.235	&	-0.507	&	0.487	&	0.993	&	1.484	&	1.971	& &	0.235	&	-0.508	&	0.485	&	0.993	&	1.484	&	1.971	\\
	&		&	\textsf{median}		& &	0.242	&	-0.491	&	0.502	&	0.997	&	1.480	&	1.958	& &	0.240	&	-0.494	&	0.492	&	0.991	&	1.481	&	1.956	\\
	&		&	\textsf{var}			& &	0.008	&	0.010	&	0.015	&	0.012	&	0.017	&	0.029	& &	0.008	&	0.011	&	0.015	&	0.012	&	0.017	&	0.029	\\
	&		&	\textsf{rmse}			& &	0.088	&	0.102	&	0.122	&	0.109	&	0.131	&	0.174	& &	0.088	&	0.105	&	0.123	&	0.110	&	0.132	&	0.173	\\
	&		&	\(t\)							& &	4.20	&	2.00	&	3.00	&	4.60	&	5.40	&	10.40	& &	4.40	&	2.20	&	2.80	&	4.40	&	5.20	&	10.00	\\
	&		&	\(J\)							& &			\multicolumn{6}{c}{1.40}									& &			\multicolumn{6}{c}{1.40}				\\				 
\cmidrule{2-17}																													
	&	1,000	&	\textsf{mean}	&&	0.241	&	-0.507	&	0.481	&	1.004	&	1.499	&	1.991	&&	0.241	&	-0.508	&	0.480	&	1.004	&	1.500	&	1.991	\\
	&		&	\textsf{median}	&&	0.240	&	-0.499	&	0.490	&	1.005	&	1.496	&	1.979	&&	0.241	&	-0.500	&	0.485	&	1.004	&	1.497	&	1.978	\\
	&		&	\textsf{var}	&&	0.005	&	0.004	&	0.008	&	0.005	&	0.008	&	0.014	&&	0.005	&	0.005	&	0.008	&	0.006	&	0.008	&	0.015	\\
	&		&	\textsf{rmse}	&&	0.069	&	0.066	&	0.090	&	0.074	&	0.090	&	0.119	&&	0.070	&	0.068	&	0.094	&	0.076	&	0.091	&	0.121	\\
	&		&	\(t\)	&&	4.20	&	3.80	&	2.40	&	4.40	&	4.40	&	4.60	&&	4.40	&	3.20	&	1.80	&	5.00	&	5.20	&	5.00	\\
	&		&	\(J\)	&&			\multicolumn{6}{c}{2.00}									&&			\multicolumn{6}{c}{2.00}	\\								
\cmidrule{2-17}																													
	&	2,000	&	\textsf{mean}	&&	0.247	&	-0.504	&	0.490	&	1.001	&	1.501	&	1.999	&&	0.247	&	-0.505	&	0.490	&	1.001	&	1.501	&	1.999	\\
	&		&	\textsf{median}	&&	0.246	&	-0.500	&	0.492	&	1.003	&	1.500	&	1.991	&&	0.243	&	-0.500	&	0.491	&	1.003	&	1.498	&	1.990	\\
	&		&	\textsf{var}	&&	0.003	&	0.002	&	0.004	&	0.003	&	0.004	&	0.008	&&	0.003	&	0.003	&	0.004	&	0.003	&	0.004	&	0.008	\\
	&		&	\textsf{rmse}	&&	0.055	&	0.046	&	0.064	&	0.052	&	0.064	&	0.087	&&	0.055	&	0.051	&	0.066	&	0.052	&	0.064	&	0.087	\\
	&		&	\(t\)	&&	4.60	&	3.80	&	2.60	&	2.80	&	5.20	&	4.60	&&	3.80	&	4.40	&	3.20	&	3.40	&	6.40	&	4.60	\\
	&		&	\(J\)	&&			\multicolumn{6}{c}{3.80}									&&			\multicolumn{6}{c}{3.60}		\\						
	\toprule
	\end{tabular}
\end{table}
\renewcommand\arraystretch{1}\normalsize
 
Tables \ref{tab1a}, \ref{tab1b}, \ref{tab2a}, and \ref{tab2b} contain Monte Carlo estimates of mean, median, and variance of the SMM estimator using 500 Monte Carlo iterations\footnote{The computations were implemented in Matlab, parallelized and performed using CHEOPS, the DFGfunded (Funding number: INST 216/512/1FUGG) High Performance Computing (HPC) system of the
Regional Computing Center at the University of Cologne (RRZK).}.  Moreover, we report rejection frequencies of two-sided $t$-tests under the null \(\theta = \theta_0\) and rejection frequencies of the test of overidentifying restrictions \ref{overJ}. Both hypothesis tests are investigated at a nominal significance level of five percent and the test statistics are equipped with the bootstrap standard error \eqref{bootSE}. We use \(B =\) 500 bootstrap replications and set the tuning parameter for the numerical derivative to \(\pi_T  = 0.05\); moreover, we use 1,000 random draws to obtain critical values for the test of overidentifying restrictions \eqref{overJ}. We report results for the feasible and the unfeasible SMM estimator that differ with respect to whether the parameters governing the estimable factors reported in Table \ref{tab:factor structure} are estimated (\textsf{feasible}) or treated as known constants (\textsf{unfeasible}). The simulation evidence reveals that, in accordance with the theory, the estimation accuracy increases with \(T\). Although some size distortions can be observed for \(T = 500\), rejection frequencies are close to the nominal significance level when \(T \geq\) 1,000. Moreover,  the feasible estimator performs almost equally well as its unfeasible counterpart.

\setlength{\tabcolsep}{1.65pt}
\renewcommand{\arraystretch}{0.6}
\begin{table}[!htbp]
\centering
\footnotesize
\vspace*{1cm}
\caption[Table: Simulation results]{Simulation results (\textsf{design 2}: \textsf{skew}-\(t\)/\textsf{normal})}
\label{tab2a}
\vspace*{-1.3cm}
\centering 
\begin{tabular}{lrrcccccccccccccccc}
\toprule
&&&&\multicolumn{6}{c}{\textsf{feasible}}&&\multicolumn{6}{c}{\textsf{unfeasible}}\vspace*{0.085cm}\\
&&&&$\zeta_0$&$\xi_0 $&$\beta_{0,1}$&$\beta_{0,2}$&$\beta_{0,3}$&$\alpha_0$&&$\zeta_0$&$\xi_0 $&$\beta_{0,1}$&$\beta_{0,2}$&$\beta_{0,3}$&$\alpha_0$\\
\cmidrule{4-10} \cmidrule{12-17}\vspace*{-.3cm}\\
$n$&$T$&&&$0.25$&$-0.5$&$0.5$&$1.0$&$1.5$&$2.0$&\hspace*{.1cm}&$0.25$&$-0.5$&$0.5$&$1.0$&$1.5$&$2.0$\vspace*{-.05cm}\\
\cmidrule{1-17}	
15	&	500	&	\textsf{mean}	&&	0.258	&	-0.544	&	0.665	&	1.162	&	1.567	&	1.978	&&	0.255	&	-0.541	&	0.656	&	1.156	&	1.564	&	1.966	\\
	&		&	\textsf{median}	&&	0.263	&	-0.515	&	0.696	&	1.159	&	1.570	&	1.906	&&	0.267	&	-0.514	&	0.682	&	1.150	&	1.555	&	1.905	\\
	&		&	\textsf{var}	&&	0.008	&	0.019	&	0.067	&	0.063	&	0.048	&	0.170	&&	0.008	&	0.018	&	0.072	&	0.061	&	0.047	&	0.155	\\
	&		&	\textsf{rmse}	&&	0.090	&	0.144	&	0.306	&	0.298	&	0.230	&	0.413	&&	0.089	&	0.141	&	0.311	&	0.292	&	0.225	&	0.395	\\
	&		&	\(t\)	&&	8.40	&	1.80	&	14.60	&	9.40	&	4.80	&	13.00	&&	8.60	&	1.60	&	12.80	&	7.80	&	3.80	&	11.00	\\
	&		&	\(J\)	&&			\multicolumn{6}{c}{1.60}									&&			\multicolumn{6}{c}{1.40}									\\
\cmidrule{2-17}																													
	&	1,000	&	\textsf{mean}	&&	0.257	&	-0.520	&	0.598	&	1.102	&	1.541	&	1.984	&&	0.258	&	-0.520	&	0.602	&	1.107	&	1.545	&	1.977	\\
	&		&	\textsf{median}	&&	0.256	&	-0.509	&	0.632	&	1.106	&	1.547	&	1.946	&&	0.258	&	-0.510	&	0.638	&	1.113	&	1.547	&	1.955	\\
	&		&	\textsf{var}	&&	0.005	&	0.006	&	0.052	&	0.038	&	0.029	&	0.066	&&	0.005	&	0.007	&	0.054	&	0.037	&	0.029	&	0.048	\\
	&		&	\textsf{rmse}	&&	0.072	&	0.083	&	0.248	&	0.219	&	0.174	&	0.257	&&	0.071	&	0.085	&	0.253	&	0.221	&	0.176	&	0.221	\\
	&		&	\(t\)	&&	7.40	&	0.80	&	10.80	&	8.60	&	4.60	&	7.00	&&	6.60	&	0.80	&	13.20	&	8.00	&	5.40	&	6.00	\\
	&		&	\(J\)	&&			\multicolumn{6}{c}{2.20}									&&			\multicolumn{6}{c}{2.80}									\\
\cmidrule{2-17}																													
	&	2,000	&	\textsf{mean}	&&	0.256	&	-0.506	&	0.547	&	1.057	&	1.521	&	1.996	&&	0.256	&	-0.508	&	0.548	&	1.059	&	1.522	&	1.997	\\
	&		&	\textsf{median}	&&	0.250	&	-0.504	&	0.567	&	1.047	&	1.526	&	1.975	&&	0.249	&	-0.503	&	0.563	&	1.058	&	1.527	&	1.972	\\
	&		&	\textsf{var}	&&	0.003	&	0.003	&	0.042	&	0.026	&	0.016	&	0.029	&&	0.003	&	0.003	&	0.043	&	0.025	&	0.016	&	0.035	\\
	&		&	\textsf{rmse}	&&	0.054	&	0.053	&	0.209	&	0.172	&	0.127	&	0.171	&&	0.054	&	0.057	&	0.212	&	0.169	&	0.128	&	0.186	\\
	&		&	\(t\)	&&	6.00	&	2.00	&	13.60	&	10.20	&	5.00	&	6.00	&&	6.40	&	3.00	&	12.40	&	9.80	&	4.00	&	6.00	\\
	&		&	\(J\)	&&			\multicolumn{6}{c}{3.20}									&&			\multicolumn{6}{c}{3.80}									\\
\cmidrule{1-17}																													
30	&	500	&	\textsf{mean}	&&	0.243	&	-0.523	&	0.587	&	1.083	&	1.530	&	1.970	&&	0.241	&	-0.527	&	0.589	&	1.085	&	1.531	&	1.971	\\
	&		&	\textsf{median}	&&	0.240	&	-0.507	&	0.622	&	1.079	&	1.530	&	1.906	&&	0.238	&	-0.508	&	0.611	&	1.083	&	1.534	&	1.905	\\
	&		&	\textsf{var}	&&	0.008	&	0.014	&	0.060	&	0.049	&	0.036	&	0.103	&&	0.008	&	0.014	&	0.058	&	0.047	&	0.036	&	0.106	\\
	&		&	\textsf{rmse}	&&	0.088	&	0.119	&	0.260	&	0.235	&	0.193	&	0.322	&&	0.088	&	0.121	&	0.258	&	0.232	&	0.192	&	0.326	\\
	&		&	\(t\)	&&	9.20	&	1.80	&	10.60	&	6.80	&	3.80	&	9.80	&&	9.20	&	1.80	&	10.20	&	6.20	&	4.20	&	9.20	\\
	&		&	\(J\)	&&			\multicolumn{6}{c}{1.40}									&&			\multicolumn{6}{c}{1.00}									\\
\cmidrule{2-17}																													
	&	1,000	&	\textsf{mean}	&&	0.249	&	-0.510	&	0.543	&	1.048	&	1.521	&	1.996	&&	0.248	&	-0.509	&	0.532	&	1.038	&	1.519	&	1.997	\\
	&		&	\textsf{median}	&&	0.241	&	-0.501	&	0.566	&	1.048	&	1.519	&	1.952	&&	0.241	&	-0.498	&	0.543	&	1.041	&	1.514	&	1.964	\\
	&		&	\textsf{var}	&&	0.005	&	0.005	&	0.046	&	0.032	&	0.020	&	0.055	&&	0.004	&	0.005	&	0.049	&	0.035	&	0.020	&	0.057	\\
	&		&	\textsf{rmse}	&&	0.068	&	0.073	&	0.218	&	0.186	&	0.142	&	0.235	&&	0.067	&	0.071	&	0.223	&	0.190	&	0.144	&	0.239	\\
	&		&	\(t\)	&&	6.80	&	3.00	&	9.20	&	7.00	&	5.00	&	6.20	&&	8.80	&	2.40	&	10.40	&	8.00	&	5.40	&	5.80	\\
	&		&	\(J\)	&&			\multicolumn{6}{c}{2.00}									&&			\multicolumn{6}{c}{2.60}									\\
\cmidrule{2-17}																													
	&	2,000	&	\textsf{mean}	&&	0.247	&	-0.502	&	0.512	&	1.019	&	1.505	&	1.990	&&	0.247	&	-0.501	&	0.509	&	1.018	&	1.504	&	1.990	\\
	&		&	\textsf{median}	&&	0.243	&	-0.499	&	0.520	&	1.020	&	1.503	&	1.981	&&	0.243	&	-0.499	&	0.516	&	1.017	&	1.499	&	1.981	\\
	&		&	\textsf{var}	&&	0.002	&	0.002	&	0.028	&	0.018	&	0.010	&	0.017	&&	0.002	&	0.002	&	0.028	&	0.018	&	0.010	&	0.016	\\
	&		&	\textsf{rmse}	&&	0.045	&	0.049	&	0.167	&	0.137	&	0.102	&	0.130	&&	0.044	&	0.046	&	0.168	&	0.135	&	0.102	&	0.126	\\
	&		&	\(t\)	&&	5.60	&	3.20	&	7.20	&	6.80	&	3.80	&	5.80	&&	4.60	&	3.60	&	7.40	&	6.60	&	3.60	&	6.00	\\
	&		&	\(J\)	&&			\multicolumn{6}{c}{4.80}									&&			\multicolumn{6}{c}{4.20}									\\
	\toprule
	\end{tabular}
\end{table}
\renewcommand\arraystretch{1}\normalsize

\setlength{\tabcolsep}{1.65pt}
\renewcommand{\arraystretch}{0.6}
\begin{table}[!htbp]
\centering
\footnotesize
\vspace*{1cm}
\caption[Table: Simulation results]{Simulation results (\textsf{design 2}: \textsf{skew}-\(t\)/\(t\))}
\label{tab2b}
\vspace*{-1.3cm}
\centering 
\begin{tabular}{lrrcccccccccccccccc}
\toprule
&&&&\multicolumn{6}{c}{\textsf{feasible}}&&\multicolumn{6}{c}{\textsf{unfeasible}}\vspace*{0.085cm}\\
&&&&$\zeta_0$&$\xi_0 $&$\beta_{0,1}$&$\beta_{0,2}$&$\beta_{0,3}$&$\alpha_0$&&$\zeta_0$&$\xi_0 $&$\beta_{0,1}$&$\beta_{0,2}$&$\beta_{0,3}$&$\alpha_0$\\
\cmidrule{4-10} \cmidrule{12-17}\vspace*{-.3cm}\\
$n$&$T$&&&$0.25$&$-0.5$&$0.5$&$1.0$&$1.5$&$2.0$&\hspace*{.1cm}&$0.25$&$-0.5$&$0.5$&$1.0$&$1.5$&$2.0$\vspace*{-.05cm}\\
\cmidrule{1-17}	
15	&	500	&	\textsf{mean}	&&	0.237	&	-0.534	&	0.613	&	1.106	&	1.519	&	1.888	&&	0.238	&	-0.533	&	0.613	&	1.106	&	1.521	&	1.883	\\
	&		&	\textsf{median}	&&	0.245	&	-0.517	&	0.614	&	1.099	&	1.521	&	1.887	&&	0.255	&	-0.516	&	0.609	&	1.100	&	1.525	&	1.896	\\
	&		&	\textsf{var}	&&	0.008	&	0.016	&	0.059	&	0.053	&	0.054	&	0.037	&&	0.008	&	0.016	&	0.062	&	0.056	&	0.054	&	0.045	\\
	&		&	\textsf{rmse}	&&	0.089	&	0.130	&	0.267	&	0.253	&	0.234	&	0.224	&&	0.089	&	0.130	&	0.273	&	0.259	&	0.234	&	0.242	\\
	&		&	\(t\)	&&	5.40	&	1.00	&	6.60	&	3.40	&	2.60	&	9.00	&&	5.60	&	0.80	&	7.40	&	3.00	&	2.00	&	9.40	\\
	&		&	\(J\)	&&			\multicolumn{6}{c}{1.20}								&&			\multicolumn{6}{c}{1.00}									\\
\cmidrule{2-17}																													
	&	1,000	&	\textsf{mean}	&&	0.241	&	-0.515	&	0.563	&	1.064	&	1.513	&	1.937	&&	0.242	&	-0.516	&	0.567	&	1.065	&	1.515	&	1.935	\\
	&		&	\textsf{median}	&&	0.247	&	-0.503	&	0.586	&	1.075	&	1.516	&	1.939	&&	0.247	&	-0.506	&	0.588	&	1.074	&	1.527	&	1.934	\\
	&		&	\textsf{var}	&&	0.005	&	0.006	&	0.040	&	0.035	&	0.032	&	0.020	&&	0.005	&	0.007	&	0.039	&	0.035	&	0.031	&	0.021	\\
	&		&	\textsf{rmse}	&&	0.072	&	0.082	&	0.211	&	0.197	&	0.178	&	0.155	&&	0.071	&	0.085	&	0.209	&	0.199	&	0.177	&	0.159	\\
	&		&	\(t\)	&&	5.80	&	1.40	&	8.00	&	3.40	&	2.80	&	5.60	&&	6.00	&	1.60	&	8.00	&	4.40	&	2.60	&	6.00	\\
	&		&	\(J\)	&&			\multicolumn{6}{c}{1.60}									&&			\multicolumn{6}{c}{1.60}									\\
\cmidrule{2-17}																													
	&	2,000	&	\textsf{mean}	&&	0.248	&	-0.506	&	0.532	&	1.034	&	1.500	&	1.968	&&	0.249	&	-0.507	&	0.536	&	1.035	&	1.500	&	1.967	\\
	&		&	\textsf{median}	&&	0.248	&	-0.506	&	0.547	&	1.029	&	1.502	&	1.966	&&	0.249	&	-0.505	&	0.556	&	1.033	&	1.505	&	1.967	\\
	&		&	\textsf{var}	&&	0.003	&	0.003	&	0.028	&	0.020	&	0.015	&	0.011	&&	0.003	&	0.003	&	0.027	&	0.019	&	0.015	&	0.011	\\
	&		&	\textsf{rmse}	&&	0.058	&	0.052	&	0.170	&	0.144	&	0.122	&	0.108	&&	0.057	&	0.053	&	0.170	&	0.144	&	0.124	&	0.109	\\
	&		&	\(t\)	&&	4.60	&	3.60	&	8.00	&	3.80	&	2.80	&	5.60	&&	4.40	&	3.20	&	7.60	&	4.40	&	2.80	&	5.40	\\
	&		&	\(J\)	&&			\multicolumn{6}{c}{1.20}									&&			\multicolumn{6}{c}{1.20}									\\
\cmidrule{1-17}																													
30	&	500	&	\textsf{mean}	&&	0.224	&	-0.523	&	0.567	&	1.053	&	1.501	&	1.914	&&	0.225	&	-0.523	&	0.564	&	1.048	&	1.501	&	1.914	\\
	&		&	\textsf{median}	&&	0.229	&	-0.510	&	0.585	&	1.060	&	1.503	&	1.915	&&	0.230	&	-0.509	&	0.577	&	1.052	&	1.496	&	1.912	\\
	&		&	\textsf{var}	&&	0.008	&	0.011	&	0.046	&	0.039	&	0.044	&	0.029	&&	0.008	&	0.012	&	0.045	&	0.039	&	0.043	&	0.030	\\
	&		&	\textsf{rmse}	&&	0.094	&	0.109	&	0.224	&	0.204	&	0.210	&	0.191	&&	0.094	&	0.111	&	0.222	&	0.203	&	0.207	&	0.194	\\
	&		&	\(t\)	&&	7.80	&	0.80	&	5.00	&	2.20	&	3.00	&	8.40	&&	7.20	&	1.00	&	4.40	&	2.00	&	2.00	&	8.00	\\
	&		&	\(J\)	&&			\multicolumn{6}{c}{1.40}									&&			\multicolumn{6}{c}{1.60}									\\
\cmidrule{2-17}																													
	&	1,000	&	\textsf{mean}	&&	0.236	&	-0.511	&	0.534	&	1.025	&	1.500	&	1.955	&&	0.234	&	-0.510	&	0.530	&	1.023	&	1.503	&	1.956	\\
	&		&	\textsf{median}	&&	0.230	&	-0.504	&	0.534	&	1.034	&	1.503	&	1.953	&&	0.232	&	-0.504	&	0.526	&	1.026	&	1.509	&	1.957	\\
	&		&	\textsf{var}	&&	0.005	&	0.005	&	0.031	&	0.027	&	0.022	&	0.017	&&	0.005	&	0.005	&	0.033	&	0.025	&	0.022	&	0.016	\\
	&		&	\textsf{rmse}	&&	0.074	&	0.069	&	0.179	&	0.166	&	0.150	&	0.136	&&	0.071	&	0.069	&	0.183	&	0.160	&	0.147	&	0.135	\\
	&		&	\(t\)	&&	7.40	&	1.80	&	5.60	&	2.40	&	2.00	&	5.80	&&	6.80	&	2.60	&	6.20	&	2.20	&	2.80	&	5.00	\\
	&		&	\(J\)	&&			\multicolumn{6}{c}{1.60}									&&			\multicolumn{6}{c}{1.60}									\\
\cmidrule{2-17}																													
	&	2,000	&	\textsf{mean}	&&	0.244	&	-0.504	&	0.518	&	1.018	&	1.494	&	1.977	&&	0.244	&	-0.504	&	0.514	&	1.014	&	1.493	&	1.978	\\
	&		&	\textsf{median}	&&	0.237	&	-0.501	&	0.538	&	1.016	&	1.500	&	1.973	&&	0.239	&	-0.502	&	0.532	&	1.013	&	1.497	&	1.976	\\
	&		&	\textsf{var}	&&	0.003	&	0.002	&	0.021	&	0.014	&	0.011	&	0.009	&&	0.003	&	0.002	&	0.022	&	0.015	&	0.012	&	0.009	\\
	&		&	\textsf{rmse}	&&	0.053	&	0.047	&	0.147	&	0.121	&	0.107	&	0.097	&&	0.054	&	0.047	&	0.150	&	0.125	&	0.108	&	0.097	\\
	&		&	\(t\)	&&	5.40	&	3.60	&	6.20	&	4.40	&	3.00	&	5.20	&&	6.40	&	3.20	&	7.20	&	4.40	&	3.20	&	5.20	\\
	&		&	\(J\)	&&			\multicolumn{6}{c}{3.00}									&&			\multicolumn{6}{c}{3.00}									\\
	\toprule
	\end{tabular}
\end{table}
\renewcommand\arraystretch{1}\normalsize

\section{Empirical Application}\label{copula_empapp}
 We apply the above to study the cross-sectional dependence between \(n = 43\) companies matching the first four largest groups of the S\&P 100 found by \citet[Table 4]{ohpa21}. Specifically, we consider daily close prices, adjusted for stock splits and dividends from January 2014 to January 2020 resulting in \(T = \) 1,461 trading days.

\setlength{\tabcolsep}{1.65pt}
\renewcommand{\arraystretch}{0.6}
\begin{table}[!htbp]
\centering
\footnotesize
\vspace*{1cm}
\caption[Table: companies]{Companies} 
\label{tab:companies}
\vspace*{-1.3cm}
\centering 
\begin{tabular}{lrlc@{\hskip .45cm}lrlc@{\hskip .45cm}lrlc@{\hskip .45cm}lrlc}
\toprule
$i$&\multicolumn{3}{c}{\textsf{Group 1} (`\textit{Pharma}')}&$i$&\multicolumn{3}{c}{\textsf{Group 2} (`\textit{Finance}')}&$i$&\multicolumn{3}{c}{\textsf{Group 3} (`\textit{Oil \& Gas}')}&$i$&\multicolumn{3}{c}{\textsf{Group 4} (`\textit{Transport}')}\\
\cmidrule{1-15}	
1&\textit{Abbott Lab.}&ABT&&14&\textit{Bank Of Am}&BAC&&25&\textit{Apache}&APA&&36&\textit{Caterpillar}&CAT&\\			
2&\textit{AbbVie}&ABBV&&15&\textit{Bank Of NY}&BK&&26&\textit{Baker Hughes}&BHI&&37&\textit{Emerson Ele}&EMR&\\						
3&\textit{Amgen}&AMGN&&16&\textit{Citigroup Inc}&C&&27&\textit{Conocophillips}&COP&&38&\textit{Fedex}&FDX&\\	
4&\textit{Baxter}&BAX&&17&\textit{Capital One}&COF&&28&\textit{Chevron}&CVX&&39&\textit{Honeywell Int}&HON&\\
5&\textit{Biogen}&BIIB&&18&\textit{Goldman Sachs}&GS&&29&\textit{Devon}&DVN&&40&\textit{3M}&MMM&\\	
6&\textit{Bristol-Myers}&BMY&&19&\textit{Jpmorgan}&JPM &&30&\textit{Halliburton}&HAL&&41&\textit{Norfolk South}&NSC&\\			
7&\textit{Gilead}&GILD&&20&\textit{Metlife}&MET &&31&\textit{Nat. Oilwell}&NOV&&42&\textit{Union Pacific}&UNP&\\			
8&\textit{Johnson \& J}&JNJ&&21&\textit{Morgan Stanley}&MS  &&32&\textit{Occidental}&OXY&&43&\textit{United Parcel}&UPS&\\	
9&\textit{Lilly Eli}&LLY&&22&\textit{Regions Fin}&RF &&33&\textit{Schlumberger}&SLB &&&&&\\		
10&\textit{Medtronic}&MDT &&23&\textit{US Bancorp}&USB  &&34&\textit{Williams Co}&WMB  &&&&&\\			
11&\textit{Merck}&MRK  &&24&\textit{Wells Fargo}&WFC  &&35&\textit{Exxon Mobil}&XOM  &&&&&\\	
12&\textit{Pfizer}&PFE   &&&&&&&&&&&&\\		
13&\textit{Unitedhealth}&UNH   &&&&&&&&&&&&\\						
	\toprule
	\end{tabular}
\end{table}
\renewcommand\arraystretch{1}\normalsize

 Since gold often acts as an hedge and/or a safe haven for stock markets, its price may convey information about the inter-dependencies between stock returns; see, e.g., \cite{Baurmc10}. Hence, we examine the extent to which information on gold prices can help us to describe the cross-sectional dependence structure among the 43 companies. The conditional mean of the (percentage) logarithmic return of the \(i\)-th stock price \(Y_{i,t}\), \(i \in \{1,\dots,43\}\), is modeled as an AR(1) process augmented with the first lag of  the (percentage) logarithmic change of the three \textit{p.m.} gold fixing price in London bullion market \(W_t\) 
\begin{equation}
Y_{i,t} =  \lambda_{1,i} + \lambda_{2,i}Y_{i,t-1} + \lambda_{3,i}W_{t-1} + \epsilon_{i,t},\;\;\epsilon_{i,t} \coloneqq \mu_{i,t}\eta_{i,t},
\end{equation}
while, similar to \cite{ohpa13, ohpa17, ohpa21}, the conditional variance \(\mu_{i,t}^2\) is assumed to follow a GJR-GARCH(1,1) model 
\begin{equation}
\begin{split}
\mu_{i,t}^2 = \lambda_{4,i}\,&  + \lambda_{5,i}\mu_{i,t-1}^2 + \lambda_{6,i}\epsilon_{i,t-1}^2 + \lambda_{7,i}\epsilon_{i,t-1}^21\{\epsilon_{i,t-1} < 0\}.
\end{split}
\end{equation}
As we clearly fail to reject the null hypothesis\footnote{The $p$-value of a Wald test with Newey-West standard errors is about  \(0.77\).} of a zero conditional mean based on an AR(1) specification with unrestricted constant, a GJR-GARCH(1,1) model is considered for the gold price 
\begin{equation}
W_t = \sigma_tZ_t,\;\;\sigma_t^2 =  \nu_1 + \nu_2 \sigma_{t-1}^2 + \nu_3 W_{t-1}^21\{W_{t-1}<0\}.
\end{equation}
Thus, the use of \(\textsf{log}|\hat{Z}_{t-1}|\) as an estimable factor is justified because, as mentioned earlier, the logarithmic transformation fits into the location specification \eqref{model-WS1}. Table \ref{tab:summary} summarizes descriptive statistics alongside the results from quasi maximum-likelihood estimation with skewed Student's $t$-distributed innovations. The stock returns are left-skewed and leptokurtic with conditional mean and variance dynamics that are similar to findings from the literature; see, e.g., \cite{boletal94}. Note that the distribution of \(W_t\) (\textsf{gold}), while also leptokurtic, is right-skewed.

\setlength{\tabcolsep}{1.65pt}
\renewcommand{\arraystretch}{0.6}
\begin{table}[!htbp]
\centering
\footnotesize
\vspace*{1cm}
\caption[Table: summary]{Summary statistics} 
\label{tab:summary}
\vspace*{-1.3cm}
\centering 
\begin{tabular}{rcccccc@{\hskip .25cm}c}
\toprule
\textsf{Mean}&	\textsf{10\%}	&	\textsf{25\%} &	\textsf{Median} &	\textsf{75\%}	&	\textsf{90\%}&	\textsf{Gold}	\\
\cmidrule{1-7}
\textsf{Mean}&-0.0399  & -0.0012    &0.0277    &0.0520   & 0.0132\\
 \textsf{Standard Dev}&	1.1518   & 1.2614   & 1.4962   & 1.6701  &  2.2298&	0.8119\\
\textsf{Skewness}&	-1.1644  & -0.8172 &  -0.3015 &  -0.0717  &  0.0902 &	0.2363\\ 
\textsf{Kurtosis}&	 5.2379  &  5.7804  &  6.4064 &  10.2722  & 16.7054 &	5.1414\\ 
\cmidrule{2-7}
\textsf{Constant}&	-0.0387  &	 -0.0005  &	  0.0243  &	  0.0503  &	  0.0617 &	\\
\textsf{AR(1)}&	-0.0172  &  0.0003  &  0.0189 &   0.0370  &  0.0453&\\
\textsf{Gold}&	-0.0138  &  0.0057  &  0.0315  &  0.0427  & 0.0645&\\
\cmidrule{2-7}
\textsf{Constant}&	 0.0217  &  0.0349  &  0.0916 &   0.1656  &  0.2758 &	0.0043\\
\textsf{ARCH}&	0.0000   & 0.0033  &  0.0171  &  0.0414  & 0.0546 &	0.0265\\
\textsf{Leverage}&	0.0555  &  0.0699   & 0.1141  &  0.1465 &   0.2021&		-0.0021\\
\textsf{GARCH}&	0.7383  &  0.8050   & 0.8766 &   0.9335  &  0.9575&	0.9680\\
$\xi$&	3.9098  &  4.2967  &  4.8267   &5.8873  &  6.5596&	5.6674\\
$1/\zeta$	& -0.1088 &  -0.0825 &  -0.0634 &  -0.0358 &  -0.0187&	0.0433\\
	\toprule
	\end{tabular}
\end{table}
\renewcommand\arraystretch{1}\normalsize

Turning to the specification of the cross-sectional distribution of the 43 companies, we use various skewed-\textit{t} factor models, inspired by \cite{ohpa13, ohpa17},  as our benchmark specifications for the copula. Based on Table \ref{tab:companies} we consider the following block-equidependent design
\begin{equation}\label{eq:emp_fact}
X_{i,t} =   \alpha_{1,j} F_t + \alpha_{2,j} F_{j,t} + \eps_{i,t},\quad i  \in \mathcal{G}_j,\, j \in \{1,2,3,4\},
\end{equation}
where \(\mathcal{G}_1 = \{1,\dots,13\}\),  \(\mathcal{G}_2 = \{14,\dots,24\}\),  \(\mathcal{G}_3 = \{25,\dots,35\}\), and \(\mathcal{G}_4 = \{36,\dots,43\}\). We assume that \(F_t \stackrel{\textsf{IID}}{\sim} t(\zeta,\xi)\), \(F_{j,t} \stackrel{\textsf{IID}}{\sim}t(\zeta)\), and \(\eps_{i,t} \stackrel{\textsf{IID}}{\sim} t(\zeta)\), while factors and idiosyncratic errors are mutually independent. We use the following four versions of Eq. \eqref{eq:emp_fact}, labeled \textsf{A1}, \textsf{A2}, \textsf{A3}, and \textsf{A4}, imposing certain restrictions on the loadings: specification \textsf{A1} is a one-factor equidependent model so that \(\alpha_{1,j} = \alpha\) and \(\alpha_{2,j} = 0\); specification \textsf{A2} is a one-factor block-equidependent model with \(\alpha_{2,j} = 0\); specification \textsf{A3} has a common factor with common loading \(\alpha_{1,j} = \alpha\) and group-specific factors with group-specific loadings; specification \textsf{A4} does not impose any restrictions and allows for a common factor and group-specific factors, both with group-specific loadings. The following competitor for the copula is generated from a factor model with estimable gold factor: 
 \begin{equation}\label{eq:emp_factZ}
X_{i,t} =  \alpha_{j}F_{j,t} + \beta_j \textsf{log} |Z_{t-1}| + \eps_{i,t},\quad i  \in \mathcal{G}_j,\, j \in \{1,2,3,4\},
\end{equation}
where \(\eps_{i,t} \stackrel{\textsf{IID}}{\sim} t(\zeta)\). Similar to Eq. \eqref{eq:emp_fact}, four versions of Eq. \eqref{eq:emp_factZ}, labeled \textsf{B1}, \textsf{B2}, \textsf{B3}, and \textsf{B4}, are considered: specification \textsf{B1} allows for group-specific loadings on a common simulable factor \(F_{j,t} = F_t \stackrel{\textsf{IID}}{\sim}t(\zeta,\xi)\) and imposes \(\beta_j = \beta\); specification \textsf{B2} imposes \(\beta_j = \beta\) but allows for group-specific simulable factors \(F_{j,t} \stackrel{\textsf{IID}}{\sim}t(\zeta,\xi)\) with group-specific loadings; specifications \textsf{B3} and \textsf{B4} allow for group-specific loadings assuming a common simulable factor \(F_{j,t} = F_{t} \stackrel{\textsf{IID}}{\sim}t(\zeta,\xi)\), \(\xi = 0\) (\textsf{B3}) and group-specific simulable factors \(F_{j,t} \stackrel{\textsf{IID}}{\sim}t(\zeta,\xi)\), \(\xi = 0\) (\textsf{B4}), respectively.

\setlength{\tabcolsep}{1.65pt}
\renewcommand{\arraystretch}{0.6}
\begin{table}[!htbp]
\centering
\footnotesize
\vspace*{1cm}
\caption[Table: Copula parameter]{Copula parameter estimates}
\label{tab_cop}
\vspace*{-1.3cm}
\centering 
\begin{tabular}{lrcccccccccccc}
 
\toprule 
\midrule
\multicolumn{2}{l}{\textsf{Panel A:} \textit{w/o gold factor}}&\multicolumn{2}{c}{\textsf{A1}}&&\multicolumn{2}{c}{\textsf{A2}}&&\multicolumn{2}{c}{\textsf{A3}}&&\multicolumn{2}{c}{\textsf{A4}}&\\
 \cmidrule{3-14} 
&&\textsf{Identity}&\textsf{Optimal}&&\textsf{Identity}&\textsf{Optimal}&&\textsf{Identity}&\textsf{Optimal}&&\textsf{Identity}&\textsf{Optimal}&\\
 \cmidrule{3-4} \cmidrule{6-7} \cmidrule{9-10} \cmidrule{12-13}
&	\multirow{2}{*}{\textsf{restrictions}}	&	\multicolumn{2}{c}{\(\alpha_{1,j} = \alpha\)}	&	&		\multicolumn{2}{c}{\(-\)}	&&		\multicolumn{2}{c}{\(\alpha_{1,j} = \alpha\)}&&	\multicolumn{2}{c}{\(-\)}	&\\
&	&	\multicolumn{2}{c}{\(\alpha_{2,j} = 0\)}	&	&	\multicolumn{2}{c}{\(\alpha_{2,j} = 0\)}	&&		\multicolumn{2}{c}{\(-\)}&	&	\multicolumn{2}{c}{\(-\)}\\
\cmidrule{3-4} \cmidrule{6-7} \cmidrule{9-10} \cmidrule{12-13}
&	$\xi$		&	 -0.2207	&	-0.2172	&	&	-0.1782	&	-0.1563	&	&	-0.3652	&	-0.2687	&	&	-0.2780	&	-0.2730	&\vspace*{-.1cm}\\
&			&	 \scriptsize\textcolor[rgb]{1,1,1}{-}\textit{0.0320}	&	\scriptsize\textcolor[rgb]{1,1,1}{-}\textit{0.0334}	&	&	\scriptsize\textcolor[rgb]{1,1,1}{-}\textit{0.0300}	&	\scriptsize\textcolor[rgb]{1,1,1}{-}\textit{0.0283}	&	&	\scriptsize\textcolor[rgb]{1,1,1}{-}\textit{0.1726}	&	\scriptsize\textcolor[rgb]{1,1,1}{-}\textit{0.1168} &	&	\scriptsize\textcolor[rgb]{1,1,1}{-}\textit{0.0976}	&	\scriptsize\textcolor[rgb]{1,1,1}{-}\textit{0.0926}	&\\
&	$\zeta$		&	0.1740	&	0.1596	&	&	0.0749	&	0.0689 &	&	0.1200	&	0.1205	&	& 0.1119	&	0.1133	&\vspace*{-.1cm}\\
&			&	 \scriptsize\textit{0.0368}	&	\scriptsize\textit{0.0384}	&	&	\scriptsize\textit{0.0263}	&	\scriptsize\textit{0.0250}	&	&	\scriptsize\textit{0.0372}	&	\scriptsize\textit{0.0340}	&	&	\scriptsize\textit{0.0418}	&	\scriptsize\textit{0.0393}	&\\
 \cmidrule{3-4} \cmidrule{6-7} \cmidrule{9-10} \cmidrule{12-13}
&	$\alpha_{1,1}$		&	\multirow{6}{*}{0.8497} &	\multirow{6}{*}{0.8282}	&	&	0.9655	&	0.9217	&	&	\multirow{6}{*}{0.8751} &	\multirow{6}{*}{0.8449}	&	&	0.9121	&	0.8632	\vspace*{-.1cm}\\
&										&	\multirow{6}{*}{\scriptsize\textit{0.0216}}	&	\multirow{6}{*}{\scriptsize\textit{0.0200}}	&	&	\scriptsize\textit{0.0246}	&	\scriptsize\textit{0.0206}	&	&	\multirow{6}{*}{\scriptsize\textit{0.1258}}	&	\multirow{6}{*}{\scriptsize\textit{0.1335}}	&	&	\scriptsize\textit{0.1308}	&	\scriptsize\textit{0.1155}	&\\
&	$\alpha_{1,2}$		&		&		&	&	1.8174	& 1.7402	&	&		&		&	&	1.3347	&	1.3255	\vspace*{-.1cm}&\\
&										&		&		&	&	\scriptsize\textit{0.0581}	&	\scriptsize\textit{0.0395}	&	&		&		&	&	\scriptsize\textit{0.2905}	&	\scriptsize\textit{0.2409}	&\\
&	$\alpha_{1,3}$		&		&		&	&	1.2697	&	1.2995	&	&		&		&	&	0.9557	&	0.9657\vspace*{-.1cm}\\
&										&		&		&	&	\scriptsize\textit{0.0339}	&	\scriptsize\textit{0.0279}	&	&		&		&	& \scriptsize\textit{0.2003}	&	\scriptsize\textit{0.1860}	&\\
&	$\alpha_{1,4}$		&		&		&	&	1.2467	&	1.1851	&	&		&		&	&	1.0764	&	1.0130	\vspace*{-.1cm}\\
&										&		&		&	&	\scriptsize\textit{0.0366}	&	\scriptsize\textit{0.0287}	&	&		&		&	&	\scriptsize\textit{0.2046}	&	\scriptsize\textit{0.1559}	&\\
 \cmidrule{3-4} \cmidrule{6-7} \cmidrule{9-10} \cmidrule{12-13}	
&	$\alpha_{2,1}$		&	 &	&	&	&	&	&	0.4007 &	0.3744	&	&	0.2747	&	0.3414	&\vspace*{-.1cm}\\
&										&	 &	&	&	&	&	&	\scriptsize\textit{0.2775}	&	\scriptsize\textit{0.2956}	&	&	\scriptsize\textit{0.4178}	&	\scriptsize\textit{0.2795}	&\\
&	$\alpha_{2,2}$		&	 &	&	&	&	&	&	1.5668 &	1.5200	&	&	1.2292	&	1.1235	&\vspace*{-.1cm}\\
&										&	 &	&	&	&	&	&	\scriptsize\textit{0.0994}	&	\scriptsize\textit{0.0891}	&	&	\scriptsize\textit{0.3208}	&	\scriptsize\textit{0.2813}&\\
&	$\alpha_{2,3}$		&	 &	&	&	&	&	&	0.9156 &	0.9470	&	&	0.8182	&	0.8397	&\vspace*{-.1cm}\\
&										&	 &	&	&	&	&	&	\scriptsize\textit{0.1302}	&	\scriptsize\textit{0.1210}	&	&	\scriptsize\textit{0.2173}	&	\scriptsize\textit{0.2027}	&\\
&	$\alpha_{2,4}$		&	 &	&	&	&	&	&	0.8729 &	0.8151	&	&	0.6090	&	0.6121	&\vspace*{-.1cm}\\
&										&	 &	&	&	&	&	&	\scriptsize\textit{0.1355}	&	\scriptsize\textit{0.1421}	&	&	\scriptsize\textit{0.3404}	&	\scriptsize\textit{0.2515}	&\\
 \cmidrule{3-4} \cmidrule{6-7} \cmidrule{9-10} \cmidrule{12-13}	
&	$J$								&	0.4961  &	15.7855		  &	& 9.8333 &	34.4458 	&	&	6.2357   &	25.6570 &	&	4.6589 	&	25.7659		&\\
&	\textsf{p-value}	& 0.0030	&	0.0030		&	&	0.0340 & 0.0025	&	&	0.1940& 0.0189	&		&0.0100	&		0.0041	&\\
 \midrule
\multicolumn{2}{l}{\textsf{Panel B:} \textit{w/\textcolor[rgb]{1,1,1}{o} gold factor}}&\multicolumn{2}{c}{\textsf{B1}}&&\multicolumn{2}{c}{\textsf{B2}}&&\multicolumn{2}{c}{\textsf{B3}}&&\multicolumn{2}{c}{\textsf{B4}}&\\
 \cmidrule{3-14} 
&&\textsf{Identity}&\textsf{Optimal}&&\textsf{Identity}&\textsf{Optimal}&&\textsf{Identity}&\textsf{Optimal}&&\textsf{Identity}&\textsf{Optimal}&\\
 \cmidrule{3-4} \cmidrule{6-7} \cmidrule{9-10} \cmidrule{12-13}
&	\multirow{3}{*}{\textsf{restrictions}}	&	\multicolumn{2}{c}{\(\beta_j = \beta\)}	&	&		\multicolumn{2}{c}{\(\beta_j = \beta\)}	&&		\multicolumn{2}{c}{\(-\)}&&	\multicolumn{2}{c}{\(-\)}	&\\
&	&	\multicolumn{2}{c}{\(F_{j,t} = F_t\)}	&	&	\multicolumn{2}{c}{\(-\)}	&&		\multicolumn{2}{c}{\(F_{j,t} = F_t\)}&	&	\multicolumn{2}{c}{\(-\)}&\\
&	& \multicolumn{2}{c}{\(-\)}&	&	\multicolumn{2}{c}{\(-\)}	&&		\multicolumn{2}{c}{\(\xi = 0\)}&	&	\multicolumn{2}{c}{\(\xi = 0\)}&\\
\cmidrule{3-4} \cmidrule{6-7} \cmidrule{9-10} \cmidrule{12-13}
&	$\xi$		&	 -0.0976	&	-0.1175	&	&	-0.0788	&	-0.0560	&	&		&	&	&		&		&\vspace*{-.1cm}\\
&			&	 \scriptsize\textcolor[rgb]{1,1,1}{-}\textit{0.0517}	&	\scriptsize\textcolor[rgb]{1,1,1}{-}\textit{0.0460}	&	&	\scriptsize\textcolor[rgb]{1,1,1}{-}\textit{0.0467}	&	\scriptsize\textcolor[rgb]{1,1,1}{-}\textit{0.0428}	&	&	&	 &	&		&		&\\
&	$\zeta$		&	0.1155&	0.0949	&	&	0.1378	&	0.1212	&	&	0.1561	&	0.1548	&	& 0.1556	& 0.1498	&\vspace*{-.1cm}\\
&			&	 \scriptsize\textit{0.0515}	&	\scriptsize\textit{0.0428}	&	&	\scriptsize\textit{0.0512}	&	\scriptsize\textit{0.0424}	&	&	\scriptsize\textit{0.0694}	&	\scriptsize\textit{0.0643}	&	&	\scriptsize\textit{0.0662}	&	\scriptsize\textit{0.0597
}	&\\
\cmidrule{3-13}
&	$\alpha_{1}$		&	0.6043 &	0.6081	&	& 0.6028 &	0.5824	&	&	0.5570 &	0.5389	&	&	0.5613	& 0.5298	\vspace*{-.1cm}\\
&										&	\scriptsize\textit{0.0724}	&	 \scriptsize\textit{0.0695}	&	&	\scriptsize\textit{0.0680}	&	\scriptsize\textit{0.0632}	&	&	\scriptsize\textit{0.0638}	&	\scriptsize\textit{0.0612}		&	&	\scriptsize\textit{0.0634}	&	\scriptsize\textit{0.0608}	&\\
&	$\alpha_{2}$		&	1.6260	&	1.5912	&	&1.6364&	1.6022	&	&	1.4083	&	1.3771	&	&	1.4243	&	1.3509	&\vspace*{-.1cm}\\
&										&	\scriptsize\textit{0.0581}	&	 \scriptsize\textit{0.0497}	&	&	\scriptsize\textit{0.0588}	&	\scriptsize\textit{0.0478}	&	&	\scriptsize\textit{0.1208}	&	\scriptsize\textit{0.1065}		&	&	\scriptsize\textit{0.1350}	&	\scriptsize\textit{0.1226}	&\\
&	$\alpha_{3}$		&	1.0245	&	1.0750	&	&	1.0197	&	1.0591	&	&	0.9353 &	0.9544	&	&	0.9404 & 0.9792 &\vspace*{-.1cm}\\
&										&	\scriptsize\textit{0.0594}	&	 \scriptsize\textit{0.0501}	&	&	\scriptsize\textit{0.0572}	&	\scriptsize\textit{0.0468}	&	&	\scriptsize\textit{0.0813}	&	\scriptsize\textit{0.0725}		&	&	\scriptsize\textit{0.0796}	&	\scriptsize\textit{0.0783}	&\\
&	$\alpha_{4}$		&	0.9903	&	0.9550	&	&	0.9901 &	0.9387 &	&	0.8504	&	0.8275	&	&	0.8521 &	0.8324 &\vspace*{-.1cm}\\
&										&	\scriptsize\textit{0.0611}	&	 \scriptsize\textit{0.0548}	&	&	\scriptsize\textit{0.0571}	&	\scriptsize\textit{0.0495}	&	&	\scriptsize\textit{0.0852}	&	\scriptsize\textit{0.0846}		&	&	\scriptsize\textit{0.0857}	&	\scriptsize\textit{0.0797}	&\\
	\cmidrule{3-13}			
&	$\beta_{1}$		&		\multirow{6}{*}{0.6538} &	\multirow{6}{*}{0.6227}&	&	\multirow{6}{*}{0.6419}&\multirow{6}{*}{0.6253}	&	&	0.6666 &	0.6512	&	&	0.6619	&	0.6613	&\vspace*{-.1cm}\\
&								&	 \multirow{6}{*}{\scriptsize\textit{0.0557}}&	\multirow{6}{*}{\scriptsize\textit{0.0556}}&	&	\multirow{6}{*}{\scriptsize\textit{0.0510}}&	\multirow{6}{*}{\scriptsize\textit{0.0489}}	&	&	\scriptsize\textit{0.0487}	&	\scriptsize\textit{0.0441}	&	&	\scriptsize\textit{0.0486}	&	\scriptsize\textit{0.0457}	&\\
&	$\beta_{2}$		&	 &	&	&	&	&	&	0.9378 &	0.9432	&	& 0.9598	&	0.9839	&\vspace*{-.1cm}\\
&										&	 &	&	&	&	&	&	\scriptsize\textit{0.1384}	&	\scriptsize\textit{0.1169}	&	&	\scriptsize\textit{0.1516}	&	\scriptsize\textit{0.1259}&\\
&	$\beta_{3}$		&	 &	&	&	&	&	&	0.7151 &	0.7442	&	&	0.7084 &	0.7246	&\vspace*{-.1cm}\\
&										&	 &	&	&	&	&	&	\scriptsize\textit{0.0814}	&	\scriptsize\textit{0.0699}	&	&	\scriptsize\textit{0.0851}	&	\scriptsize\textit{0.0822}	&\\
&	$\beta_{4}$		&	 &	&	&	&	&	&	0.7675 &	0.7412	&	&	0.7779 &	0.7345	&\vspace*{-.1cm}\\
&										&	 &	&	&	&	&	&	\scriptsize\textit{0.0769}	&	\scriptsize\textit{0.0710}	&	&	\scriptsize\textit{0.0746}	&	\scriptsize\textit{0.0732}	&\\
	\cmidrule{3-13}	
&	$J$								&	4.3537 &	22.2472		  &	& 4.7058 &	23.5429	&	&	3.9102 &	21.8012		&	&4.1245 	&	 22.1737		&\\
&	\textsf{p-value}	& 0.3350	&	 0.0516			&	&	0.2680 & 0.0356		&	& 0.0925& 0.0260	&		&0.0805	&		0.0225 	&\\
\midrule
\bottomrule
\multicolumn{14}{p{12.5cm}}{\textit{Note:} SMM point estimates with standard errors below in italics using \(S = 25\). Standard errors}\\ 
\multicolumn{14}{p{12.5cm}}{and the \(p\)-values for the overidentifying restrictions test are based on \(B = \) 2,000 and \(\pi_T = 0.05\).}\\ 
\multicolumn{14}{p{12.5cm}}{The value of the latter test statistic is labeled \(J \coloneqq J_{T,S} =  T\hat{A}_{T,S}\); see Eq. \eqref{overJ} .}
	\end{tabular}
\end{table}
\renewcommand\arraystretch{1}\normalsize

\begin{figure}[!htbp]
\caption[Figure: Density]{Histogram and time-series plot of \(\textsf{log}|\hat{Z}_t|\)}\vspace*{-.25cm} 
\includegraphics[width=.8\textwidth]{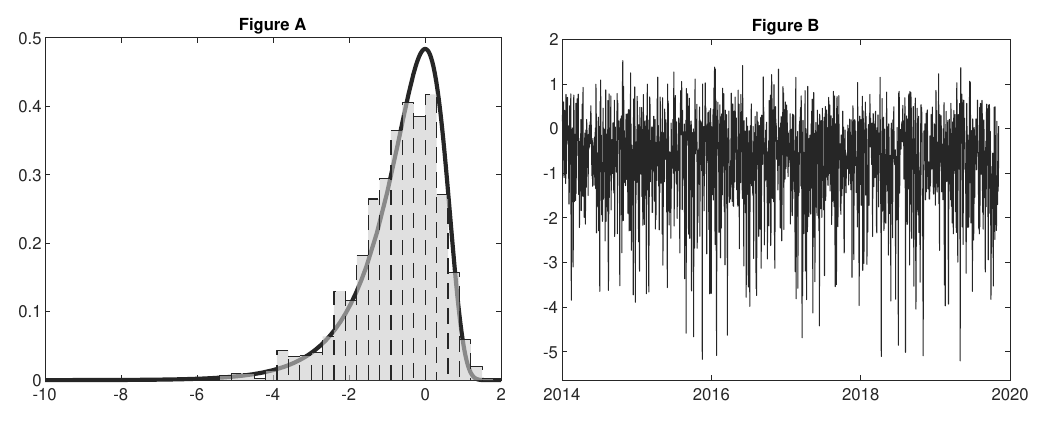}\label{fig_log_abs}
\end{figure}

Table \ref{tab_cop} summarizes the SMM estimation results for the benchmark specification \eqref{eq:emp_fact} as well as for the counterpart with estimable gold factor \eqref{eq:emp_factZ} based on the same rank-based dependence measures used in the Monte Carlo study. We set \(S = 25\), \(B = \) 2,000, and \(\pi_T = 0.05\) and report estimation results using identity weighting \(L_{T,S} = I_{\bar{p}}\) and `optimal' weighting \(L_{T,S} = \hat{\Sigma}^{-1}_{T,S}\). The point estimates for the benchmark specifications in the upper panel are in line with the values reported in \citet[Table 3]{ohpa17} and suggest significant (negative) asymmetric dependence and significant tail dependence. As can be deduced from the results of the overidentifying restrictions test \eqref{overJ}, all specifications but \textsf{A3} are clearly rejected by the data. Moving to the lower panel, we find results for our competitors with estimable factor. The specifications improve the performance of the benchmark models and cannot (at least for the identity weight matrix) be rejected by the data. Interestingly, conditionally on the estimable gold factor, the asymmetry parameter is smaller in size and no longer statistically significant different from zero. This can be explained by the fact that the estimable gold factor already accounts for (some) asymmetry. To illustrate, Figure \ref{fig_log_abs} depicts the distribution of the logarithmic absolute residuals, which is seen to be left-skewed; for comparison, the density of \(\textsf{log}|Z_t^\St|\), \(Z_t^\St \sim \mathcal{N}(0,1)\), is depicted as the solid line in panel \textsf{A}.\footnote{If \(Z^\St \sim\mathcal{N}(0,1)\), then \(Z \coloneqq \textsf{log}|Z^\St|\) has absolutely continuous density \(\ff\)
\[\ff(z) \coloneqq \sqrt{\frac{2\,\textsf{exp}\{2z - \textsf{exp}\{2z\}\}}{\pi}}, \;\;z \in \mathbb{R}, \]
with mean  \(\Ex[Z] = (-\gamma-\textsf{log}\, 2)/2\) and variance \(\var[Z] = \pi^2/8\), where \(\gamma \approx 0.5772\) is the Euler‐Mascheroni constant. The density given by the preceding display is depicted as the solid line in Figure \ref{fig_log_abs}.}


%

%
\section{Conclusion}\label{copula_con}
We derive the asymptotic properties of an SMM estimator of the unknown parameter vector governing a factor copula model with estimable factors and show how to estimate its limiting variance-covariance matrix consistently. The asymptotic theory is derived from primitive conditions, thereby complementing also the earlier work of \cite{ohpa13}, whose model is nested in our framework.
 One avenue for future resarch that can be pursued is to consider quasi-Bayesian estimation to alleviate the difficulties of having to deal with a non-smooth objective function. For example, the sample criterion function considered may be shown to fulfill the regularity conditions needed for Laplace-type estimation in \citet[section 4.1]{cheho03}; see also \cite{hongetal21} for a recent application of this idea to an SMM objective function with overlapping simulation draws.

{}

\clearpage
\appendix

\setcounter{equation}{0}
\renewcommand{\theequation}{\Alph{section}.\arabic{equation}}

\section[A Technical Appendix]{Technical Appendix}

\textbf{Preliminaries:} To begin with, suppose \(i,j \in \mathcal{G}_q\) for \(1\leq i < j \leq n\) and some \(q \in \{1,\dots,Q\}\). Next, define the empirical copulae
\begin{equation}\label{Chat0}
\begin{split}
\hCTS(u_i&,u_j;\,\theta_q,\nu) \\
 \coloneqq & \, \frac{1}{TS}\sum_{t \eq 1}^T\sum_{s \eq 1}^S1\{\hat{X}_{i,t,s}(\theta_q,\nu) \leq \hGTSi^{-}(u_i;\theta_q,\nu), \hat{X}_{j,t,s}(\theta_q,\nu) \leq \hGTSj^{-}(u_j;\theta_q,\nu)\}
\end{split}
\end{equation}
and
\begin{equation}\label{Chat0a}
\hCT(u_i,u_j;\lambda) \coloneqq \, \frac{1}{T}\sum_{t \eq 1}^T1\{\hat{\eta}_{i,t}(\lambda) \leq \hFTi^{-}(u_i;\lambda), \hat{\eta}_{j,t}(\lambda) \leq \hFTj^{-}(u_j;\lambda)\}, 
\end{equation} 
where \(\theta_q = (\alpha_q',\beta_q',\gamma',\delta')'\in \Theta\), \(u_i,u_j \in [0,1]\), while \(\nu  \in \mathcal{U}(\nu_0)\) and \(\lambda \in \mathcal{U}(\lambda_0)\), with \(\mathcal{U}(\cdot)\) denoting neighborhoods around a given parameter as defined by Assumption \ref{asu-E}. Below, we make frequently use of the identities [see, e.g., \cite{tak05} and \cite{seg12}]:
\begin{equation}\label{copid}
\begin{split}
\hCTS(u_i,u_j;\,\theta_q,\nu) = \,&\tCC_{T,S,i,j}\{\tCC_{T,S,i}^{-}(u_i;\theta_q,\nu),\tCC_{T,S,j}^{-}(u_j;\theta_q,\nu);\theta_q,\nu\}\\
\hCT(u_i,u_j;\,\lambda) = \,& \tCC_{T,i,j}\{\tCC_{T,i}^{-}(u_i;\lambda),\tCC_{T,j}^{-}(u_j;\lambda);\lambda\},
\end{split}
\end{equation}
 where
\begin{equation}\label{Ctilde0}
\begin{split}
\hspace*{-.25cm}\tCC_{T,S,i,j}(u_i,u_j;\theta_q,\nu)\coloneqq  \,& \frac{1}{TS}\sumt\sums 1\{\hat{X}_{i,t,s}(\theta_q,\nu) \leq \GG_q^{-1}(u_i;\theta_q),\hat{X}_{j,t,s}(\theta_q,\nu) \leq \GG_q^{-1}(u_j;\theta_q)\} \\
\tCC_{T,i,j}(u_i,u_j;\lambda) \coloneqq    \,& \frac{1}{T}\sumt 1\{\hat{\eta}_{i,t}(\lambda) \leq \FF_i^{-1}(u_i),\hat{\eta}_{j,t}(\lambda) \leq \FF_j^{-1}(u_j)\},
\end{split}
\end{equation}
and $\tCC_{T,S,k}(u_i;\theta_q,\nu) = \tCC_{T,S,i,j}(\vec{u}_k;\theta_q,\nu)$, $\tCC_{T,k}(u_k;\lambda) = \tCC_{T,i,j}(\vec{u}_k;\lambda)$, with $\vec{u}_i = (u_i,1)'$, $\vec{u}_j = (1,u_j)'$. Moreover, define the non-centered processes
\begin{equation}\label{Chat}
\begin{split}
\hat{\mathbb{C}}_{T,S,i,j}(u_i,u_j;\theta_q,\nu) \coloneqq \,& \sqrt{T}(\hCTS(u_i,u_j;\theta_q,\nu) - \CC_q(u_i,u_j;\theta_q)) \\
\hat{\mathbb{C}}_{T,i,j}(u_i,u_j;\lambda) \coloneqq \,& \sqrt{T}(\hCT(u_i,u_j;\lambda) - \CC_q(u_i,u_j)),
\end{split}
\end{equation}
and the centered processes
\begin{equation}\label{Ctilde}
\begin{split}
\tilde{\mathbb{C}}_{T,S,i,j}(u_i,u_j;\theta_q,\nu) \coloneqq \,& \sqrt{T}(\tCC_{T,S,i,j}(u_i,u_j;\theta_q,\nu) - \Ex[\tCC_{T,S,i,j}(u_i,u_j;\theta_q,\nu)]) \\
\tilde{\mathbb{C}}_{T,i,j}(u_i,u_j;\lambda) \coloneqq \,& \sqrt{T}(\tCC_{T,i,j}(u_i,u_j;\lambda) - \Ex[\tCC_{T,i,j}(u_i,u_j;\lambda)]).
\end{split}
\end{equation}

\subsection[A.1 Proof of Proposition \ref{prop-1}]{Proof of Proposition \ref{prop-1}}
\textbf{Proof of Proposition \ref{prop-1} \ref{prop-1-a}:} Recall the definition of the population rank statistics from Eq. \eqref{pop_gq} and note that 
\[\psi_{q,k} = \int_{[0,1]^2}\varphi_k(u_i,u_j)\dd\CC_{q}(u_i,u_j),\;\;\psi_{q,k}(\theta) = \int_{[0,1]^2}\varphi_k(u_i,u_j)\dd\CC_{q}(u_i,u_j;\theta_q)\]
for any \(i,j \in \mathcal{G}_q\), \(q \in \{1,\dots,Q\}\). Hence, one gets, in view of  Lemma \ref{lemmaA3}  and Lemma \ref{lemmaA4}, the following representation
\begin{equation}\label{hatCT0}
\sqrt{T}(\hat{\psi}_{T,i,j,k}(\hlambda)-\psi_{q,k}) = \int_{[0,1]^2}\hat{\mathbb{C}}_{T,i,j}(u_i,u_j)\dd\varphi_k(u_i,u_j) + o_p(1),
\end{equation}
with $\hat{\mathbb{C}}_{T,i,j}(u_i,u_j) \coloneqq \sqrt{T}(\hCT(u_i,u_j)-\CC_{q}(u_i,u_j))$, and, uniformly in $\theta \in \Theta$, 
\begin{equation}\label{hatCTS0}
\sqrt{T}(\hat{\psi}_{T,S,i,j,k}(\theta_q,\hnu)-\psi_{q,k}(\theta)) = \int_{[0,1]^2}\hat{\mathbb{C}}_{T,S,i,j}(u_i,u_j;\theta_q)\dd\varphi_k(u_i,u_j) + o_p(1),
\end{equation}
with $\hat{\mathbb{C}}_{T,S,i,j}(u_i,u_j;\theta_q) \coloneqq \sqrt{T}(\hCTS(u_i,u_j;\theta_q)-\CC_{q}(u_i,u_j;\theta_q))$, $\hCT(u_i,u_j) \coloneqq \hCT(u_i,u_j;\lambda_{0})$, $\hCTS(u_i,u_j;\theta_q) \coloneqq \hCTS(u_i,u_j;\theta,\nu_0)$; see Eqs. \eqref{Chat0} and \eqref{Chat0a}. Thus, the claim is due to the weak convergence of $\hat{\mathbb{C}}_{T,i,j}$ and $\hat{\mathbb{C}}_{T,S,i,j}$. To see this, note that the functional delta method [cf. \citet[Theorem 3.9.4]{vw96}] in conjunction with Assumption \ref{asu-C} and \citet[Theorem 2.4.]{buvo13} yields
\begin{equation}
\hat{\mathbb{C}}_{T,i,j}(u_i,u_j)= \tilde{\mathbb{C}}_{T,i,j}(u_i,u_j)- \sum_{k \inn \{i,j\}} \partial_k\CC_{q}(u_i,u_j)\tilde{\mathbb{C}}_{T,i,j}(\vec{u}_k) + R(u_i,u_j),
\end{equation}
and
\begin{equation}\label{Ctitayl}
\begin{split}
\hat{\mathbb{C}}_{T,S,i,j}(u_i,u_j;\theta_q)= \,& \tilde{\mathbb{C}}_{T,S,i,j}(u_i,u_j;\theta_q) \\
\,& - \sum_{k \inn \{i,j\}} \partial_k\CC_{q}(u_i,u_j;\theta_q)\tilde{\mathbb{C}}_{T,S,i,j}(\vec{u}_k;\theta_q)  + R_1(u_i,u_j;\theta_q),
\end{split}
\end{equation}
where $\ssup\limits_{u_i,u_j \inn [0,1]}|R(u_i,u_j)| = o_p(1)$, $\ssup\limits_{\theta_q \inn \Theta}\ssup\limits_{u_i,u_j \inn [0,1]}|R_1(u_i,u_j;\theta_q)| = o_p(1)$, and $\tilde{\mathbb{C}}_{T,i,j}(u_i,u_j) \coloneqq \tilde{\mathbb{C}}_{T,i,j}(u_i,u_j;\lambda_{0})$, $\tilde{\mathbb{C}}_{T,S,i,j}(u_i,u_j;\theta_q) \coloneqq \tilde{\mathbb{C}}_{T,S,i,j}(u_i,u_j;\theta_q,\nu_0)$; see Eq. \eqref{Ctilde}. The claim follows from Lemma \ref{lemmaV} and Lemma \ref{lemmaU}.

\noindent \textbf{Proof of Proposition \ref{prop-1} \ref{prop-1-b}:} Taking Eq. \eqref{Psik}, Lemma \ref{lemmaA3},  and Lemma \ref{lemmaA4} into account, one gets
\begin{equation}\label{PsikA}
\sqrt{T}\hat{\Psi}_{T,S,i,j,k}(\theta_q,\hlambda,\hnu) = \int_{[0,1]^2}\hat{\mathbb{B}}_{T,S,i,j}(u_i,u_j)\,\dd\varphi_k(u_i,u_j) + o_p(1), 
\end{equation}
where $\hat{\mathbb{B}}_{T,S,i,j}(u_i,u_j) \coloneqq \hat{\mathbb{B}}_{T,S,i,j}(u_i,u_j;\theta_{0,q},\lambda_{0},\nu_0)$. We show next that $\hat{\mathbb{B}}_{T,S,i,j}$ converges weakly to the tight Gaussian process $\mathbb{B}_S$ concentrated on \(\mathbb{D}_0 \coloneqq \{\alpha \in C[0,1]^n: \alpha(1,\dots,1) = 0,\; \alpha(x) = 0 \text{ if some of the components of \(x\) are equal to zero}\}\) by establishing (1)  asymptotic tightness and (2) finite dimensional (`\textit{fidi}', henceforth) convergence. \underline{(1) \textit{Stochastic equicontinuity:}} The functional delta method yields
\begin{equation}
\hat{\mathbb{B}}_{T,S,i,j}(u_i,u_j) = \tilde{\mathbb{B}}_{T,S,i,j}(u_i,u_j)- \sum_{k \inn \{i,j\}}\partial_k\CC_{q}(u_i,u_j)\tilde{\mathbb{B}}_{T,S,i,j}(\vec{u}_k) + o_p(1),
\end{equation}
where 
\begin{equation}
\tilde{\mathbb{B}}_{T,S,i,j}(u_i,u_j) 
= \frac{1}{\sqrt{T}}\sum_{t \eq 1}^T\frac{1}{S}\sum_{s \eq 1}^S(1\{V_{i,t} \leq u_i, V_{j,t} \leq u_j\}-1\{U_{i,t,s} \leq u_i, U_{j,t,s} \leq u_j\}),
\end{equation}
with $V_{k,t} \coloneqq \FF_k(\eta_{k,t})$, $U_{k,t,s} \coloneqq \GG_q(X_{k,t,s}(\theta_{0,q});\theta_{0,q})$ for $k \in \{i,j\}$. Let $\xi_{i,j,t,S} = (V_{i,t},V_{j,t},\UU_{S:i,j,t}')'$, with
$\UU_{S:i,j,t} \coloneqq (U_{i,t,1},U_{j,t,1},\dots,U_{i,t,S},U_{j,t,S})'$. We can view $\tilde{\mathbb{B}}_{T,S,i,j}$ as an empirical process indexed by $\bar{f} \in \bar{\mathcal{F}}$:
\begin{equation}
\bar{\mathcal{F}} \coloneqq \{\xi_{S:i,j,t} \mapsto \bar{f}(\xi_{S:i,j,t})\coloneqq \frac{1}{S}\sum_{s \eq 1}^S(f(V_{i,t},V_{j,t}) - f(U_{i,t,s},U_{j,t,s})): f \in \mathcal{F}\},
\end{equation}
where $\mathcal{F} \coloneqq \big\{(x_1,x_2) \mapsto  1\{x_1 \leq  u_1, x_2 \leq u_2\}, \,u_1,u_2 \in [0,1]\big\}$. Clearly, $\bar{\mathcal{F}}$ has envelope 1. We use Theorem \ref{thmap-94} below to establish asymptotic equicontinuity. Specifically, the bracketing number $\mathcal{N}_{[\,]}(\epsi,\bar{\mathcal{F}} ,\rho(\cdot))$, with $\rho(\bar{f}) \coloneqq \ssup_{t,T}\norm{\bar{f}(\xi_{i,j,t,S})}_2$,  shall be determined; see the discussion surrounding theorem \ref{thmap-94} for details. It is well-known [see, e.g, \citet[Example 19.6]{vv94}] that $\mathcal{N}_{[\,]}(\epsi, \mathcal{F}_k,\norm{\cdot}_2) = O(\eps^{-2})$, with  $\mathcal{F}_k \coloneqq \{x_k \mapsto 1\{x_k \leq u_k\}, u_k \in [0,1]\}$. Since $\mathcal{F} \subseteq \mathcal{F}_1 \cdot \mathcal{F}_2$, one gets $\mathcal{N}_{[\,]}(\epsi, \mathcal{F},\norm{\cdot}_2) = O(\eps^{-4})$; see, e.g., \citet[Lemma 9.25]{ko08}. Suppose $[l_k,u_k]$, $k = 1,\dots, m \coloneqq \mathcal{N}_{[\,]}(\epsi, \mathcal{F},\norm{\cdot}_2)$, represent the brackets needed to cover $\mathcal{F}$. We can then cover $\bar{\mathcal{F}}$ with $[\bar{l}_k,\bar{u}_k]$, $k = 1,\dots, m$, where
\begin{equation}
\begin{split}
\bar{l}_k(\xi_{S:i,j,t}) \coloneqq \,&\frac{1}{S}\sums (l_k(V_{i,t},V_{j,t})-u_k(U_{i,t,s},U_{j,t,s}))\\
\bar{u}_k(\xi_{S:i,j,t}) \coloneqq \,&\frac{1}{S}\sums (u_k(V_{i,t},V_{j,t})-l_k(U_{i,t,s},U_{j,t,s})).
\end{split}
\end{equation}
Note, that $\rho(\bar{u}_k-\bar{l}_k) \leq 2\epsi$. Thus, $\mathcal{N}_{[\,]}(\epsi,\bar{\mathcal{F}},\rho(\cdot)) = O(\epsi^{-4})$. Now, since $\{\xi_{i,j,t,S}: t \geq 1\}$ is $i.i.d.$ and $\bar{\mathcal{F}}$ is uniformly bounded, the conditions of Theorem \ref{thmap-94} are satisfied.  \underline{(2) \textit{`\textit{Fidi}'-convergence:}} By the Cram\'{e}r-Wold device [see, e.g., \citet[Proposition 5.1]{white01}], it suffices to fix some $c \coloneqq (c_1,\dots,c_m)' \in \mathbb{R}^m$, with $\norm{c} = 1$, $(\{u_{1},v_{1}\}\dots,\{u_{m},v_{m}\})' \in [0,1]^{2m}$, and to consider
\begin{equation}
Z_{T,S}(m) \coloneqq \sum_{l \eq 1}^m c_l \tilde{\mathbb{B}}_{T,S,i,j}(u_{l},v_{l})  = \frac{1}{\sqrt{T}}\sum_{t \eq 1}^T B_{t,S}(m),
\end{equation}
where $B_{t,S}(m) \coloneqq \sum_{l \eq 1}^m c_l \zeta_{t,S}(u_{l},v_{l})$, with
\begin{equation}
\zeta_{t,S}(u_{l},v_{l}) \coloneqq 1\{V_{i,t} \leq u_{l}, V_{j,t} \leq v_{l}\}-\frac{1}{S}\sum_{s \eq 1}^S1\{U_{i,t,s} \leq u_{l}, U_{j,t,s} \leq v_{l}\}.
\end{equation}
The sequence $\{B_{t,S}(m): t \geq 1\}$ is $i.i.d$, bounded, and, by Assumption \ref{asu-A}, centered. It thus follows from \citet[Theorem 5.11]{white01} that $Z_{T,S}(m)\stackrel{d}{\longrightarrow} \mathcal{N}(0,\sigma_S^2(m))$, with $\sigma_S^2(m) \coloneqq \lim\limits_{T \rightarrow \infty} \sigma_{T,S}^2(m)$,  provided $\iinf\limits_{T,S \,\geq\, 1}\sigma_{T,S}^2(m) > 0$ for $\sigma_{T,S}^2(m) \coloneqq \var[Z_{T,S}(m)]$. Now,
\begin{equation}
\sigma_{T,S}^2(m)= \sum_{k,l \eq 1}^m c_kc_l \hat{\gamma}_{T,S}(k,l),\;\;
 \hat{\gamma}_{T,S}(k,l) \coloneqq \frac{1}{T}\sum_{t,h \eq 1}^T\cov[\zeta_{t,S}(w_k),\zeta_{t,S}(w_l)],
\end{equation}
where $w_j \eqqcolon (u_{j},v_{j})' \in [0,1]^2$, $j \in \{1,\dots,m\}$. Since, $\{\zeta_{t,S}(w_k): t \geq 1\}$ is $i.i.d.$, one gets
\(
\hat{\gamma}_{1,S}(k,l) = \cov[\zeta_{1,S}(w_k),\zeta_{1,S}(w_l)].
\)
Next, for any \(i,j \in \mathcal{G}_q\), \(q \in \{1,\dots,Q\}\), one gets
\begin{equation}
\begin{split}
\cov[\zeta_{1,S}(w_k),\zeta_{1,S}(w_l) \mid Z_1] 
= \,& \CC_{q,1}(u_k \wedge u_l,v_k \wedge v_l) - \CC_{q,1}(u_k,v_k)\CC_{q,1}(u_l,v_l) \\
\,& + [\mathcal{C}_{q,1}(u_k \wedge u_l,v_k \wedge v_l) - \mathcal{C}_{q,1}(u_k,v_k)\mathcal{C}_{q,1}(u_l,v_l)]/S, 
\end{split}
\end{equation}
where the penultimate equality uses that \(\{V_{i,t}, V_{j,t}\} \perp \{U_{i,t,s},U_{j,t,s}\} \mid Z_t\) and  \(\{U_{i,t,r},U_{j,t,r}\} \perp \{U_{i,t,s},U_{j,t,s}\} \mid  Z_t\), \(r \neq s\). Since \(\Ex[\zeta_{1,S}(w_k) \mid Z_1] = \CC_{q,1}(w_k) - \mathcal{C}_{q,1}(w_k)\), we get, by the law of total covariance, 
\begin{equation}
\begin{split}
\cov[\zeta_{1,S}(w_k),\zeta_{1,S}(w_l)] = \,& \Ex[\CC_{q,t}(w_k \wedge w_l) -\CC_{q,t}(w_k)\CC_{q,t}(w_l)] \nn\\
 \,& + \Ex[(\CC_{q,t}(w_k)-\mathcal{C}_{q,t}(w_k))(\CC_{q,t}(w_l)-\mathcal{C}_{q,t}(w_l))]  \\
  \,& + \Ex[\mathcal{C}_{q,t}(w_k \wedge w_l) -\mathcal{C}_{q,t}(w_k)\mathcal{C}_{q,t}(w_l)]/S \\
	\eqqcolon \,& \gamma^{(1)}_q(k,l)+\gamma^{(2)}_q(k,l)+\gamma^{(3)}_q(k,l)/S,
\end{split}
\end{equation}
say. We are left with showing 
\[
\sigma_{T,S}^2(m)  \coloneqq \sigma_{S}^2(m) = c'(\gamma^{(1)}_q(k,l))_{1\leq k,l \leq m} c + c'(\gamma^{(2)}_q(k,l))_{1\leq k,l \leq m} c/S + c'(\gamma^{(3)}_q(k,l))_{1\leq k,l \leq m} c> 0.
\]
 Clearly, if \(\CC_{q,t}(w_1),\dots,\CC_{q,t}(w_m)\) are all distinct \(a.s.\) and \(\mathcal{C}_{q,t}(w_1),\dots,\mathcal{C}_{q,t}(w_m)\) are all distinct \(a.s.\) with values in \((0,1)\) \(a.s.\), then the claim follows by Assumption \ref{asu-F}. For the cases where not all are distinct or take values in \(\{0,1\}\) the claim follows from the argument used in the proof of \citet[Lemma 9.6]{boist17}. Therefore, combining (1) and (2) yields $\tilde{\mathbb{B}}_{T,S,i,j}(u,v)  \rightsquigarrow  \mathbb{C}_{q,S}(u,v)$, so that 
\[
\sqrt{T}\hat{\Psi}_{T,S,i,j,k}(\theta,\hlambda,\hnu) \stackrel{d}{\longrightarrow} \mathcal{N}(0,  \sigma_{0,S}(q,q\mid k,k)),
\]
with
\[
\sigma_{0,S}(q,q\mid k,k)\coloneqq \int_{[0,1]^2}\int_{[0,1]^2}\Ex[\mathbb{C}_{q,S}(u_1,v_1)\mathbb{C}_{q,S}(u_2,v_2)]\dd \varphi_k(u_1,v_1)\dd\varphi_k(u_2,v_2);
\]
see also \citet[Theorem 3.3]{betal17}. To conclude from here, use that $\{\hat{\mathbb{B}}_{T,S,i,j}(u_i,u_j): 1 \leq i < j \leq n\}$ are jointly normal as $T \rightarrow \infty$. To see this, note that $\hat{\mathbb{B}}_{T,S,i,j}(u_i,u_j) = \hat{\mathbb{B}}_{T,S}(\vec{u}_{i,j})$, where $\vec{u}_{i,j}$ denotes the $n$-dimensional vector of ones with $u_i$ ($u_j$) at $i$-th ($j$-th) position and $\tilde{\mathbb{B}}_{T,S}(u_1,\dots,u_n) = \hat{\mathbb{B}}_{T,S}(u_1,\dots,u_n;\theta_0,\lambda_0,\nu_0)$, with 
\begin{equation}\nn
\begin{split}
\hat{\mathbb{B}}_{T,S}(u_1,&\dots,u_n;\theta,\lambda,\nu) \\
 \coloneqq  \,&\frac{1}{\sqrt{T}}\sumt \big[1\{\hat{\eta}_{1,t}(\lambda_1) \leq \FF_{1,T}^{-}(u_1;\lambda_1),\dots,\hat{\eta}_{n,t}(\lambda_n) \leq \FF_{n,T}^{-}(u_n;\lambda_n)\}\\
\,& \hspace*{.75cm}- \frac{1}{S}\sums 1\{\hat{X}_{1,t,s}(d_1,\hnu) \leq G_1^{-1}(u_1;d_1,\nu),\dots,\hat{X}_{n,t,s}(d_n,\nu) \leq  G_n^{-1}(u_n;d_n,\nu)\}\big].
\end{split}
\end{equation}
Weak convergence of $\tilde{\mathbb{B}}_{T,S}(u_1,\dots,u_n)$ follows by the same arguments used to establish weak convergence of $\tilde{\mathbb{B}}_{T,S,i,j}(u_i,u_j)$. Since $n$ is finite, the claim follows.

\noindent \textbf{Proof of Proposition \ref{prop-1} \ref{prop-1-c}:}  
Note that we can restrict the event inside the probability to the case where  $\hat{\nu}_T$ lies in a \(\sqrt{T}\) neighborhood of \(\nu_0\). To see this, recall from Assumption \ref{asu-E} that $\sqrt{T}\norm{\hat{\nu}_T - \nu_0} = O_p(1)$; i.e., there exists some constant $K \coloneqq K(\eps) < \infty$, independent of \(T\), such that $\suplim\limits_{T \rightarrow \infty}\PP(\hnu \notin \mathcal{V}_{K,T}) < \epsi$ for any $\epsi > 0$, where 
\begin{equation}\label{eq:VKT}
 \mathcal{V}_{K,T} \coloneqq \{\nu \in \mathcal{V}_0: \norm{\nu-\nu_0}\leq K/\sqrt{T}\}.
\end{equation}
Clearly, for sufficiently large \(T\), \( \mathcal{V}_{K,T}  \subseteq \mathcal{U}(\lambda_0)\), where the neighborhood has been defined in Assumption \ref{asu-E}. Hence,
\begin{equation}
\begin{split}
\suplim\limits_{T \rightarrow \infty}& \PP\bigg[\ssup\limits_{\theta,\ttheta \inn \Theta: \norm{\theta-\ttheta} \lleq \delta} \sqrt{T}\norm{\hat{\psi}_{T,S}(\theta,\hat{\nu}_T)-\psi(\theta)-\hat{\psi}_{T,S}(\ttheta,\hnu)+\psi(\ttheta)} > \eta\bigg] \nn \\
&\leq \suplim\limits_{T \rightarrow \infty} \PP(\hat{\nu}_T \notin  \mathcal{V}_{K,T}) \nn \\
& \hspace*{.5cm} + \suplim\limits_{T \rightarrow \infty}\PP\bigg[\Big\{\ssup\limits_{\theta_1,\theta_2 \inn \Theta: \norm{\theta-\ttheta} \lleq \delta} \sqrt{T}\norm{\hat{\psi}_{T,S}(\theta,\nu_T)-\psi(\theta)-\hat{\psi}_{T,S}(\ttheta,\nu_T)+\psi(\ttheta)} > \eta \Big\} \,\cap\, \big\{\hnu \in  \mathcal{V}_{K,T}\big\}\bigg] \\
&\leq \epsi + \suplim\limits_{T \rightarrow \infty}\PP\bigg[\ssup\limits_{\nu \inn  \mathcal{V}_{K,T},\,\theta,\ttheta \inn \Theta: \norm{\theta-\ttheta} \lleq \delta} \sqrt{T}\norm{\hat{\psi}_{T,S}(\theta,\nu)-\psi(\theta)-\hat{\psi}_{T,S}(\ttheta,\nu)+\psi(\ttheta)} > \eta\bigg].
\end{split}
\end{equation}
Since, by Lemma \ref{lemmaA3} and Lemma \ref{lemmaA4}, one has
\[ \suplim\limits_{T \rightarrow \infty}\PP\bigg[\ssup\limits_{\nu \inn  \mathcal{V}_{K,T}}\ssup\limits_{\theta \inn \Theta} \sqrt{T}\norm{\hat{\psi}_{T,S}(\theta,\nu)-\hat{\psi}_{T,S}(\theta,\nu_0)} > \eta\bigg] < \epsi, \]
it suffices to show that
\begin{equation}
\suplim\limits_{T \rightarrow \infty}\PP\bigg[\ssup\limits_{\theta,\ttheta \inn \Theta: \norm{\theta-\ttheta} \lleq \delta} \sqrt{T}\norm{\hat{\psi}_{T,S}(\theta,\nu_0)-\psi(\theta)-\hat{\psi}_{T,S}(\ttheta,\nu_0)+\psi(\ttheta)} > \eta\bigg] <  \epsi. \label{sest2}
\end{equation}
To begin with, recall that \(\theta_q = (\alpha_{q}',\beta_q',\gamma',\delta')'\), \(q = 1,\dots,Q\). Thus, by a slight abuse of notation, \(\theta = (\alpha_1',\beta_1',\dots,\alpha_Q',\beta_Q',\gamma',\delta')' =  \cup_{q \eq 1}^Q \theta_q\). Therefore, by the triangle inequality 
\begin{equation}\nn 
\begin{split}
\norm{&\hat{\psi}_{T,S}(\theta,\nu_0)-\psi(\theta)-\hat{\psi}_{T,S}(\ttheta,\nu_0)+\psi(\ttheta)} \\
&\leq \sum_{q \eq 1}^Q\frac{1}{\displaystyle{|\mathcal{G}_q| \choose 2}}\mathop{\sum\sum}\limits_{\substack{\ijn\\i,j \inn \mathcal{G}_q}}\sum_{k \eq 1}^\ell |\hat{\psi}_{T,S,i,j,k}(\theta_{q},\nu_0)-\psi_{q,k}(\theta_{q})-\hat{\psi}_{T,S,i,j,k}(\ttheta_{q},\nu_0)+\psi_{q,k}(\ttheta_q)|.
\end{split}
\end{equation}
Hence,
\begin{equation} \nn
\begin{split}
\PP\bigg[&\ssup\limits_{\theta,\ttheta \inn \Theta: \norm{\theta-\ttheta} \lleq \delta} \sqrt{T}\norm{\hat{\psi}_{T,S}(\theta,\nu_0)-\psi(\theta)-\hat{\psi}_{T,S}(\ttheta,\nu_0)+\psi(\ttheta)} > \eta\bigg] \\
& \hspace*{.25cm}\leq  \sum_{q \eq 1}^Q\mathop{\sum\sum}\limits_{\substack{\ijn\\i,j \inn \mathcal{G}_q}} \sum_{k \eq 1}^\ell \PP\bigg[\ssup\limits_{\theta_{q},\ttheta_{q} \inn \Theta: \norm{\theta-\ttheta} \lleq \delta}|\hat{\psi}_{T,S,i,j,k}(\theta_{q},\nu_0)-\psi_{q,k}(\theta_{q})\\
&  \hspace*{7.5cm}-\hat{\psi}_{T,S,i,j,k}(\ttheta_{q},\nu_0)+\psi_{q,k}(\ttheta_q)| > \frac{\eta}{Q\ell}\bigg].
\end{split}
\end{equation}
Now, suppose \(i,j \in \mathcal{G}_q\) for some \(q \in \{1,\dots,Q\}\). Moreover, let us recall from Eq. \eqref{hatCTS0} that uniformly in $\theta_q \in \Theta$ 
\begin{equation}
\begin{split}
\sqrt{T}|\hat{\psi}_{T,S,i,j,k}(\theta_q,\nu_0)-&\psi_{q,k}(\theta_q)| \\
 &  \leq \, \left\vert \int_{[0,1]^2} \dd\varphi_k(u_i,u_j) \right\vert \ssup\limits_{u_i,u_j \inn [0,1]}|\hat{\mathbb{C}}_{i,j,T,S}(u_i,u_j;\theta_q)| + |R(\theta_q)|
\end{split}
\end{equation}
for $1 \leq i < j \leq n$ and $k \in \{1,\dots,\ell\}$, where $\ssup\limits_{\theta \inn \Theta}|R(\theta_q)| = o_p(1)$. Since $n$ and $\ell$ are fixed, we are left with showing that for any $\epsi,\eta>0$, there exists some $\delta > 0$ such that
\begin{equation}
\begin{split}
\suplim\limits_{T \rightarrow \infty} \PP\bigg[\ssup\limits_{\theta_q,\ttheta_q \inn \Theta: \norm{\theta_q-\ttheta_q} \lleq \delta}\ssup\limits_{u_i,u_j \inn [0,1]}|\hat{\mathbb{C}}_{T,S,i,j}(u_i,u_j;\theta_q)-\hat{\mathbb{C}}_{T,S,i,j}(u_i,u_j;\ttheta_q)|> \eta \bigg] < \epsi. 
\end{split}
\end{equation}
In view of Eqs. \eqref{copid} and \eqref{Ctilde0}, we see that 
\begin{equation}\nn
\begin{split}
\hat{\mathbb{C}}_{T,S,i,j}(u_i,u_j;\theta_q) 
= \,& \sqrt{T}(\tCC_{T,S,i,j}\{\tCC_{T,S,i}^{-}(u_i;\theta_q),\tCC_{T,S,j}^{-}(u_j;\theta_q);\theta_q\}-\CC_{i,j}\{\tCC_{T,S,i}^{-}(u_i;\theta_q),\tCC_{T,S,j}^{-}(u_j;\theta_q);\theta_q\}) \\
\,& + \sqrt{T}(\CC_{i,j}\{\tCC_{T,S,i}^{-}(u_i;\theta_q),\tCC_{T,S,j}^{-}(u_j;\theta_q);\theta_q\}-\CC_{q}(u_i,u_j)). 
\end{split}
\end{equation}
with $\tCC_{T,S,i,j}(u_i,u_j;\theta_q) \coloneqq \tCC_{T,S,i,j}(u_i,u_j;\theta,\nu_0)$ and $\tCC_{T,S,k}^{-}(u_k;\theta_q) \coloneqq \tCC_{T,S,k}^{-}(u_k;\theta_q,\nu_0)$ for $k \in \{i,j\}$. Therefore, using an argument similar that in \citet[Appendix B]{tak05}, one obtains 
\begin{equation}
\begin{split}
\PP\bigg[&\ssup\limits_{\theta_q,\ttheta_q \inn \Theta: \norm{\theta_q-\ttheta_q} \lleq \delta}\ssup\limits_{u_i,u_j \inn [0,1]}|\hat{\mathbb{C}}_{T,S,i,j}(u_i,u_j;\theta_q)-\hat{\mathbb{C}}_{T,S,i,j}(u_i,u_j;\ttheta_q)| > \eta \bigg] \\
 & \leq  \,\PP\bigg[\ssup\limits_{\theta_q,\ttheta_q \inn \Theta: \norm{\theta_q-\ttheta_q} \lleq \delta}\ssup\limits_{u_i,u_j \inn [0,1]}|\tilde{\mathbb{C}}_{T,S,i,j}(u_i,u_j;\theta_q)-\tilde{\mathbb{C}}_{T,S,i,j}(u_i,u_j;\ttheta_q)| > \frac{\eta}{2}\bigg]\\
&  \hspace*{.2cm} + \sum_{k \in \{i,j\}} \PP\bigg[\ssup\limits_{\theta_q,\ttheta_q \inn \Theta: \norm{\theta_q-\ttheta_q} \lleq \delta}\ssup\limits_{u_k \inn [0,1]}|\tilde{\mathbb{C}}_{T,S,i,j}(\vec{u}_k;\theta_q)-\tilde{\mathbb{C}}_{T,S,i,j}(\vec{u}_k;\ttheta_q)|> \frac{\eta}{4}\bigg].
\end{split}
\end{equation}
The claim thus follows from part \ref{lemmaU-c} of Lemma \ref{lemmaU}.

\label{app:A1}

\subsection[A.2 Proof of Corollary \ref{cor-1}]{Proof of Corollary \ref{cor-1}}
\noindent \textbf{Proof of Corollary} \ref{cor-1}: As argued in \citet[Lemma 1]{cheng15}, it suffices to show that the conditional distribution of \(\sqrt{T}(\hat{\Psi}_{T,S}^{(b)}(\htheta,\hlambda,\hnu)-\hat{\Psi}_{T,S}(\htheta,\hlambda,\hnu))\) converges  in probability to the limiting distribution of \(\sqrt{T}\hat{\Psi}_{T,S}(\theta_0,\hlambda,\hnu)\) given in Proposition \ref{prop-1}. 
Specifically, define for any \(\ijn\), with \(i,j \in \mathcal{G}_q\), \(q \in \{1,\dots,Q\}\):
\begin{equation}\nn
\begin{split}
\sqrt{T}(\hat{\Psi}_{T,S,i,j}^{(b)}(\hthetaq,\hlambda,\hnu)   -& \hat{\Psi}_{T,S,i,j}(\hthetaq,\hlambda,\hnu)) \\ 
= & \,    \hat{\xi}_{T,S,i,j}^{(b)}(\theta_{0,q},\hlambda,\hnu)  \\
& \,+ \hat{\xi}_{T,S,i,j}^{(b)}(\hthetaq,\hlambda,\hnu) - \hat{\xi}_{i,j,T,S}^{(b)}(\theta_{0,q},\hlambda,\hnu),
\end{split}
\end{equation}
where
\begin{equation}\label{xib}
\begin{split}
 \hat{\xi}_{T,S,i,j}^{(b)}(\hthetaq,\hlambda,\hnu) \coloneqq \sqrt{T}(\hat{\Psi}_{T,S,i,j}^{(b)}(\hthetaq,\hlambda,\hnu) - \hat{\Psi}_{i,j,T,S}(\hthetaq,\hlambda,\hnu)) 
\end{split}
\end{equation}
represents the bootstrap analogue of 
\begin{equation}\label{xi}
\hat{\xi}_{T,S,i,j}(\hthetaq,\hlambda,\hnu) = \sqrt{T}(\hat{\Psi}_{T,S,i,j}(\hthetaq,\hlambda,\hnu) - \Psi_{T,S,i,j}(\hthetaq)).
\end{equation}
 We proceed in two steps: 1) the conditional distribution of $\hat{\xi}_{T,S,i,j}^{(b)}(\theta_{0,q},\hlambda,\hnu)$ converges in probability to that of $\hat{\xi}_{T,S,i,j}(\theta_{0,q},\hlambda,\hnu) = \sqrt{T}\hat{\Psi}_{T,S,i,j}(\theta_{0,q},\hlambda,\hnu)$; 2) $\hat{\xi}_{T,S,i,j}^{(b)}(\hthetaq,\hlambda,\hnu) - \hat{\xi}_{T,S,i,j}^{(b)}(\theta_{0,q},\hlambda,\hnu) = o_p(1)$. \underline{Step 1):} Mimicking the derivation of Eq. \eqref{Psik}, note that the $k$-th element of $\hat{\xi}_{T,S,i,j}^{(b)}(\theta_{0,q},\hlambda,\hnu)$ can be written as:
\[
\int_{[0,1]^2}(\hat{\mathbb{B}}_{T,S,i,j}^{(b)}-\hat{\mathbb{B}}_{T,S,i,j})(u_i,u_j;\theta_{0,q},\lambda_{0},\nu_0)\,\dd\varphi_k(u_i,u_j) + o_p(1), 
\]
where $\hat{\mathbb{B}}_{T,S,i,j}^{(b)}$ is the bootstrap analogue of $\hat{\mathbb{B}}_{T,S,i,j}$ defined in Eq. \eqref{Bemp}. We are thus left with showing that the conditional distribution of \(\hat{\mathbb{B}}_{T,S,i,j}^{(b)}-\hat{\mathbb{B}}_{T,S,i,j}\) converges weakly in probability to the same limiting process that governs the weak limit of $\hat{\mathbb{B}}_{T,S,i,j}$. But, by the functional delta method, this is due the weak convergence of $\hat{\mathbb{B}}_{T,S,i,j}$; see the proof of Proposition \ref{prop-1} in conjunction with \citet[Theorem 5]{fetal04}. \underline{Step 2):} We can deduce the bootstrap stochastic equicontinuity of \eqref{xib} from that of \eqref{xi} established in Proposition \ref{prop-1} \ref{prop-1-c}; see also \cite{gizi90}, \cite{brwe02},  or \cite{chetal03} for a similar argument.  The claim then follows because, by Proposition \ref{prop-3}, $\sqrt{T}\norm{\htheta-\theta_0} = O_p(1)$.

\setcounter{equation}{0}
\section[B Auxiliary results]{Auxiliary results}

This section contains auxiliary results. We make frequently use of the following stochastic equicontinuity result due to \citet[Theorem 2.2]{anpo94}: 

\begin{theorem}\label{thmap-94}\normalfont For any $T$, let $\{\xi_{t,T}: t = 1,\dots,T;\, T \in \mathbb{N}_1\}$ be a strong mixing triangular array whose mixing coefficients satisfy
\begin{equation}
\sum_{i \eq 1}^\infty i^{p-2}\alpha(i)^{\theta/(p+\theta)} < \infty \tag{$i$}\label{AP-c1}
\end{equation}
for $p \geq 2$, $p \in \mathbb{N}$, and $\theta > 0$, and let $\mathcal{F}$ be a uniformly bounded class of real-valued functions whose bracketing number satisfies
\begin{equation}
\int_0^1 x^{-\theta/(2+\theta)}\mathcal{N}_{[\,]}(x,\mathcal{F},\rho(\cdot))^{1/p} \,\dd x< \infty,\tag{$ii$}\label{AP-c2}
\end{equation}
for the same $p$ and $\theta$, while $\rho(f) \coloneqq \ssup_{t,T}\norm{f(\xi_{t,T})}_2$. Then for each $\epsilon_0 > 0$ there is a $\epsilon_1 > 0$ such that
\begin{equation}
\suplim\limits_{T \rightarrow \infty} \left\lVert \ssup_{\rho(f-g) < \epsilon_1}\left\vert\frac{1}{\sqrt{T}}\sum_{t\eq 1}^T[f(\xi_{t,T})-g(\xi_{t,T})]\right\vert \right\rVert_p <  \epsilon_0. \tag{$\ast\ast$} \label{AP-ec}
\end{equation}
\end{theorem} 

\begin{remark}\label{remark:bracket} Note that $\int_0^1 1/x^c \text{d}x < \infty$ for $c < 1$. Thus, if $\alpha(i) = O(i^{-a})$ and $\mathcal{N}_{[\,]}(x,\mathcal{F},\rho(\cdot)) = O(i^{-b})$ for $a,b > 0$, then \eqref{AP-c1} and \eqref{AP-c2} are satisfied if $p > b(2+\theta)/2$ and $a > (p-1)(p+\theta)/\theta$ so that $p \geq 2$ and $p \in \mathbb{N}$.
\end{remark}

\noindent Recall that the bracketing number $\mathcal{N}_{[\,]}(\eps,\mathcal{F},\rho)$ denotes the  minimum number of $\eps$-brackets needed to cover a class $\mathcal{F}$ of functions $f: \mathcal{X} \mapsto \mathbb{E}$ equipped with a metric $\rho$:
\begin{equation}\label{bracket}
\mathcal{N}_{[\,]}(\eps,\mathcal{F},\rho) \coloneqq \mmin\{m: \exists \,(\{l_j,u_j\})_{j = 1}^m \text{ s.th. } \rho(u_j,l_j) \leq \eps,\, \mathcal{F} \subseteq \bigcup_{j \eq 1}^m[l_j,u_j]\},
\end{equation}
 where the bracket $[l_j,u_j]$ is defined by
\[
[l_j,u_j] \coloneqq \{f \in \mathcal{F}: l_j(x) \leq h(x) \leq u_j(x),  \forall x \in \mathcal{X}\},
\]
see, e.g. \cite{vw96}.

 \begin{remark}  Let \(k\) and \(K\) be constants independent of \(T\), and define
\begin{equation}\label{eq:VKT}
 \mathcal{V}_{k,T} \coloneqq \{\nu \in \mathcal{V}_0: \norm{\nu-\nu_0}\leq k/\sqrt{T}\},\quad \Lambda_{K,T} \coloneqq \{\lambda \in \Lambda_0: \norm{\lambda-\lambda_0}\leq K/\sqrt{T}\}
\end{equation}
Then, by Assumption \ref{asu-E}, for \(T\) sufficiently large and any \(\epsi > 0\), there exist \(k \coloneqq k(\epsi)\), \(K \coloneqq K(\epsi)\), such that
$\normalfont\suplim\limits_{T \rightarrow \infty}\PP(\hnu \notin \mathcal{V}_{k,T}) < \epsi$ and $\normalfont\suplim\limits_{T \rightarrow \infty}\PP(\hlambda \notin \Lambda_{K,T}) < \epsi$.
For brevity, we write \(\mathcal{V}_{T} \coloneqq \mathcal{V}_{k,T}\) and \(\Lambda_{T} \coloneqq \Lambda_{K,T}\).
\end{remark}

\begin{lemma}\label{lemmaV}\normalfont For any $1\leq i < j \leq n$ with \(i,j \in \mathcal{G}_q\), \(q \in \{1,\dots,Q\}\), 
\begin{enumerate}[label= ($\alph*$)]
\item\label{lemmaV-a} $\{\tilde{\mathbb{C}}_{T,i,j}(u_i,u_j;\lambda):\,u_i,u_j \in [0,1],\,  \lambda \in \Lambda_{T}\}$ is stochastically equicontinuous;
\item\label{lemmaV-b} Let $\tilde{\mathbb{V}}_{T,i,j}(u_i,u_j;\lambda)\coloneqq \sqrt{T}(\tCC_{T,i,j}(u_i,u_j;\lambda)-\tCC_{T,i,j}(u_i,u_j;\lambda_{0}))$. Then, \[
\tilde{\mathbb{V}}_{T}(u_i,u_j;\lambda) = \bar{\mathbb{V}}_{T}(u_i,u_j;\lambda) + o_p(1),
\] uniformly in $u_i,u_j \in [0,1], \lambda \in \Lambda_{T}$, with
\begin{equation}
\begin{split}
\bar{\mathbb{V}}_T&(u_i,u_j;\lambda) \nn  \\
 & =\,  \sum_{k \inn \{\ii\}} \partial_k\CC_{q}(u_i,u_j)\ff_k\{\FF_k^{-1}(u_k)\}\Ex[\tau_k(R_1,u_k,\lambda_{0})]\sqrt{T}(\lambda-\lambda_{0})  + o_p(1), \nn
\end{split}
\end{equation}
with \(\tau_k(R_t;u_k,\lambda_0) = \tau_k(\mathcal{F}_t;u_k,\lambda_0)\), where for any \(\bar{\lambda} \in \mathcal{V}_T\) 
\[
\tau_k(\mathcal{F}_t;u_k,\bar{\lambda}) \coloneqq \frac{\FF_k^{-1}(u_k)\nabla_{\hspace*{-.05cm}\lambda} \mu_{2,k}(R_t(\lambda),\lambda)\at[\big]{\lambda \eq \bar{\lambda}}+\nabla_{\hspace*{-.05cm}\lambda} \mu_{1,k}(R_t(\lambda),\lambda)\at[\big]{\lambda \eq \bar{\lambda}}}{\mu_{k,2}(R_t,\lambda_{0})},\;\; k \in \{i,j\},
\] and we recall that \(R_t = R_t(\lambda_0)\).
\item\label{lemmaV-c} $\{\tilde{\mathbb{C}}_{T,i,j}(u_i,u_j;\lambda_{0}): u_i,u_j \in [0,1]\}$ converges weakly to a tight Gaussian process in $\ell^\infty([0,1]^2)$.
\end{enumerate}
\end{lemma}

\begin{lemma}\label{lemmaU}\normalfont For any $1\leq i < j \leq n$ with \(i,j \in \mathcal{G}_q\), \(q \in \{1,\dots,Q\}\),
\begin{enumerate}[label= ($\alph*$)]
\item\label{lemmaU-a} $\{\tilde{\mathbb{C}}_{T,S,i,j}(u_i,u_j;\theta_q,\nu):u_i,u_j \in [0,1], \theta_q \in \Theta, \nu \in \mathcal{V}_0\}$ is stochastically equicontinuous;
\item\label{lemmaU-b} Let $\tilde{\mathbb{U}}_{T,S,i,j}(u_i,u_j;\theta_q,\nu)\coloneqq \sqrt{T}(\tCC_{T,S,i,j}(u_i,u_j;\theta_q,\nu)-\tCC_{T,S,i,j}(u_i,u_j;\theta_q,\nu_0))$. Then, $\tilde{\mathbb{U}}_{T,S}(u_i,u_j;\theta_q,\nu) = \bar{\mathbb{U}}_{T}(u_i,u_j;\theta_q,\nu) + o_p(1)$, uniformly in $u_i,u_j \in [0,1], \theta_q \in \Theta, \nu \in \mathcal{V}_T$, with
\begin{equation}
\begin{split}
\bar{\mathbb{U}}_T(u_i,u_j;&\theta_q,\nu) \nn  \\
 & =\,  \beta\Ex[\nabla_\nu \sigma(M_t,\nu_0)]\sqrt{T}(\nu-\nu_0)\sum_{k \inn \{\ii\}} \partial_k\CC_{q}(u_i,u_j;\theta_q)\gg_q\{\GG_q^{-1}(u_k;\theta_q);\theta_q\}  + o_p(1); \nn
\end{split}
\end{equation}
\item\label{lemmaU-c} $\{\tilde{\mathbb{C}}_{T,S,i,j}(u_i,u_j;\theta,\nu_0): u_i,u_j \in [0,1], \theta_q \in \Theta\}$ converges weakly to a tight Gaussian process in $\ell^\infty([0,1]^2 \times \Theta)$.
\end{enumerate}
\end{lemma}
 
\begin{lemma}\label{lemmaA3}\normalfont Set \(\hat{\mathbb{V}}_{T,i,j}(u_i,u_j;\lambda)\coloneqq \sqrt{T}(\hCT(u_i,u_j;\lambda)-\hCT(u_i,u_j;\lambda_{0}))\), and define \(\hat{\mathbb{U}}_{T,S,i,j}(u_i,u_j;\theta_q,\nu)\coloneqq \sqrt{T}(\hCTS(u_i,u_j;\theta_q,\nu)-\hCTS(u_i,u_j;\theta_q,\nu_0))\). Then, 
\[
\textnormal{($a$) }\,\,\ssup\limits_{u_i,u_j \inn [0,1]}|\hat{\mathbb{V}}_{T,i,j}(u_i,u_j;\hlambda)| = o_p(1),\;\; \textnormal{($b$) }\,\,\ssup\limits_{\theta_q \inn \Theta}\ssup\limits_{u_i,u_j \inn [0,1]}|\hat{\mathbb{U}}_{i,j,T,S}(u_i,u_j;\theta_q,\hnu)| = o_p(1).
\]
\end{lemma}

\begin{lemma}\label{lemmaA4}
\normalfont For any $k \in \{1,\dots,\ell\}$ and \(i,j \in \mathcal{G}_q\), \(q \in \{1,\dots,Q\}\), we have
\begin{equation}\tag{$a$}
\begin{split}
\sqrt{T}(\hat{\psi}_{T,i,j,k}&(\hlambda)-\psi_{q,k})\\
\,&  = \int_{[0,1]^2} \sqrt{T}(\hCT(u_i,u_j;\hlambda)-\CC_{q}(u_i,u_j))\, \dd \varphi_k(u_i,u_j) + o_p(1).
\end{split}
\end{equation}
and, uniformly in $\theta_q \in \Theta$,
\begin{equation}\tag{$b$}
\begin{split}
\sqrt{T}(\hat{\psi}_{T,S,i,j,k}&(\theta_q,\hnu)-\psi_{q,k}(\theta_q)) \\ 
\,& = \int_{[0,1]^2} \sqrt{T}(\hCTS(u_i,u_j;\theta_q,\hnu)-\CC_{q}(u_i,u_j;\theta_q))\, \dd \varphi_k(u_i,u_j) + o_p(1).
\end{split}
\end{equation}
\end{lemma}

\subsection{Proof  of Lemma \ref{lemmaV}}
\textbf{Proof of Lemma \ref{lemmaV}-\ref{lemmaV-a}:} 
To begin with, note that
\begin{equation}
\begin{split}
1 \{&\hat{\eta}_{i,t}(\lambda)\leq \FF_i^{-1}(u_i),\hat{\eta}_{j,t}(\lambda)  \leq \FF_j^{-1}(u_j)\}  \\
&  = 1\{\eta_{i,t} \leq \FF_i^{-1}(u_i)a_{2,i}^\dagger(\mathcal{F}_t,\lambda)+a_{1,i}^\dagger(\mathcal{F}_t,\lambda),\,\eta_{j,t}\leq  \FF_j^{-1}(u_j)a_{2,j}^\dagger(\mathcal{F}_t,\lambda)+a_{1,j}^\dagger(\mathcal{F}_t,\lambda)\} \\
& =  g(\eta_{i,t},\eta_{j,t},\mathcal{F}_t),
\end{split}
\end{equation}
where \(g\) can be viewed as a an element of the function class
\begin{equation}\nn 
\begin{split}
\mathcal{G} \coloneqq \Big\{\eta_{i,t},\eta_{j,t},\mathcal{F}_{t} \mapsto 1\{&\eta_{i,t} \leq  \FF_i^{-1}(u_i)a_{2,i}^\dagger(\mathcal{F}_t,\lambda)+a_{1,i}^\dagger(\mathcal{F}_t,\lambda),\\
\,& \eta_{j,t} \leq \FF_j^{-1}(u_j)a_{2,j}^\dagger(\mathcal{F}_t,\lambda)+a_{1,j}^\dagger(\mathcal{F}_t,\lambda)\}: u_i,u_j \in [0,1], \lambda \in \Lambda_{T}\Big\},
\end{split}
\end{equation}
with
\[
a_{1,k}^\dagger(\mathcal{F}_t,\lambda) \coloneqq \, \frac{\mu_{1,k}(R_{t}(\lambda),\lambda)-\mu_{1,k}(R_{t},\lambda_0)}{\mu_{2,k}(R_{t},\lambda_{0})}\, \text{ and } \,a_{2,k}^\dagger(\mathcal{F}_t,\lambda) \coloneqq \, \frac{\mu_{2,k}(R_{t}(\lambda),\lambda)}{\mu_{2,k}(R_{t},\lambda_{0})},\;\;\; k \in \{i,j\}.
\]
Therefore, $\tilde{\mathbb{C}}_{T,i,j}(u_i,u_j;\lambda)$ can be viewed as an empirical process indexed by $\mathcal{G}$, i.e.,
\begin{equation}\label{emp-V}
\begin{split}
\tilde{\mathbb{C}}_{T,i,j}: \mathcal{G} \longrightarrow &\,\mathbb{R} \\
g \in \mathcal{G},\,g  \longmapsto &\,\tilde{\mathbb{C}}_{T,i,j}(g) \coloneqq \frac{1}{\sqrt{T}}\sumt \big\{g(\eta_{i,t},\eta_{j,t},\mathcal{F}_t) - \Ex[g(\eta_{i,t},\eta_{j,t},\mathcal{F}_t)]\big\}.
\end{split}
\end{equation}
Instead of showing stochastic equicontinuity of \(\{\tilde{\mathbb{C}}_{T,i,j}(g): g \in \mathcal{G}\}\) directly, we show below stochastic equicontinuity of
\begin{equation}\label{emp-Vcheck}
\check{\mathbb{C}}_{T,i,j}(g_1) \coloneqq \frac{1}{\sqrt{T}}\sumt \big\{g_1(\eta_{i,t},\eta_{j,t},R_t) - \Ex[g_1(\eta_{i,t},\eta_{j,t},R_t)]\big\},\;\;g_1 \in \mathcal{G}_1
\end{equation}
for
\begin{equation}\nn
\begin{split}
\mathcal{G}_1 \coloneqq \Big\{\eta_{i,t},\eta_{j,t},R_{t} \mapsto 1\{&\eta_{i,t} \leq  \FF_i^{-1}(u_i)a_{2,i}(R_t,\lambda)+a_{1,i}(R_t,\lambda), \\
\,&\eta_{j,t} \leq \FF_j^{-1}(u_j)a_{2,j}(R_t,\lambda)+a_{1,j}(R_t,\lambda)\}: u_i,u_j \in [0,1], \lambda \in \Lambda_{T}\Big\},
\end{split}
\end{equation}
with
\[
a_{1,k}(R_t, \lambda) \coloneqq \,\frac{\nabla \mu_{2,k}(R_{t}, \lambda_0)}{\mu_{2,k}(R_{t},\lambda_{0})}(\lambda-\lambda_{0})\,\text{ and }\,a_{2,k}(R_t, \lambda) \coloneqq \,1+\frac{\nabla \mu_{2,k}(R_{t}, \lambda_0)}{\mu_{2,k}(R_{t},\lambda_{0})}(\lambda-\lambda_{0}),\;\;\; k \in \{i,j\}. 
\]
Observe that a first order Taylor approximation yields for \(k \in \{i,j\}\) and \(l \in \{1,2\}\)
\begin{equation}
\begin{split}
 \mu_{l,k}(R_{t}(\lambda), \lambda) =  \mu_{l,k}(R_{t}, \lambda_0) \, + \,& \nabla \mu_{l,k}(R_{t}, \lambda_0) (\lambda-\lambda_{0})  \\
 \, + \,& \frac{1}{2} (\lambda-\lambda_{0})'\nabla_{\lambda}^2 \mu_{l,k}(R_t(\bar{\lambda}), \bar{\lambda})(\lambda-\lambda_{0}),
\end{split}
\end{equation}
where \(\bar{\lambda} = \lambda c + \lambda_{0}(1-c) \in \Lambda_T\) for some \(c \in (0,1)\) and, by Assumption \ref{asu-E}, 
\[
\ssup\limits_{1 \lleq t \lleq T}\ssup\limits_{\lambda \inn \Lambda_T}\norm{\nabla_{\lambda}^2 \mu_{l,k}(R_{t}(\lambda), \lambda)/\mu_{2,k}(R_t,\lambda_0)} = O_p(1).
\]
Therefore, we obtain
\begin{equation}\label{eq:rouven}
a_{l,k}^\dagger(\mathcal{F}_t,\lambda)  =  a_{l,k}(R_t, \lambda) + o_p(T^{-1/2}),\; k \in \{i,j\}, \,l \in \{1,2\}.
\end{equation}
Similar to the initial argument of the proof of Lemma 1 in \citet[p. 565]{avk01}, which is also employed in the proof of Lemma 1 in \citet[pp. 794-796]{detal09}, conclude from Eq. \eqref{eq:rouven} that \(\textsf{P}(g(\eta_{i,t},\eta_{j,t},\mathcal{F}_t) \in  \mathcal{G}_1) \rightarrow 1\) for any \(g \in \mathcal{G}\). More specifically, by monotonicity of the indicator function,  one gets
\begin{equation}\label{eq:monotonic}
\begin{split}
1\{x \leq ya_2+a_1\}\geq \,& 1\{x \leq y(b_2-|b_2-a_2|\textsf{sign}(y))+b_1+|b_1-a_1|\}\\
1\{x \leq ya_2+a_1\}\leq \,& 1\{x \leq y(b_2+|b_2-a_2|\textsf{sign}(y))+b_1-|b_1-a_1|\}
\end{split}
\end{equation}
for any \(x,y,a_1,a_2,b_1,b_2 \in \mathbb{R}\). Adopting an argument used  by \citet[pp. 151-153]{neumetal19} in their Appendix A.2, we get from Eq. \eqref{eq:rouven} in conjunction with Eq. \eqref{eq:monotonic} on an event with probability converging to one 
\begin{equation}
\begin{split}
\frac{1}{\sqrt{T}}\sumt[\check{g}_1(\eta_{i,t},\eta_{j,t},R_{t},-\gamma_T)-\check{g}_1(\eta_{i,t},\eta_{j,t},R_{t},0)] \leq & \\
\frac{1}{\sqrt{T}}\sumt [g(\eta_{i,t},\eta_{j,t},\mathcal{F}_t) -& g_1(\eta_{i,t},\eta_{j,t},R_t)] \\
\leq \frac{1}{\sqrt{T}}\sumt[\check{g}_1&(\eta_{i,t},\eta_{j,t},R_{t};\gamma_T)-\check{g}_1(\eta_{i,t},\eta_{j,t},R_{t};0)] 
\end{split}
\end{equation}
for any \(g \in \mathcal{G}\), \(g_1 \in \mathcal{G}_1\), where \(\gamma_T = o(T^{-1/2})\) is a deterministic sequence and
\begin{equation}
\begin{split}
\check{g}_1(\eta_{i,t},\eta_{j,t},R_{t},\gamma) \coloneqq 1\{&\eta_{i,t} \leq  \FF_i^{-1}(u_i)(a_{2,i}(R_t,\lambda)+\gamma\textsf{sign}(\FF_i^{-1}(u_i)))+a_{1,i}(R_t,\lambda)+\gamma,\\
&\eta_{j,t} \leq \FF_j^{-1}(u_j)(a_{2,i}(R_t,\lambda)+\gamma\textsf{sign}(\FF_i^{-1}(u_i)))+a_{1,j}(R_t,\lambda)+\gamma\}.
\end{split}
\end{equation}
Since \(\check{g}_1(\eta_{i,t},\eta_{j,t},R_{t},0) \in \mathcal{G}_1\) and \(\textsf{P}(\check{g}(\eta_{i,t},\eta_{j,t},R_{t},\pm\gamma_T) \in \mathcal{G}_1) \rightarrow 1\), it readily follows from the stochastic equicontinuity of \(\{\check{\mathbb{C}}_{T,i,j}(g_1): g_1 \in \mathcal{G}_1\}\) (shown below) that, by the same arguments used in \cite{neumetal19},
\[
\frac{1}{\sqrt{T}}\sumt[\check{g}_1(\eta_{i,t},\eta_{j,t},R_{t},\pm\gamma_T)-\check{g}_1(\eta_{i,t},\eta_{j,t},R_{t},0)] = o_p(1),
\]
which holds uniformly over \(u_i,u_j \in [0,1]\), \(\lambda \in \Lambda_T\), so that
\[
\frac{1}{\sqrt{T}}\sumt [g(\eta_{i,t},\eta_{j,t},\mathcal{F}_t) -g_1(\eta_{i,t},\eta_{j,t},R_t)] = o_p(1)
\]
uniformly in \(g \in \mathcal{G}\), \(g_1 \in \mathcal{G}_1\). 

\noindent \emph{Stochastic equicontinuity of \(\{\check{\mathbb{C}}_{T,i,j}(g_1): g_1 \in \mathcal{G}_1\}\).} To make use of Theorem \ref{thmap-94}, the bracketing number $\mathcal{N}(\epsilon,\mathcal{G}_1,\norm{\cdot}_2)$ needs to be determined. Since $\mathcal{G}_1 \subset \mathcal{H}_i\cdot\mathcal{H}_j$ and $\mathcal{N}_{[\,]}(2\epsi,\mathcal{G}_1,\norm{\cdot}_2) \leq \mathcal{N}_{[\,]}(\epsi,\mathcal{H}_i,\norm{\cdot}_2)\mathcal{N}_{[\,]}(\epsi,\mathcal{H}_j,\norm{\cdot}_2)$ [see, e.g., see, e.g., \citet[Lemma 9.25 ($ii$)]{ko08}], it remains to compute $\mathcal{N}_{[\,]}(\epsi,\mathcal{H}_k,\norm{\cdot}_2)$, where 
\[
\mathcal{H}_k \coloneqq \Big\{\eta_{k,t},R_{t} \mapsto 1\{\eta_{k,t} \leq  ya_{2,k}(R_t,\lambda)+a_{1,k}(R_t,\lambda)\}: y \in \mathbb{R}, \lambda \in \Lambda_{T}\Big\},\;\;k \in \{i,j\}.
\]
An element $h \in \mathcal{H}_k$ can be uniquely identified by $\{y,\lambda\}$; thus, we use the notation $h(\eta_{k,t},R_{t}) = h(\eta_{k,t},R_{t};y,\lambda)$ to refer to a specific member of $\mathcal{H}_k$. 

Begin by determining $\mathcal{N}(\epsilon,\mathcal{H}_k,\norm{\cdot}_2)$. In doing so, we mimic the proof of Lemma 1 in \cite{avk01}. For brevity, let us drop the index $k \in \{i,j\}$. Since $\Lambda_0 \subset \mathbb{R}^{r}$ is compact, we can fix a grid $\{\lambda^{(1)},\dots,\lambda^{(N)}\}$, with $N = O(\eps^{-2r})$, to cover $\Lambda_{T}$ with $N$ balls $\{\mathcal{B}_1,\dots,\mathcal{B}_{N}\}$, each of radius $\epsi^2$; i.e., for each $\lambda$, there exists at least one integer $i \in \{1,\dots,N\}$ such that $\lambda \in \mathcal{B}_{i} \coloneqq \{\tilde{\lambda} \in \Lambda_{T}: \norm{\lambda-\lambda^{(i)}} \leq \epsi^2\}$. Now, for such an integer $i$, Assumption \ref{asu-E} ensures
$a_j(R_t,\lambda) \leq a_j(R_t,\lambda^{(i)}) + \epsi^2\dot{\mu}(R_t),\;\;j \in \{1,2\}.$ Therefore, $\{a_j(R_t,\lambda): \lambda \in \Lambda_T\}$ can be covered by $[l_{j,i}(R_t),u_{j,i}(R_t)]$ for $j \in \{1,2\}$, where
$u_{j,i}(R_t) \coloneqq a_j(R_t,\lambda^{(i)}) + \epsi^2\dot{\mu}(R_t)$, $l_{j,i}(R_t) \coloneqq  a_j(R_t,\lambda^{(i)}) - \epsi^2\dot{\mu}(R_t)$.
Hence, by monotonicity of the indicator function, we can cover each $h \in \mathcal{H}$ for a fix $y \in \mathbb{R}$ by 
\[
1\{\eta_{t} \leq  yl_{2,j}(R_t)+l_{1,i}(R_t)\} \leq h(\eta_t,R_t;y,\lambda) \leq 1\{\eta_{t} \leq  yu_{2,j}(R_t)+u_{1,i}(R_t)\},\;\;i,j \in \{1,\dots,N\}.
\]
Next, we can partition the real line by a total of $N_1 = O(\epsi^{-2})$ points for given $\{i,j\}$; i.e., $-\infty = \bar{y}_{i,j,1} \leq \bar{y}_{i,j,2} \leq \dots \leq \bar{y}_{i,j,N_1}  = \infty$
such that for $2 \leq k \leq N_1$
\[\FF(\bar{y}_{i,j,k}\,u_{2,j}(R_t)+u_{1,i}(R_t)) - \FF(\bar{y}_{i,j,k-1}\,u_{2,j}(R_t)+u_{1,i}(R_t)) \leq K\epsi^2.\]
Similarly, for some $N_2 = O(\epsi^{-2})$ there exists a partition  $-\infty = \munderbar{y}_{i,j,1} \leq \munderbar{y}_{i,j,2} \leq \dots \leq \munderbar{y}_{i,j,N_2}  = \infty$ such that for $2 \leq k \leq N_2$
 \[\FF(\munderbar{y}_{i,j,k}\,l_{2,j}(R_t)+l_{1,i}(R_t)) - \FF(\munderbar{y}_{i,j,k-1}\,l_{2,j}(R_t)+l_{1,i}(R_t)) \leq K\epsi^2.\]
Now, define the following brackets for $y$ by $\munderbar{y}_{i,j}^\dagger  \leq y \leq \bar{y}_{i,j}^\dagger$, with
 \[ \munderbar{y}_{i,j}^\dagger  \coloneqq \mmax\limits_{1\,\leq\,k\,\leq\,N_1}\{\munderbar{y}_{i,j,k}: \munderbar{y}_{i,j,k} \leq y\},\;\; \bar{y}_{i,j}^\dagger \coloneqq  \mmin\limits_{1\,\leq\,k\,\leq\,N_2}\{\bar{y}_{i,j,k}: \bar{y}_{i,j,k} \geq y\}.\]
Thus, with $L_{i,j}(\eta_t,R_t) \coloneqq 1\{\eta_{t} \leq  \munderbar{y}_{i,j}^\dagger l_{2,j}(R_t)+l_{1,i}(R_t)\}$, $U_{i,j}(\eta_t,R_t) \coloneqq 1\{\eta_{t} \leq  \bar{y}_{i,j}^\dagger u_{2,j}(R_t)+u_{1,i}(R_t)\}$, we obtain a total of $N \coloneqq O(\epsi^{-4(r+1)})$ brackets $[L_{i,j}(\eta_t,R_t),U_{i,j}(\eta_t,R_t)]$ covering $\mathcal{H}$ and whose $\norm{\cdot}_2$ length is $O(\epsi)$. To see this, note that
\begin{equation}
\begin{split}
\hspace*{-.25cm}\norm{U_{i,j}(\eta_t,R_t)-L_{i,j}(\eta_t,R_t)}_2^2 = \,& \Ex[\FF(\bar{y}_{i,j}^\dagger\,u_{2,j}(R_t)+u_{1,i}(R_t))] - \Ex[\FF(\munderbar{y}_{i,j}^\dagger\,l_{2,j}(R_t)+l_{1,i}(R_t))] \\
\leq  \,& \Ex[\FF(y\,u_{2,j}(R_t)+u_{1,i}(R_t))]\\
\,& - \Ex[\FF(y\,l_{2,j}(R_t)+l_{1,i}(R_t))] + K\epsi^2.
\end{split}
\end{equation}
Next, set $b_{q,i}(R_t;z) \coloneqq a_{q}(R_t,\lambda^{(i)}) + z \dot{\mu}(R_t)$ and note that $l_{q,i}(R_t) = b_{q,i}(R_t;-\epsi^2)$ and $u_{q,i}(R_t) = b_{q,i}(R_t;\epsi^2)$ for $q \in \{1,2\}$. By the mean-value theorem, there exists $x \in (-\epsi^2,\epsi^2)$ such that 
\begin{equation}\label{bracket-distV}
\begin{split}
\hspace*{-.25cm}\Ex[\FF(\,&y\,u_{2,j}(R_t)+u_{1,i}(R_t))] -  \Ex[\FF(y\,l_{2,j}(R_t)+l_{1,i}(R_t))]\\
= \,& 2\epsi^2\Ex[\ff\{yb_{2,i}(R_t;x)+b_{1,j}(R_t;x)\}(1+y)\dot{\mu}(R_t)] \\
= \,& 2\epsi^2\Ex[\ff\{yb_{2,i}(R_t;x)+b_{1,j}(R_t;x)\}[yb_{2,i}(R_t;x)+b_{1,j}(R_t;x)]\dot{\mu}(R_t)/b_{2,i}(R_t;x)] \\
\,& - 2\epsi^2\Ex[\ff\{yb_{2,i}(R_t;x)+b_{1,j}(R_t;x)\}\dot{\mu}(R_t)b_{1,i}(R_t;x)/b_{2,i}(R_t;x)] \\
\,& + 2\epsi^2\Ex[\ff\{yb_{2,i}(R_t;x)+b_{1,j}(R_t;x)\}\dot{\mu}(R_t)] \\
\leq \,& 2\epsi^2(\munderbar{b}\Ex[\dot{\mu}(R_1)]\ssup\limits_{y \inn \mathbb{R}} |\ff(y)y|  + \munderbar{b}\Ex[\dot{\mu}(R_1)^2](\epsi^2+K_\lambda/\sqrt{T})\ssup\limits_{y \inn \mathbb{R}} \ff(y)  + \Ex[\dot{\mu}(R_1)]\ssup\limits_{y \inn \mathbb{R}} \ff(y)),
\end{split}
\end{equation}
where the last equality uses that, by Assumption \ref{asu-E}, there exists a $\munderbar{b} \in (0, \infty)$ such that $b_{2,i}(R_t;x) > 1/\munderbar{b}$ and $|b_{1,i}(R_t;x)| \leq \dot{\mu}(R_t)(|x|+\norm{\lambda^{(i)}-\lambda_0})$. Therefore, 
\begin{equation}
\mathcal{N}_{[\,]}(\epsilon,\mathcal{G}_1,\norm{\cdot}_2) \leq \mathcal{N}_{[\,]}(\epsi/2,\mathcal{H}_i,\norm{\cdot}_2)\mathcal{N}_{[\,]}(\epsi/2,\mathcal{H}_j,\norm{\cdot}_2) =  O(\epsi^{-8(1+r)}),
\end{equation}
 and the claim follows from Assumption \ref{asu-E0} by setting \(b = 8(1+r)\) in Remark \ref{remark:bracket} below Theorem \ref{thmap-94}.  

\noindent\textbf{Proof of Lemma \ref{lemmaV}-\ref{lemmaV-b}:} First, it is shown that $\bar{\mathbb{V}}_{T,i,j}(u_i,u_j;\lambda)  = \Ex[\tilde{\mathbb{V}}_{T,i,j}(u_i,u_j;\lambda)] + o(1)$ uniformly in $u_i,u_j \in [0,1]$, $\lambda \in \Lambda_{T}$. Set $y_k(\mathcal{F}_t; u_k,\lambda) \coloneqq \FF_k^{-1}(u_k)a_{2,k}^\dagger(\mathcal{F}_t,\lambda) + a_{1,k}^\dagger(\mathcal{F}_t,\lambda)$ and define the map $\lambda \mapsto \mathcal{T}(u_i,u_j,\lambda) \coloneqq  \FF_{i,j}\{y_i(\mathcal{F}_t; u_i,\lambda), y_j(\mathcal{F}_t; u_j,\lambda)\}$. \textbf{For brevity, suppose that \(r = 1\); i.e., \(\lambda\)  is a scalar.} Using \citet[Property 12 (b)]{res99}, a second-order Taylor-series expansion with Lagrange remainder at \(\lambda = \lambda_{0}\) yields
\begin{equation}
\begin{split}
\sqrt{T}\Ex[\bar{\mathbb{V}}_{T,i,j}(u_i,u_j;\lambda) ] = \,& \sqrt{T}\Ex[\mathcal{T}(u_i,u_j,\lambda) - \CC_{i,j}(u_i,u_j)] \nn \\
= \,&    \sum_{k \inn \{\ii\}} \partial_k \FF_{i,j}\{\FF_i^{-1}(u_i),\FF_j^{-1}(u_j)\}\Ex[\tau_k(R_1;u_k,\lambda_{0})]\sqrt{T}(\lambda-\lambda_{0}) \nn \\
\,&  + \frac{1}{\sqrt{T4}} \Ex[\textsf{R}_T(u_i,u_j,\bar{\lambda})],
\end{split}
\end{equation}
where \(\bar{\lambda}_k = \lambda c -\lambda_0 (1-c) \in \Lambda_T\) for some \(c \in (0,1)\) and for any \(\tilde{\lambda} \in \Lambda_T\)
\[
\tau_k(\mathcal{F}_t;u_k,\tilde{\lambda}) = \frac{\partial}{\partial \lambda} y_k(\mathcal{F}_t; u_k,\lambda)\at[\big]{\lambda = \tilde{\lambda}},
\] 
while we note that \(\tau_k(R_t;u_k,\lambda_0) = \tau_k(\mathcal{F}_t;u_k,\lambda_0)\), which has been already defined in Lemma \ref{lemmaV}. Since 
\[\partial_i \FF_{i,j}(x_i,x_j) = \PP(\eta_j \leq x_j \mid \eta_i = x_i)\ff_i(x_i) = \partial_i \CC_{i,j}\{\FF_i(x_i),\FF_j(x_j)\}\ff_i(x_i),\]
the claim follows if \(\Ex[\textsf{R}_T(u_i,u_j,\bar{\lambda})] = O(1)\), uniformly in $u_i,u_j \in [0,1]$ and \(\lambda \in \Lambda_T\). To see this, note that
\begin{equation}\label{partial-TV}
\begin{split}
\textsf{R}_T(u_i,u_j,\bar{\lambda})  = \,&  T(\lambda-\lambda_{0})^2 \big( \sum_{l,k \inn \{\ii\}} \partial_l\partial_k \FF_{i,j}\{y_i(\mathcal{F}_t;u_i,\bar{\lambda}),y_j(\mathcal{F}_t;u_j,\bar{\lambda})\}\\
\,& \hspace*{2cm} \times \tau_l(\mathcal{F}_t;u_l,\bar{\lambda})\tau_k(\mathcal{F}_t;u_k,\bar{\lambda})  \\
\,&  \hspace*{2cm} +   \sum_{k \inn \{\ii\}}\partial_k \FF_{i,j}\{y_i(\mathcal{F}_t;u_i,\bar{\lambda}),y_j(\mathcal{F}_t;u_j,\bar{\lambda})\}\partial_k\tau_k(\mathcal{F}_t;u_k,\bar{\lambda})\big) \\
 \eqqcolon \,&  T(\lambda-\lambda_{0})^2(\sum_{l,k \inn \{\ii\}} A_{k,l}   + \sum_{k \inn \{\ii\}} B_k),
\end{split}
\end{equation}
say. Expanding terms, one gets 
\begin{equation}
\begin{split}
A_{k,l} = \,& \partial_l\partial_k \FF_{i,j}\{y_i(\mathcal{F}_t;u_i,\bar{\lambda}),y_j(\mathcal{F}_t;u_j,\bar{\lambda})\}\\
\,& \times \Bigg[y_k(\mathcal{F}_t;u_k,\bar{\lambda})y_l(\mathcal{F}_t;u_l,\bar{\lambda})\bar{a}_{2,k}^{\dagger(1)}(\mathcal{F}_t,\bar{\lambda})\bar{a}_{2,l}^{\dagger(1)}(\mathcal{F}_t,\bar{\lambda}) \\
\,& \hspace*{.6cm}+y_k(\mathcal{F}_t;u_k,\bar{\lambda})\bar{a}_{2,k}^{\dagger(1)}(\mathcal{F}_t,\bar{\lambda})\bigg(a_{1,l}^{\dagger(1)}(\mathcal{F}_t,\bar{\lambda}) - a_{1,l}^\dagger(\mathcal{F}_t,\bar{\lambda})\bar{a}_{2,l}^{\dagger(1)}(\mathcal{F}_t,\bar{\lambda})\bigg)\\
\,& \hspace*{.6cm}+y_l(\mathcal{F}_t;u_l,\bar{\lambda})\bar{a}_{2,l}^{\dagger(1)}(\mathcal{F}_t,\bar{\lambda})\bigg(a_{1,k}^{\dagger(1)}(\mathcal{F}_t,\bar{\lambda}) - a_{1,k}^\dagger(\mathcal{F}_t,\bar{\lambda})\bar{a}_{2,k}^{\dagger(1)}(\mathcal{F}_t,\bar{\lambda})\bigg)\\
\,& \hspace*{.6cm}+\bar{a}_{2,k}^{\dagger(1)}(\mathcal{F}_t,\bar{\lambda})a_{1,k}(\mathcal{F}_t,\bar{\lambda})\bigg(\bar{a}_{2,l}^{\dagger(1)}(\mathcal{F}_t,\bar{\lambda})a_{1,l}^\dagger(\mathcal{F}_t,\bar{\lambda}) - a_{1,l}^{\dagger(1)}(\mathcal{F}_t,\bar{\lambda})\bigg)\\
\,& \hspace*{.6cm}+ a_{1,k}^{\dagger(1)}(\mathcal{F}_t,\bar{\lambda})\bigg(a_{1,l}^{\dagger(1)}(\mathcal{F}_t,\bar{\lambda}) - a_{1,l}^\dagger(\mathcal{F}_t,\bar{\lambda})\bar{a}_{2,l}^{\dagger(1)}(\mathcal{F}_t,\bar{\lambda})\bigg)\Bigg],
\end{split}
\end{equation}
with $\bar{a}_{q,k}^{\dagger(1)}(\mathcal{F}_t,\lambda) \coloneqq a_{q,k}^{\dagger(1)}(\mathcal{F}_t,\lambda)/a_{q,k}^\dagger(\mathcal{F}_t,\lambda)$, where $a_{q,k}^{\dagger(1)}(\mathcal{F}_t,\lambda) \coloneqq \partial  a^\dagger_{q,k}(\mathcal{F}_t,\lambda)/(\partial \lambda)$ for $q \in \{1,2\}$, $k \in \{i,j\}$ and it has been used that 
\[
\tau_k(\mathcal{F}_t;u_k,\lambda) = \FF_k^{-1}(u_k)a_{2,k}^{\dagger(1)}(\mathcal{F}_t,\lambda) + a_{1,k}^{\dagger(1)}(\mathcal{F}_t,\lambda).
\]
Therefore, the triangle inequality, Assumption \ref{asu-E}, and the fact that $\norm{\lambda-\lambda_{0}} \leq K/\sqrt{T}$, \(K \in (0,\infty)\), for any \(\lambda \in \Lambda_T\) yields
\begin{equation}\label{taylorsecA}
\begin{split}
\Ex[|A_{k,l}|] \leq \,& \mmax\limits_{k,l \inn \{\ii\}}\ssup\limits_{x_i,x_j \inn \mathbb{R}}  |\partial_l\partial_k \FF_{i,j}(x_i,x_j)x_kx_l| \munderbar{b}^2\Ex[\dot{\mu}(R_1)^2] \\
\,& +  \mmax\limits_{k,l \inn \{\ii\}}\ssup\limits_{x_i,x_j \inn \mathbb{R}}  |\partial_l\partial_k \FF_{i,j}(x_i,x_j)x_k|\munderbar{b}\, \{\Ex[\dot{\mu}(R_1)^2]+\munderbar{b}\Ex[|\dot{\mu}(R_1)|^3]K/\sqrt{T}\} \\
\,& +  \mmax\limits_{k,l \inn \{\ii\}}\ssup\limits_{x_i,x_j \inn \mathbb{R}}  |\partial_l\partial_k \FF_{i,j}(x_i,x_j)x_l|\munderbar{b}\, \{\Ex[\dot{\mu}(R_1)^2]+\munderbar{b}\Ex[|\dot{\mu}(R_1)|^3]K/\sqrt{T}\} \\
\,& + \munderbar{b}^2\Ex[\dot{\mu}(R_1)^4]K^2/T + 2\munderbar{b}\Ex[|\dot{\mu}(R_1)|^3]K/\sqrt{T} + \Ex[\dot{\mu}(R_1)^2] < \infty.
\end{split}
\end{equation}
Similarly, it can be shown that $\Ex[|B_k|] < \infty$. Therefore, $\Ex[\tilde{\mathbb{V}}_{T,i,j}(u_i,u_j;\lambda)] = \sqrt{T}\Ex[\mathcal{T}(u_i,u_j,\lambda) - \CC_{i,j}(u_i,u_j)] = \bar{\mathbb{V}}_{T,i,j}(u_i,u_j;\lambda) + O(T^{-1/2})$.

Next, in view of Eq. \eqref{emp-V}, one gets 
\[
\tilde{\mathbb{V}}_{T,i,j}(u_i,u_j;\lambda) = \tilde{\mathbb{C}}_{T,i,j}(g-g_0) + \bar{\mathbb{V}}_{T,i,j}(u_i,u_j;\lambda),\;\; g,g_0 \in \mathcal{G},
\]
with $g_0$ imposing the restriction $\lambda = \lambda_{0}$. By part \ref{lemmaV-a} of this lemma, $\{\tilde{\mathbb{C}}_{T,i,j}(g), g \in \mathcal{G}\}$ is stochastically equicontinuous. Thus, it remains to be shown that $\norm{g-g_0}_2 \rightarrow 0$. To see this, note that the mean-value theorem yields 
\begin{equation}
\begin{split}
\norm{g(\eta_{i,t},&\eta_{j,t},\mathcal{F}_t)-g_0(\eta_{i,t},\eta_{i,t},\mathcal{F}_t)}_2^2 \\
\leq \,&  \sum_{k \inn \{\ii\}} \norm{1\{\eta_{k,t} \leq \FF_k^{-1}(u_k)a_{2,k}^\dagger(\mathcal{F}_t,\lambda) + a_{1,k}^\dagger(\mathcal{F}_t,\lambda)\} -1\{\eta_{k,t} \leq \FF_k^{-1}(u_k)\}}_{2}^2 \\
= \,& \sum_{k \inn \{\ii\}} \{\Ex[\FF_k(\FF_k^{-1}(u_k)a_{2,k}^\dagger(\mathcal{F}_t,\lambda_k) + a_{1,k}^\dagger(\mathcal{F}_t,\lambda_k))] +u_k \\
  \,& \hspace*{2.5cm}-2\Ex[\FF_k(\mmin\{u_k,\FF_k^{-1}(u_k)a_{2,k}^\dagger(\mathcal{F}_t,\lambda_k) + a_{1,k}^\dagger(\mathcal{F}_t,\lambda_k)\})]\} \\
 \leq \,&  \sum_{k \inn \{\ii\}} \Ex[|\ff_k\{\FF_k^{-1}(u_k)a_{2,k}^\dagger(\mathcal{F}_t,\bar{\lambda}_k)+a_{1,k}^\dagger(\mathcal{F}_t,\bar{\lambda}_k)\}\tau_k(\mathcal{F}_t,u_k,\bar{\lambda})|]|\lambda-\lambda_{0}|
\end{split}
\end{equation}
where $\bar{\lambda} = \lambda c + \lambda_0(1-c) \in \Lambda_T$  for some \(c \in (0,1)\). By similar arguments as just used to prove \eqref{taylorsecA}, one gets
\begin{equation}
\begin{split}
\ff_k\{&\FF_k^{-1}(u_k)a_{2,k}(\mathcal{F}_t,\bar{\lambda})+a_{1,k}(\mathcal{F}_t,\bar{\lambda})\}\tau_k(\mathcal{F}_t,u_k,\bar{\lambda})\nn \\
\,& = \ff_k\{\FF_k^{-1}(u_k)a_{2,k}(\mathcal{F}_t,\bar{\lambda})+a_{1,k}(\mathcal{F}_t,\bar{\lambda})\}\{\FF_k^{-1}(u_k)a_{2,k}(\mathcal{F}_t,\bar{\lambda})+a_{1,k}(\mathcal{F}_t,\bar{\lambda})\}\bar{a}_{2,k}^{(1)}(R_1,\tilde{\lambda}_k) \nn\\
\,& \hspace*{2.5cm}- \ff_k\{\FF_k^{-1}(u_k)a_{2,k}(\mathcal{F}_t,\bar{\lambda})+a_{1,k}(\mathcal{F}_t,\bar{\lambda})\}\bar{a}_{2,k}^{(1)}(\mathcal{F}_t,\bar{\lambda}) a_{1,k}(\mathcal{F}_t,\bar{\lambda})\nn \\
\,& \hspace*{2.5cm} + \ff_k\{\FF_k^{-1}(u_k)a_{2,k}(\mathcal{F}_t,\bar{\lambda})+a_{1,k}(\mathcal{F}_t,\bar{\lambda})\} \bar{a}_{2,k}^{(1)}(\mathcal{F}_t,\bar{\lambda}) \nn \\
\,& \leq \dot{\mu}(R_1)\{\ssup\limits_{y \inn \mathbb{R}} |\ff(y)y|\munderbar{b}+\ssup\limits_{y \inn \mathbb{R}} \ff(y) \dot{\mu}(R_1)K/\sqrt{T}\munderbar{b} + \ssup\limits_{y \inn \mathbb{R}} \ff(y)K/\sqrt{T}\},\nn
\end{split}
\end{equation}
which, in turn, implies $\norm{g(\eta_{i,t},\eta_{j,t},R_t)-g_0(\eta_{i,t},\eta_{i,t},R_t)}_2^2 = O(T^{-1/2})$, thereby completing the proof.

\noindent\textbf{Proof of Lemma \ref{lemmaV}-\ref{lemmaV-c}:} Note that $\Ex[\tCC_{T,i,j}(u_i,u_j;\lambda_{0})]= \CC_{i,j}(u_i,u_j)$. The claim thus follows from part \ref{lemmaV-a} of this lemma and the `fidi'-convergence of $\tilde{\mathbb{C}}_{T,i,j}(u_i,u_j;\lambda_{0})$ which is readily obtained by the CLT for $i.i.d.$ data.

\subsection{Proof  of Lemma \ref{lemmaU}}
\noindent\textbf{Proof of Lemma \ref{lemmaU}-\ref{lemmaU-a}:} Throughout, suppose that \(i,j \inn \mathcal{G}_q\) for some \(q \in \{1,\dots,Q\}\). Recall, that \(\eps_{k,t,s}(\delta) = \DD_\eps^{-1}(\eps_{k,t,s}^\st;\delta)\), \(k \in \{\ii\},\) and \(F_{t,s}(\gamma) = \DD_{F}^{-1}(F_{t,s}^\St;\gamma)\), with
\(\DD_{F}^{-1}(F_{t,s}^\St;\gamma) \coloneqq  (\DD_{F,1}^{-1}(F_{t,s,1}^\St;\gamma_1),\dots,\DD_{F,p_\alpha}^{-1}(F_{t,s,p_\alpha}^\St;\gamma_{p_\alpha}))';\)
\(\eps_{k,t,s}^\st\) and \(F_{t,s}^\St \coloneqq (F_{t,s,1}^\St,\dots,F_{t,s,p_\alpha}^\St)'\) denote mutually independent \textit{i.i.d.} draws from the standard uniform distribution. Moreover, observe that 
\[
\hat{X}_{k,t,s}(\theta_q,\nu) = X_{k,t,s}(\theta_q)-\beta_q'\Delta^\dagger(\mathcal{W}_t,\nu),\;\;\Delta^\dagger(\mathcal{W}_t,\nu) \coloneqq  \sigma(\mathcal{W}_t,\nu) - \sigma(M_t,\nu_0),
\]
where \(\sigma(\mathcal{W}_t,\nu) \coloneqq (\sigma_1(\mathcal{W}_t,\nu),\dots,\sigma_{p_\beta}(\mathcal{W}_t,\nu))'\) is a \(p_\beta \times 1\) vector for \(\nu  \in \mathcal{V}_T \in \mathbb{R}^{m}\). Using that \(G_i = G_j = \GG_{q}\) for any \(i,j \inn \mathcal{G}_q\), we get
\begin{equation}\nn
\begin{split}
1\{\hat{X}_{i,t,s}(\theta_q,\nu) &\leq \GG_q^{-1}(u_i;\theta_q),\hat{X}_{j,t,s}(\theta_q,\nu)  \leq \GG_q^{-1}(u_j;\theta_q)\}  \\
= \ & 1\{X_{i,t,s}(\theta_q) \leq \GG_q^{-1}(u_i;\theta_q)+\beta_q'\Delta^\dagger(\mathcal{W}_t,\nu),X_{j,t,s}(\theta) \leq \GG_q^{-1}(u_j;\theta)+\beta_q'\Delta^\dagger(\mathcal{W}_t,\nu)\} \\
= \,& 1\big\{\eps_{i,t,s}^\st \leq  \DD_\eps\big[\GG_q^{-1}(u_i;\,\theta_q)-\alpha_q' \DD_F^{-1}(F_{t,s}^\St;\,\gamma) + \beta_q' (\Delta^\dagger(\mathcal{W}_t,\nu)-Z_t);\,\delta\big],  \\
\,&\;\;\;\,\eps_{j,t,s}^\st \leq  \DD_\eps\big[\GG_q^{-1}(u_j;\,\theta_q)-\alpha_q' \DD_F^{-1}(F_{t,s}^\St;\,\gamma) + \beta_q' (\Delta^\dagger(\mathcal{W}_t,\nu)-Z_t);\,\delta\big]\big\} \\
\eqqcolon \,& g(\eps_{i,t,s}^\st,\eps_{j,t,s}^\st,F_{t,s}^{\St},Z_t,\mathcal{W}_t),
\end{split}
\end{equation}
where $g$ can be viewed as an element of the function class
\begin{equation}
\begin{split}
\mathcal{G} \coloneqq \Big\{\eps_{i,t,s}^\st,\eps_{j,t,s}^\st,F_{t,s}^{\St},Z_t,\mathcal{W}_t \mapsto 1\Big\{&\eps_{i,t,s}^\st \leq  \DD_\eps\big[\GG_q^{-1}(u_i;\,\theta_q)-\alpha_q'\DD_F^{-1}(F_{t,s}^\St;\,\gamma) + \beta_q'(\Delta^\dagger(\mathcal{W}_t,\nu)-Z_t);\,\delta\big], \nn \\
& \eps_{j,t,s}^\st \leq  \DD_\eps\big[\GG_q^{-1}(u_j;\,\theta_q)-\alpha_q'\DD_F^{-1}(F_{t,s}^\St;\,\gamma) + \beta_q'(\Delta^\dagger(\mathcal{W}_t,\nu)-Z_t);\,\delta\big]\Big\}:\nn \\
& u_i,u_j \in [0,1],\theta_q \in \Theta, \nu \in \mathcal{V}_T\Big\}.
\end{split}
\end{equation}
Since, by Assumption \ref{asu-E2}, a first order Taylor-expansion around \(\nu = \nu_0\) yields \(\Delta^\dagger(\mathcal{W}_t,\nu) = \Delta (M_t,\nu) + o_p(T^{-1/2})\), with  \(\Delta (M_t,\nu) \coloneqq \nabla_{\nu} (M_t,\nu_0)(\nu - \nu_0)\), it follows that  on an event with probability converging to one, \(g \in \mathcal{G}_1\) for any \(g \in \mathcal{G}\), where
\begin{equation}
\begin{split}
\mathcal{G}_1 \coloneqq \Big\{\eps_{i,t,s}^\st,\eps_{j,t,s}^\st, \mathcal{X}_{t,s} \mapsto 1\Big\{&\eps_{i,t,s}^\st \leq  \DD_\eps\big[\GG_q^{-1}(u_i;\,\theta_q)-\alpha_q'\DD_F^{-1}(F_{t,s}^\St;\,\gamma) + \beta_q'(\Delta(M_t,\nu)-Z_t);\,\delta\big], \nn \\
& \eps_{j,t,s}^\st \leq  \DD_\eps\big[\GG_q^{-1}(u_j;\,\theta_q)-\alpha_q'\DD_F^{-1}(F_{t,s}^\St;\,\gamma) + \beta_q'(\Delta(M_t,\nu)-Z_t);\,\delta\big]\Big\}:\nn \\
& u_i,u_j \in [0,1],\theta_q \in \Theta, \nu \in \mathcal{V}_T\Big\}.
\end{split}
\end{equation}
with \(\mathcal{X}_{t,s} \coloneqq ({F_{t,s}^{\St}}',Z_t',M_t')'\). Formally, an element $g_1 \in \mathcal{G}_1$ can be uniquely identified by the quadruple $\{u_i,u_j,\theta_q,\nu\}$. Hence, in order to refer to a specific $g_1 \in \mathcal{G}_1$, the notation $g_1(\eps_{i,t,s}^\st,\eps_{j,t,s}^\st,\mathcal{X}_{t,s}) = g_1(\eps_{i,t,s}^\st,\eps_{j,t,s}^\st,\mathcal{X}_{t,s};u_i,u_j,\theta_q,\nu)$ is used. Moreover, define
\begin{equation}\label{class-Fbar}
\bar{\mathcal{G}}_1 \coloneqq \Big\{\xi_{i,j,t,S} \mapsto \frac{1}{S}\sum_{s \eq 1}^S g_1(\eps_{i,t,s}^\st,\eps_{j,t,s}^\st,\mathcal{X}_{t,s}),\; g_1 \in \mathcal{G}_1\Big\},
\end{equation}
where 
\[
\xi_{i,j,t,S} \coloneqq (\eps_{i,t,1}^\st,\eps_{j,t,1}^\st,{F_{t,1}^\St}',\dots,\eps_{i,t,S}^\st,\eps_{j,t,S}^\st,{F_{t,S}^\St}',Z_t',M_t')',\; t \in \{1,\dots,T\}.
\]
For the case \(S = S(T) \rightarrow \infty\), the  process \(\{\xi_{i,j,t,S}: t \geq 1, S \geq 1\}\) is an \(\alpha\)-mixing triangular array, that inherits the \(\alpha\)-mixing size from the process \(\{M_t: t\geq 1\}\) stated in Assumption \ref{asu-E0}. Importantly, the \(p_{\textsf{M}} \times 1\) vector \(M_t\) is mutually independent of the remaining \(i.i.d.\) components of \(\xi_{i,j,t,S}\).\footnote{Note that, by Assumption \ref{asu-B} and \ref{asu-E}-\ref{asu-E0}, \(\{\eps_{i,t,s}^\st,F_{t,s}^\st,Z_{t},M_t\}\)  are mutually independent for any \(s\in \{1,\dots,S\}\), \(t \in \{1,\dots,T\}\) and \(i \in \{1,\dots,n\}\).} Following the same arguments used at the beginning of the proof of Lemma \ref{lemmaV}-\ref{lemmaV-a}, we conclude that in order to establish stochastic equicontinuity of \(\{\tilde{\mathbb{C}}_{i,j,T,S}(u_i,u_j,\theta_q,\nu), u_i,u_j \in [0,1],\,\theta_q \in \Theta,\,\nu \in \mathcal{V}_T\}\), defined in Eq. \eqref{Ctilde}, it suffices to show stochastic equicontinuity of the process \(\{\check{\mathbb{C}}_{i,j,T,S}(\bar{g}_1), \bar{g}_1 \in \mathcal{G}_1\}\) given by
\begin{equation}\label{emppro-G}
\begin{split}
\check{\mathbb{C}}_{i,j,T,S}: \bar{\mathcal{G}}_1 \longrightarrow &\,\mathbb{R} \\
\bar{g}_1 \in \bar{\mathcal{G}}_1,\,\bar{g}_1  \longmapsto &\,\check{\mathbb{C}}_{i,j,T,S}(\bar{g}_1\,) \coloneqq \frac{1}{\sqrt{T}}\sumt \big\{\bar{g}_1(\xi_{i,j,t,S}) - \Ex[\bar{g}_1(\xi_{i,j,t,S})]\big\},
\end{split}
\end{equation}
Again, a specific element $\bar{g}_1 \in \bar{\mathcal{G}}_1$ will be identified by the quadruple $\{u_i,u_j,\theta_q,\nu\}$ for which we write $\bar{g}_1(\xi_{i,j,t,S}) = \bar{g}_1(\xi_{i,j,t,S};u_i,u_j,\theta_q,\nu)$. Since the class \(\bar{\mathcal{G}}_1\) is uniformly bounded and \(\xi_{i,j,t,S}\) an \(\alpha\)-mixing triangular array, we have cast \(\{\check{\mathbb{C}}_{i,j,T,S}(\bar{g}_1), \bar{g}_1 \in \mathcal{G}_1\}\) in terms of Theorem \ref{thmap-94}.  We thus show next that $\mathcal{N}_{[\,]}(\epsi,\bar{\mathcal{G}}_1,\rho) = O(\epsi^{-4(m+p+2)}),$
which then yields, in view of Theorem \ref{thmap-94}, the result. To verify the preceding display, suppose first we can cover $\mathcal{G}_1$ with brackets 
$L_k(\eps_{i,t,s}^\st,\eps_{j,t,s}^\st,\mathcal{X}_{t,s}) \leq  g_1(\eps_{i,t,s}^\st,\eps_{j,t,s}^\st,\mathcal{X}_{t,s}) \leq U_k(\eps_{i,t,s}^\st,\eps_{j,t,s}^\st,\mathcal{X}_{t,s}),\; g_1 \in \mathcal{G}_1,$ $k = 1,\dots,N\coloneqq\mathcal{N}_{[\,]}(\epsi,\mathcal{G}_1,\norm{\cdot}_2)$. We can now construct $\mathcal{N}_{[\,]}(\epsilon,\bar{\mathcal{G}}_1,\rho)$ brackets for $\bar{\mathcal{G}}_1$ by setting
\begin{equation}
\begin{split}
\bar{L}_k(\xi_{i,j,t,S}) \coloneqq \,& \frac{1}{S}\sum_{s \eq 1}^SL_k(\eps_{i,t,s}^\st,\eps_{j,t,s}^\st,\mathcal{X}_{t,s}), \\
\bar{U}_k(\xi_{i,j,t,S}) \coloneqq \,& \frac{1}{S}\sum_{s \eq 1}^SU_k(\eps_{i,t,s}^\st,\eps_{j,t,s}^\st,\mathcal{X}_{t,s}),\;\; k = 1,\dots,N.
\end{split}
\end{equation}
Note that, by the triangle inequality and the stationarity of $\{\xi_{i,j,t,S}: t \geq 1\}$, one gets
 \[
\ssup\limits_{1\, \leq\, t \, \leq \, T}\norm{\bar{U}_k(\xi_{i,j,t,S}) -\bar{L}_k(\xi_{i,j,t,S})}_2 \leq \norm{U_k(\eps_{i,t,s}^\st,\eps_{j,t,s}^\st,\mathcal{X}_{t,s})-L_k(\eps_{i,t,s}^\st,\eps_{j,t,s}^\st,\mathcal{X}_{t,s})}_{2} \leq \epsi.
\] 
Therefore, $\mathcal{N}_{[\,]}(\epsi,\bar{\mathcal{G}}_1,\rho) =  \mathcal{N}_{[\,]}(\epsi,\mathcal{G}_1,\norm{\cdot}_2)$. To proceed, note that $\mathcal{G}_1 \subseteq \mathcal{H}_i \cdot \mathcal{H}_j$, where
\begin{align}\label{H-class}
\mathcal{H}_k \coloneqq \Big\{\eps_{k,t,s}^\st,\mathcal{X}_{t,s} \mapsto 1\big\{\eps_{k,t,s}^\st \leq  \DD_\eps\big[\GG_q^{-1}(u_k;\,\theta_q)-&\alpha_q' \DD_F^{-1}(F_{t,s}^\St;\,\gamma) + \beta_q' (\Delta(M_t,\nu)-Z_t);\,\delta\big]\big\}: \nn \\
&u_k \in [0,1],\theta_q \in \Theta, \nu \in \mathcal{V}_T\Big\}, \;k \in \{\ii\}.
\end{align} 
 Since $\mathcal{N}_{[\,]}(2\epsi,\mathcal{G}_1,\norm{\cdot}_2) \leq \mathcal{N}_{[\,]}(\epsi,\mathcal{H}_i,\norm{\cdot}_2)\,\mathcal{N}_{[\,]}(\epsi,\mathcal{H}_j,\norm{\cdot}_2)$, it remains to compute $\mathcal{N}_{[\,]}(\epsi,\mathcal{H}_k,\norm{\cdot}_2)$ for $k \in \{\ii\}$; see, e.g., \citet[Lemma 9.25 ($ii$)]{ko08}. We will show that $\mathcal{N}_{[\,]}(\epsi,\mathcal{H}_k,\norm{\cdot}_2) = O(\epsi^{2(p+m+2)})$. For a given $\{\eps_{k,t,s}^\st,\mathcal{X}_{t,s}\}$, we can formally identify a specific element $h_k \in \mathcal{H}_k$ through its arguments $\{u_k,\alpha_q,\beta_q,\gamma,\delta, \nu\}$ \textit{via} $h_k(\eps_{k,t,s}^\st,\mathcal{X}_{t,s}) = h_k(\eps_{k,t,s}^\st,\mathcal{X}_{t,s};u_k,\theta_q, \nu)$; with
\begin{equation}\label{hfunction}
h_k(\eps_{k,t,s}^\st,\mathcal{X}_{t,s};u_k,\theta_q, \nu) \coloneqq 1\big\{\eps_{k,t,s}^\st \leq  \textsf{D}_k(\mathcal{X}_{t,s};u_k,\theta_q,\nu) \big\},
\end{equation}
where
\begin{equation}\label{Deps-u}
\textsf{D}_{k}(\mathcal{X}_{t,s};u_k,\alpha_q,\beta_q,\gamma,\delta,\nu) \coloneqq \DD_\eps\big[\GG_q^{-1}(u_k;\theta_q)-\alpha_q'\DD_F^{-1}(F_{t,s}^\St;\,\gamma) + \beta_q'(\Delta(M_t,\nu)-Z_t);\,\delta\big],
\end{equation}
for $k \in \{i,j\}$, and we reiterate here that $\theta_q = (\alpha_q',\beta_q',\gamma',\delta')' \in \Theta \subset \mathbb{R}^p$. Since, by Assumption, $\Theta$ is compact,  we can assume that $\alpha_q \in [-K_\alpha,K_\alpha]^{p_\alpha}$, $\beta_q \in [-K_\beta,K_\beta]^{p_\beta}$, $\gamma \in [-K_\gamma,K_\gamma]^{p_\alpha p_\gamma}$, and $\delta \in [-K_\delta,K_\delta]^{p_\delta}$, for finite, non-negative constants $K_\alpha, K_\beta, K_\gamma$ and \(K_\delta\). To keep the notational complexity tractable, we drop the subscript $k \in \{\ii\}$.
  
Similar to the proof of part \ref{lemmaV-a} of Lemma \ref{lemmaV}, we begin, for a given $u \in [0,1]$, with the construction of brackets by successively covering
\begin{enumerate}[label= ($\alph*$)]
\item\label{lb-a} $\DD_\eps(\cdot;\delta)$, $\delta \in  [-K_\delta,K_\delta]^{p_\delta}$,
\item\label{lb-b} $\alpha_q'\DD_F^{-1}(\,\cdot\,;\,\gamma)$, $\alpha_q \in [-K_\alpha,K_\alpha]^{p_\alpha},\,\gamma \in [-K_\gamma,K_\gamma]^{p_\alpha p_\gamma}$, and
\item\label{lb-c} $\beta_q'\Delta(\,\cdot\,;\nu)$, $\beta_q \in [-K_\beta,K_\beta]^{p_\beta}$, $\nu \in \mathcal{V}_0$.
\end{enumerate}

\noindent \underline{Step \ref{lb-a}:} By compactness, we can fix a grid $\{\delta^{(1)},\dots,\delta^{(N_1)}\}$ of length $N_1 = O(\eps^{-2p_\delta})$, to cover $[-K_\delta,K_\delta]^{p_\delta}$ with $N_1$ balls $\{\mathcal{B}_1(\delta),\dots,\mathcal{B}_{N_1}(\delta)\}$, each of radius $\epsi^2$; i.e., for each $\delta$, there is at least one $i_1 \in \mathcal{J}_1 \coloneqq \{1,\dots,N_1\}$ such that $\delta \in \mathcal{B}_{i_1}(\delta) \coloneqq \{\delta: \norm{\delta-\delta^{(i_1)}} \leq \epsi^2\}$. Now, for such a number $i_1 \in \mathcal{J}_1$, Assumption \ref{asu-D} yields $\DD_\eps(x;\delta) \leq \DD_\eps(x;\delta^{(i_1)}) + K_1 \epsi^2$, $K_1 \coloneqq \ssup\limits_{\delta \inn \Theta}\ssup\limits_{x \inn \mathbb{R}} \dd_\eps(x;\delta)$. 
Thus, for any $x \in \mathbb{R}$, $\{D_\eps(x;\delta): \delta \in [-K_\delta,K_\delta]^{p_\delta}\}$ can be covered by ${N_1}$ brackets \[[D_\eps(x;\delta^{(i_1)}) - K_1 \epsi^2, D_\eps(x;\delta^{(i_1)}) + K_1 \epsi^2],\, i_1 \in \mathcal{J}_1.\]

\noindent \underline{Step \ref{lb-b}:} Analogously, fix a grid $\{\gamma^1,\dots,\gamma^{(N_2)}\}$ to cover $[-K_\gamma,K_\gamma]^{p_\alpha p_\gamma}$ with $N_2 = O(\epsi^{-2 p_\alpha p_\gamma})$ balls $\{\mathcal{B}_1(\gamma),\dots,\mathcal{B}_{N_2}(\gamma)\}$ each of radius $\epsi^2$. Hence, any $\gamma$ is in at least one ball $\mathcal{B}_{i_2}(\gamma)$, ${i_2} \in \mathcal{J}_2 \coloneqq \{1,\dots,{N_2}\}$. For each such $i_2 \in \mathcal{J}_2$, fix a grid \(\alpha_q^{(1)},\dots,\alpha_q^{(N_3)}\) to cover the rectangle $[-K_\alpha,K_\alpha]^{p_\alpha}$ with ${N_3}\coloneqq \floor{\epsi^{-2p_\alpha}}$ balls $\{\mathcal{B}_1(\alpha),\dots,\mathcal{B}_{N_3}(\alpha)\}$ each of radius $\epsi^2$. Thus, for any $\gamma \in \mathcal{B}_{i_2}(\gamma)$, ${i_2} \in \mathcal{J}_2$ and any $\alpha \in [-K_\alpha,K_\alpha]^{p_\alpha}$, we have $\norm{\gamma - \gamma^{(i_2)}} \leq \epsi^2$ and $\norm{\alpha_q - \alpha_q^{(i_3)}} \leq \epsi^2$ for some $\alpha_q^{(i_3)}$, $i_3 \in \mathcal{J}_3 \coloneqq \{1,\dots,N_3\}$. Hence, by Assumption \ref{asu-D}, we have
\begin{equation}
\begin{split}
\alpha_q \DD_F^{-1}({F_{t,s}^\St};\,\gamma) = \,& {\alpha_q^{(i_3)}}'\DD_F^{-1}(F_{t,s}^\St;\,\gamma^{(i_2)}) + [\alpha_q'\DD_F^{-1}(F_{t,s}^\St;\,\gamma)-{\alpha_q^{(i_3)}}'\DD_F^{-1}(F_{t,s}^\St;\,\gamma^{(i_2)})]  \nn \\
\leq \,&  {\alpha_q^{(i_3)}}'\DD_F^{-1}(F_{t,s}^\St;\,\gamma^{(i_2)}) + \epsi^2 \,\zeta_1(F_{t,s}^\St), 
\end{split}
\end{equation}
with $\zeta_1(F_{t,s}^\st) \coloneqq \dot{Q}(F_{t,s}^\St)(1+K_\alpha).$ Therefore, we can cover $\{\alpha_q'\DD_F^{-1}(F_{t,s}^\St;\,\gamma): \alpha_q \in [-K_\alpha,K_\alpha]^{p_\alpha}, \gamma \in [-K_\alpha,K_\alpha]^{p_\alpha p_\gamma}\}$ with brackets of the form
\begin{equation}
[{\alpha_q^{(i_3)}}'\DD_F^{-1}(F_{t,s}^\St;\,\gamma^{(i_2)}) - \epsi^2\zeta_1(F_{t,s}^\St),{\alpha_q^{(i_3)}}'\DD_F^{-1}(F_{t,s}^\St;\,\gamma^{(i_2)}) + \epsi^2\zeta_1(F_{t,s}^\St)],\;\; i_2 \in \mathcal{J}_2,i_3 \in \mathcal{J}_3.
\end{equation}

\noindent \underline{Step \ref{lb-c}:} Similar to the two preceding steps, consider a grid $\{\nu^{(1)},\dots,\nu^{(N_4)}\}$ to cover $\mathcal{V}_T = \{\nu: \sqrt{T}\norm{\nu - \nu_0} \leq K_\nu\}$ with ${N_4} = O(\eps^{-2mp_\beta})$ balls $\{\mathcal{B}_1(\nu_0),\dots,\mathcal{B}_{N_4}(\nu_0)\}$ each of radius $\eps^2/\sqrt{T}$.  For each $i_4 \in \mathcal{J}_4 \coloneqq \{1,\dots,{N_4}\}$, we can fix a grid \(\beta_q^{(1)},\dots,\beta_q^{(N_5)}\) to cover the rectangle $[-K_\beta,K_\beta]^{p_\beta}$  with ${N_5}\coloneqq \floor{\epsi^{-2p_\beta}}$ balls $\{\mathcal{B}_1(\beta),\dots,\mathcal{B}_{N_5}(\beta)\}$ each of radius $\epsi^2$. Then, $\{\beta_q'[\Delta(M_t,\nu)-Z_t]: \beta_q \in [-K_\beta,K_\beta]^{p_\beta}, \nu \in \mathcal{V}_0\}$ can be covered by
\[
\big[{\beta_q^{(i_5)}}'[\Delta(M_t,\nu^{(i_4)})-Z_t] - \epsi^2\zeta_2(M_t,Z_t),{\beta_q^{(i_5)}}'[\Delta(M_t,\nu^{(i_4)})-Z_t] +\epsi^2\zeta_2(M_t,Z_t)\big], i_4 \in \mathcal{J}_4, i_5 \in \mathcal{J}_5,
\]
with  $\zeta_2(M_t,Z_t) \coloneqq \dot{\sigma}(M_t)(K_\nu + K_\beta)/\sqrt{T}-Z_t. $

Combining steps \ref{lb-a}, \ref{lb-b}, and \ref{lb-c}, we cover \eqref{Deps-u} uniformly in $\theta_q = (\alpha_q',\beta_q',\delta_q',\gamma')'$ and $\nu$ for a fix $y \coloneqq \GG_q^{-1}(u;\,\theta_q)$ \textit{via}
\begin{equation}\label{brackets-D}
[\munderbar{\DD}_{i_1,i_2,i_3,i_4,i_5}(y,\mathcal{X}_{t,s})-K_1 \epsi^2,\bar{\DD}_{i_1,i_2,i_3,i_4,i_5}(y,\mathcal{X}_{t,s})+K_1 \epsi^2],\; i_j \in \mathcal{J}_j, j = 1,\dots,5,
\end{equation}
where
\begin{equation}\nn
\begin{split}
\bar{\DD}_{i_1,i_2,i_3,i_4,i_5}(y,\mathcal{X}_{t,s}) = \,& \DD_\eps\big(y -  {\alpha_q^{(i_3)}}'\DD_F^{-1}(F_{t,s}^\St;\,\gamma^{(i_2)}) + {\beta_q^{(i_5)}}'[\Delta(M_t,\nu^{(i_4)})-Z_t] +\epsi^2 \zeta_3(\mathcal{X}_{t,s}) ;\delta^{(i_1)}\big),\\
\munderbar{\DD}_{i_1,i_2,i_3,i_4,i_5}(y,\mathcal{X}_{t,s}) = \,& \DD_\eps\big(y -  {\alpha_q^{(i_3)}}'\DD_F^{-1}(F_{t,s}^\St;\,\gamma^{(i_2)}) + {\beta_q^{(i_5)}}'[\Delta(M_t,\nu^{(i_4)})-Z_t]  -\epsi^2 \zeta_3(\mathcal{X}_{t,s}) ;\delta^{(i_1)}\big),
\end{split}
\end{equation}
with
\[
\zeta_3(\mathcal{X}_{t,s}) \coloneqq \zeta_1(F_{t,s}^\St)+\zeta_2(M_t,Z_t).
\]
Importantly, the indices $i_j \in \mathcal{J}_j$, $j = 1,\dots,5$, can be chosen such that for any $\theta_q = (\alpha_q',\beta_q',\gamma_q',\delta_q')'$ and $\nu$, the distances $\norm{\delta-\delta^{(i_1)}}$, $\norm{\gamma-\gamma^{(i_2)}}$, $\norm{\alpha_q-\alpha_q^{(i_3)}}$, $\norm{\nu-\nu^{(i_4)}}$ and $\norm{\beta_q-\beta_q^{(i_5)}}$ are all bounded from above by $\epsi^2$. Next, recall that $\GG_q^{-1}(u;\,\theta_q) = \GG_q^{-1}(u;\,\alpha_q,\beta_q,\gamma,\delta)$. Since the brackets \eqref{brackets-D} are uniform in $\theta_q$, it suffices to fix brackets for $\GG_q^{-1}(u;\,\theta_q)$ uniformly in $u \in [0,1]$  for $\bar{\DD}_{i_1,\dots,i_5}$ and $\munderbar{\DD}_{i_1,\dots,i_5}$, respectively. Specifically, and similar to the proof of Lemma B.1 in \cite{avk01},  we can partition the real line by a total of $N_6$ points for given $\{i_1,\dots,i_5\}$:
\[-\infty = \bar{y}_{i_1,i_2,i_3,i_4,i_5,1} \leq \bar{y}_{i_1,i_2,i_3,i_4,i_5,i_5,2} \leq \dots \leq \bar{y}_{i_1,i_2,i_3,i_4,i_5,	N_6}  = \infty,\;\;N_6 = O(\epsi^{-2}),\]
such that 
\[\bar{\DD}_{i_1,i_2,i_3,i_4,i_5}(\bar{y}_{i_1,i_2,i_3,i_4,i_5,i_6},\mathcal{X}_{t,s}) -  \bar{\DD}_{i_1,i_2,i_3,i_4,i_5}(\bar{y}_{i_1,i_2,i_3,i_4,i_5,i_6-1},\mathcal{X}_{t,s}) \leq K_2\epsi^2,\; i_6 \in \mathcal{J}_6 \coloneqq \{1,\dots,N_6\},\]and some $K_2 \in (0,\infty)$; see, e.g., \citet[example 19.6]{vv94}. Similarly, we can let 
\[-\infty = \munderbar{y}_{i_1,i_2,i_3,i_4,i_5,1} \leq \munderbar{y}_{i_1,i_2,i_3,i_4,i_5,2} \leq \dots \leq \munderbar{y}_{i_1,i_2,i_3,i_4,i_5,N_7}  = \infty,\;\;N_7 = O(\epsi^{-2}),\]
such that 
\[\munderbar{\DD}_{i_1,i_2,i_3,i_4,i_5}(\munderbar{y}_{i_1,i_2,i_3,i_4,i_5,i_7},\mathcal{X}_{t,s}) -  \munderbar{\DD}_{i_1,i_2,i_3,i_4,i_5}(\munderbar{y}_{i_1,i_2,i_3,i_4,i_5,i_7-1},\mathcal{X}_{t,s}) \leq K_2\epsi^2,\; i_7 \in \mathcal{J}_7 \coloneqq \{1,\dots,N_7\},\] and some $K_2 \in (0,\infty)$.  Now, define the following brackets for $\GG_q^{-1}(u;\theta_q) \in \mathbb{R}$ by 
\[\munderbar{y}_{i_1,i_2,i_3,i_4,i_5}^\dagger  \leq \GG_q^{-1}(u;\theta_q) \leq \bar{y}_{i_1,i_2,i_3,i_4,i_5}^\dagger,\] with
\begin{equation}
\begin{split}
 \munderbar{y}_{i_1,i_2,i_3,i_4,i_5}^\dagger  \coloneqq \,& \mmax\limits_{i_6 \inn \mathcal{J}_6}\{\munderbar{y}_{i_1,i_2,i_3,i_4,i_5,i_6}: \munderbar{y}_{i_1,i_2,i_3,i_4,i_5,i_6} \leq \GG_q^{-1}(u;\theta_q)\}\\
 \bar{y}_{i_1,i_2,i_3,i_4,i_5}^\dagger \coloneqq \,&  \mmin\limits_{i_7 \inn \mathcal{J}_7}\{\bar{y}_{i_1,i_2,i_3,i_4,i_5,i_7}: \bar{y}_{i_1,i_2,i_3,i_4,i_5,i_7} \geq \GG_q^{-1}(u;\theta_q)\}.
\end{split}
\end{equation}
Hence, by the monotonicity of the indicator function, we obtain a total of 
\[
N \coloneqq \prod_{j \eq 1}^7 |\mathcal{J}_j|= O\left(\epsi^{-2(p_\alpha(1+ p_\gamma) + p_\beta(1+ m)+ p_\delta+2)}\right)
\]
brackets 
\[L_{i_1,i_2,i_3,i_4,i_5}(\eps_{t,s}^\st,\mathcal{X}_{t,s}) \leq  h(\eps_{t,s}^\st,\mathcal{X}_{t,s};u,\theta,\nu) \leq U_{i_1,i_2,i_3,i_4,i_5}(\eps_{t,s}^\st,\mathcal{X}_{t,s}),\,i_j \in \mathcal{J}_j,\,j = 1,\dots,5,\] defined via
\begin{equation}
\begin{split}
L_{i_1,i_2,i_3,i_4,i_5}(\eps_{t,s}^\st,\mathcal{X}_{t,s}) \coloneqq \,&1\{\eps_{t,s}^\st \leq  \munderbar{\DD}_{i_1,i_2,i_3,i_4,i_5}(\munderbar{y}_{i_1,i_2,i_3,i_4,i_5}^\dagger,\mathcal{X}_{t,s})-K_1 \epsi^2\}  \\
U_{i_1,i_2,i_3,i_4,i_5}(\eps_{t,s}^\st,\mathcal{X}_{t,s}) \coloneqq \,&1\{\eps_{t,s}^\st \leq  \bar{\DD}_{i_1,i_2,i_3,i_4,i_5}(\bar{y}_{i_1,i_2,i_3,i_4,i_5}^\dagger,\mathcal{X}_{t,s})+K_1 \epsi^2\}, 
\end{split}
\end{equation}
whose $\norm{\cdot}_2$ length is $O(\epsi)$. To see this, note that 
\begin{equation}\nn
\begin{split}
\hspace*{-.15cm}\norm{(U_{i_1,i_2,i_3,i_4,i_5}-&L_{i_1,i_2,i_3,i_4,i_5})(\eps_{t,s}^\st,\mathcal{X}_{t,s})}_2^2 \\
 = \,& \Ex[U_{i_1,i_2,i_3,i_4,i_5}(\eps_{t,s}^\st,\mathcal{X}_{t,s})]-\Ex[L_{i_1,i_2,i_3,i_4,i_5}(\eps_{t,s}^\st,\mathcal{X}_{t,s})] \\
= \,& \Ex[\bar{\DD}_{i_1,i_2,i_3,i_4,i_5}(\bar{y}_{i_1,i_2,i_3,i_4,i_5}^\dagger,\mathcal{X}_{t,s})-\munderbar{\DD}_{i_1,i_2,i_3,i_4,i_5}(\munderbar{y}_{i_1,i_2,i_3,i_4,i_5}^\dagger,\mathcal{X}_{t,s})] + 2 K_1 \epsi^2  \\
= \,& \Ex[\bar{\DD}_{i_1,i_2,i_3,i_4,i_5}(\GG^{-1}(u;\theta_q),\mathcal{X}_{t,s})-\munderbar{\DD}_{i_1,i_2,i_3,i_4,i_5}(\GG^{-1}(u;\theta_q),\mathcal{X}_{t,s})]  \\
\,&    + \Ex[\bar{\DD}_{i_1,i_2,i_3,i_4,i_5}(\bar{y}_{i_1,i_2,i_3,i_4,i_5}^\dagger,\mathcal{X}_{t,s})-\bar{\DD}_{i_1,i_2,i_3,i_4,i_5}(\GG^{-1}(u;\theta_q),\mathcal{X}_{t,s})]\\
\,&    + \Ex[\munderbar{\DD}_{i_1,i_2,i_3,i_4,i_5}(\GG^{-1}(u;\theta_q),\mathcal{X}_{t,s})-\munderbar{\DD}_{i_1,i_2,i_3,i_4,i_5}(\munderbar{y}_{i_1,i_2,i_3,i_4,i_5}^\dagger,\mathcal{X}_{t,s})] + 2 K_1 \epsi^2 \\
\leq \,& \Ex[\bar{\DD}_{i_1,i_2,i_3,i_4,i_5}(\GG^{-1}(u;\theta_q),\mathcal{X}_{t,s})-\munderbar{\DD}_{i_1,i_2,i_3,i_4,i_5}(\GG^{-1}(u;\theta_q),\mathcal{X}_{t,s})] + 2 (K_1+ K_2) \epsi^2 \\
\leq \,&K_3\epsi^2,\;\;\text{ with }\;\; K_3 \coloneqq 2 \{K_1(1+\Ex[\zeta_3(\mathcal{X}_{t,s})])+ K_2\}.
\end{split}
\end{equation}
Therefore, $\mathcal{N}(\epsilon,\bar{\mathcal{G}}_1,\rho(\cdot)) = O(\epsi^{-4(p+m+2)})$ and the claim follows by Theorem \ref{thmap-94} in conjunction with Assumption \ref{asu-B}.

\noindent\textbf{Proof of lemma \ref{lemmaU}-\ref{lemmaU-b}:}
First, it is shown that $\bar{\mathbb{U}}_{i,j,T,S}(u_i,u_j;\theta_q,\nu)  = \Ex[\tilde{\mathbb{U}}_{i,j,T,S}(u_i,u_j;\theta_q,\nu)] + o(1)$ uniformly in $u_i,u_j \in [0,1]$, $\nu \in \mathcal{V}_T$, and $\theta_q \in \Theta$ for any \(i,j \in \mathcal{G}_q\), \(q \in \{1,\dots,Q\}\). Set $y_k(\mathcal{W}_t;u_k,\theta_q,\nu) \coloneqq \GG_k^{-1}(u_k;\theta_q) + \beta\Delta^\dagger(\mathcal{W}_t,\nu)$ and define the map 
\[
\nu \mapsto \mathcal{T}(u_i,u_j,\theta_q,\nu) \coloneqq \GG_{i,j}(y_i(\mathcal{W}_t;u_i,\theta_q,\nu),y_j(\mathcal{W}_t;u_j,\theta_q,\nu)).
\]
\textbf{For brevity, suppose that $\nu$ is a scalar.} A second-order Taylor-series expansion around $\nu = \nu_0$ yields
\begin{equation}
\begin{split}
\sqrt{T}\Ex[\mathcal{T}(u_i,u_j,\theta_q,\nu) - \CC_{i,j}(u_i,u_j;\theta_q)] = \,&    \beta \Ex[\nabla_\nu \sigma(M_1,\nu_0)]\sqrt{T}(\nu-\nu_0)\\
\,& \hspace*{1cm} \times \sum_{k \inn \{\ii\}} \partial_k \GG_{i,j}\{\GG_i^{-1}(u_i;\theta_q),\GG_j^{-1}(u_j;\theta_q);\theta_q\} \nn \\
\,&  + [\sqrt{T}(\nu_0-\nu)]^2 \frac{1}{\sqrt{T4}} \Ex\bigg[\frac{\partial^2}{\partial \nu^2}\mathcal{T}(u_i,u_j,\theta_q,\bar{\nu})\bigg], \nn
\end{split}
\end{equation}
where $\bar{\nu} = \nu c + \nu_0(1-c) \in \mathcal{V}_T$ for some \(c \in (0,1)\). In view of the discussion below Eq. \eqref{partial-TV}, one readily verifies that the second-order term is asymptotically negligible. 
Therefore, $\Ex[\tilde{\mathbb{U}}_{i,j,T,S}(u_i,u_j;\theta_q,\nu)]  = \sqrt{T}(\Ex[\mathcal{T}(u_i,u_j,\theta_q,\nu)]-\CC_{i,j}(u_i,u_j;\theta_q)]) = \bar{\mathbb{U}}_{i,j,T,S}(u_i,u_j;\theta_q,\nu) + o(1)$, where it has been used that
\[
\partial_i \GG_{i,j}(x_i,x_j;\theta_q) 
=  \partial_i \CC_{i,j}\{\GG_i(x_i;\theta_q),\GG_j(x_j;\theta_q);\theta\}\gg_i(x_i;\theta_q).
\]
Next, by the definition of \eqref{emppro-G}, we have $\tilde{\mathbb{V}}_{i,j,T,S}(u_i,u_j;\lambda_q,\nu) = \tilde{\mathbb{C}}_{i,j,T,S}(\bar{g}-\bar{g}_0) + \bar{\mathbb{V}}_{i,j,T,S}(u_i,u_j;\theta_q,\nu)$,
where $\bar{g},\bar{g}_0 \in \bar{\mathcal{G}}$, with $\bar{g}_0$ is such that the restriction $\nu = \nu_0$ is imposed. By part \ref{lemmaU-a} of this lemma, $\tilde{\mathbb{C}}_{i,j,T,S}$ is stochastically equicontinuous. By Theorem \ref{thmap-94}, it remains to be shown that $\rho(\bar{g}-\bar{g}_0) = o(1)$. To see this, note that the triangle-inequality yields
 \begin{equation}\nn
\begin{split}
\ssup\limits_{1 \leq t \leq T,T \geq 1}\norm{&\bar{g}(\xi_{i,j,t,S})-\bar{g}_0(\xi_{i,j,t,S})}_2 \\
\leq \,&  \sum_{k \inn \{\ii\}}  \norm{1\{X_{k,t,s}(\theta_q) \leq \GG_k^{-1}(u_k;\theta_q) + \beta\Delta^\dagger(\mathcal{W}_{t},\nu)\} -1\{X_{k,t,s}(\theta_q) \leq \beta \GG_k^{-1}(u_k;\theta_q)\}}_{2}.   
\end{split}
\end{equation}
Now, set $\hat{x}_k \coloneqq \GG_k^{-1}(u_k;\theta_q) + \beta\Delta^\dagger(\mathcal{W}_{t},\nu)$ and $x_k \coloneqq \GG_k^{-1}(u_k;\theta_q)$, so that we have (by the mean-value theorem)  for $k \in \{\ii\}$
 \begin{equation}
\begin{split}
 \norm{1\{X_{k,t,s}(\theta_q) \leq \hat{x}_k\} -1\{X_{k,t,s}(\theta_q) \leq x_k\}}_{2} = \,& \Ex[\GG_k(\hat{x}_k;\theta_q)]+\GG_k(x_k;\theta_q)-2\Ex[\GG_k(\mmin\{x_k,\hat{x}_k\};\theta_q)] \nn \\
 \leq \,&  K_\beta\ssup\limits_{\theta \inn \Theta}\ssup\limits_{y \inn \mathbb{R}}\gg_k(y;\theta_q) \Ex[|\Delta^\dagger(\mathcal{W}_t,\nu)|] \nn \\
\leq \,&  K_\beta\ssup\limits_{\theta \inn \Theta}\ssup\limits_{y \inn \mathbb{R}}\gg_k(y;\theta_q) \Ex[\dot{\sigma}(M_1)]\norm{\nu-\nu_0} = O(1/\sqrt{T}).\nn
\end{split}
\end{equation}
 This shows $\rho(\bar{g}-\bar{g}_0) = O(T^{-1/2})$ and, in view of Theorem \ref{thmap-94}, completes the proof. 

\noindent\textbf{Proof of Lemma \ref{lemmaU}-\ref{lemmaU-c}:} Due to part \ref{lemmaU-a} of the Lemma and \citet[Corollary 2.3]{anpo94}, we are left with establishing the `fidi'-convergence. Similar to step (2) of the proof of Proposition \ref{prop-1} \ref{prop-1-b}, this follows from  \citet[Theorem 5.11]{white01} in conjunction with the Cram\'{e}r-Wold device and Assumption \ref{asu-F}.

\subsection{Proof  of Lemma \ref{lemmaA3}}
\noindent \textbf{Proof of Lemma \ref{lemmaA3} ($a$):} The functional delta method [cf. \citet[Theorem 3.9.4]{vw96}] in conjunction with Assumption \ref{asu-C} and \citet[Theorem 2.4.]{buvo13} yields
for each \(i,j \in \mathcal{G}_q\), \(q \in \{1,\dots,Q\}\):
\begin{equation} \nn
\begin{split}
\hat{\mathbb{V}}_{T,i,j}(u_i,u_j;\hlambda) = \,& \tilde{\mathbb{V}}_{T,i,j}(u_i,u_j;\hlambda)\\
\,& - \partial_i\CC_{q}(u_i,u_j)\tilde{\mathbb{V}}_{i,j,T}(u_i,1;\hlambda)\\
\,&-\partial_j\CC_{q}(u_i,u_j)\tilde{\mathbb{V}}_{T,i,j}(1,u_j;\hlambda) + o_p(1) \\
\stackrel{(1)}{=} \,& \bar{\mathbb{V}}_{T,i,j}(u_i,u_j;\hlambda)\\
\,& - \partial_i\CC_{q}(u_i,u_j)\bar{\mathbb{V}}_{T,i,j}(u_i,1;\hlambda)\\
\,&-\partial_j\CC_{q}(u_i,u_j)\bar{\mathbb{V}}_{T,i,j}(1,u_j;\hlambda) + o_p(1) \stackrel{(2)}{=}   o_p(1).
\end{split}
\end{equation}
\textit{Explanations:} (1) For $T$ sufficiently large, one has, by part \ref{asu-E2} of Assumption \ref{asu-E},  for any $\epsi > 0$ and any $\delta > 0$
\begin{equation} \nn 
\begin{split}
\PP\bigg[\ssup\limits_{u_i,u_j \inn [0,1]}&| \tilde{\mathbb{V}}_{i,j,T}(u_i,u_j;\hlambda)-\bar{\mathbb{V}}_{i,j,T}(u_i,u_j;\hlambda)| >  \epsi\bigg] \\
 \leq \,& \PP(\{\hlambda \notin  \Lambda_{T}\}) \\
\,& + \PP\bigg[\bigg\{\ssup\limits_{u_i,u_j \inn [0,1]}| \tilde{\mathbb{V}}_{i,j,T}(u_i,u_j;\hlambda)-\bar{\mathbb{V}}_{i,j,T}(u_i,u_j;\hlambda)| > \epsi\bigg\} \,\cap\, \{\hlambda \in  \Lambda_{T}\}\bigg] \\
 \leq \,& \frac{\delta}{2} + \PP\bigg[\ssup\limits_{\lambda \inn \Lambda_{T},\,k \inn \{i,j\}}\ssup\limits_{u_i,u_j \inn [0,1]}| \tilde{\mathbb{V}}_{i,j,T}(u_i,u_j;\hlambda)-\bar{\mathbb{V}}_{i,j,T}(u_i,u_j;\hlambda)| > \epsi\bigg].
\end{split}
\end{equation}
By part part \ref{lemmaV-b} of Lemma \ref{lemmaV},  the probability on the right-hand side of the preceding display can be made smaller than $\delta/2$, which proves that $\tilde{\mathbb{V}}_{T,i,j}(u_i,u_j;\hlambda)-\bar{\mathbb{V}}_{T,i,j}(u_i,u_j;\hlambda)= o_p(1)$ uniformly in $u_i,u_j \in [0,1]$. (2) Here, we use the definition of $\bar{\mathbb{V}}_{i,j,T}$, the fact that $\partial_i\CC_{q}(u_i,1) = \partial_j\CC_{q}(1,u_j) = 1$, and Assumption \ref{asu-B}. \textbf{Proof of Lemma \ref{lemmaA3} ($b$):} Follows by similar arguments. 

\subsection{Proof  of Lemma \ref{lemmaA4}}
The proof follows by combining Lemma \ref{lemmaV}, \ref{lemmaU}, and \ref{lemmaA3} with  \citet[Lemma 7.2]{buse13} and the integration by parts formula in \citet[p. 854]{fetal04} [see also \cite{radetal17} and \cite{betal17}] in conjunction with Assumption \ref{asu-G}.

 \subsection{Remark on Assumption \ref{asu-D3}}
The following verifies Assumption \ref{asu-D3} for \(t\)-distributed factors. To fix ideas, suppose a scalar factor \(F\) is distributed according to Student's standardized \(t\)-distribution, denoted by \(\DD(\gamma_0)\), where \(\gamma_0 > 2\) is the degrees of freedom (shape) parameter; i.e.,
\[
\PP(F \leq x) = \DD(x; \gamma_0) = \frac{1}{\sqrt{(\gamma_0-2)\pi}}\frac{\Gamma((\gamma_0+1)/2)}{\Gamma(\gamma_0/2)}\int_{(-\infty,x]}(1+t^2/(\gamma_0-2))^{-(\gamma_0+1)/2}\,\dd t, \quad x \in \mathbb{R},
\]
with \(\Gamma(x)\) denoting the gamma function. Next, choose \(\ubar{\gamma},\bar{\gamma} \in \mathbb{R}\) such that \(2 < \ubar{\gamma} \leq \gamma_0 \leq \bar{\gamma} < \infty\). It is not difficult to see that part \ref{A-bracket1} is satisfied given that \(\gamma_0 \in [\ubar{\gamma},\bar{\gamma}] \subset (2,\infty)\). Turning to part  \ref{A-bracket2}, note that the inverse distribution function can be represented for any \(\gamma \in [\ubar{\gamma},\bar{\gamma}]\) as
\[
\DD^{-1}(u; \gamma) = \begin{cases}-\displaystyle\sqrt{\gamma-2}\sqrt{\frac{1}{I^{-1}_{2u\,}(\gamma/2,1/2)}-1} & u \in (0,1/2) \\
\displaystyle\sqrt{\gamma-2}\sqrt{\frac{1}{I^{-1}_{2(1-u)}(\gamma/2,1/2)}-1} & u \in (1/2,1), 
\end{cases}
\]
with \(\DD^{-1}(1/2; \gamma) \coloneqq 0\), \(\DD^{-1}(0; \gamma) \coloneqq -\infty\), and	 \(\DD^{-1}(1; \gamma) \coloneqq \infty\); here, \(I^{-1}_x(a,b)\) is the inverse of the regularized \(\beta\)-function
\[
I_x(a,b) \coloneqq B_x(a,b)/B(a,b), 
\]
where \(B(a,b)\) and \(B_x(a,b)\) denote the ordinary and incomplete \(\beta\)-function, respectively; see  \cite{shaw06}. Since the absolute quantile function of the non-standardized \(t\)-distribution \(\tilde{\DD}(u;\gamma) = \sqrt{\gamma/(\gamma-2)}\DD(u; \gamma)\) is \textit{de}creasing in \(\gamma\) and bounded on \([\ubar{\gamma},\bar{\gamma}]\), and \(\gamma \mapsto \sqrt{(\gamma-2)/\gamma}\) is \textit{in}creasing and bounded on \([\ubar{\gamma},\bar{\gamma}]\), one obtains the upper bound \(|\DD^{-1}(u; \gamma)| \leq \sqrt{(\bar{\gamma}-2)/\bar{\gamma}}|\tilde{\DD}^{-1}(u; \ubar{\gamma})|\). Therefore,
\[
\int_{[0,1]} \ssup\limits_{\gamma \inn [\ubar{\gamma},\bar{\gamma}]}|\DD^{-1}(u; \gamma) |  \dd u \leq \sqrt{(\bar{\gamma}-2)/\bar{\gamma}} \int_{[0,1]} |\tilde{\DD}^{-1}(u; \ubar{\gamma})| \dd u  = \sqrt{(\bar{\gamma}-2)/\bar{\gamma}} \sqrt{\frac{4\ubar{\gamma}}{\pi(\ubar{\gamma}-1)^2}}\frac{\Gamma\left(\frac{\ubar{\gamma}+1}{2}\right)}{\Gamma\left(\frac{\ubar{\gamma}}{2}\right)},
\]
where the final equality uses the definition of the expected value of the half \(t\)-distribution. Turning to the first derivative, suppose first \(u \in (0,1/2)\) and note that
\[
\frac{\partial}{\partial \gamma}\DD^{-1}(u; \gamma) = R_1(u; \gamma) - R_2(u;\gamma),\quad R_2(u;\gamma) \coloneqq \sqrt{(1/I^{-1}_{2u}(\gamma/2,1/2)-1)/4(\gamma-2)},
\]
with 
\begin{equation} \nn
\begin{split}
R_1(u; \gamma)  \coloneqq \,& \sqrt{\frac{(\gamma-2)(1-I^{-1}_{2u}(\gamma/2,1/2))}{16(1/I^{-1}_{2u}(\gamma/2,1/2)-1)}} \frac{1}{(I^{-1}_{2u}(\gamma/2,1/2))^{\gamma/2+1}}\\
\,& \times \bigg[\Gamma^2(\gamma/2)(I^{-1}_{2u}(\gamma/2,1/2))^{\gamma/2} {}_3 F_2\left(\gamma/2,\gamma/2,1/2;\gamma/2+1,\gamma/2+1; I^{-1}_{2u}(\gamma/2,1/2)\right)\\
\,& \quad - 2uB(\gamma/2,1/2)\left(\textsf{log}(I^{-1}_{2u}(\gamma/2,1/2))+\psi(\gamma/2+1/2)-\psi(\gamma/2)\right)\bigg],
\end{split}
\end{equation}
where \({}_q F_p(a_1,\dots,a_q;b_1,\dots,b_p;z)\) denotes the regularized hypergeometric function. Some tedious algebra reveals \(|R_1(u; \gamma)| < |R_2(u; \gamma)|\). Note that \(|R_2(u;\gamma)| = |\tilde{D}^{-1}(u;\gamma)|/\sqrt{4\gamma(\gamma-2)}\). Moreover, using similar arguments as above reveals that \(R_2(u;\gamma)\) is monotonically \textit{de}creasing in \(\gamma\) uniformly over \(u\). Similar arguments apply for the case \(u \in (1/2,1)\). Hence,
\begin{equation} 
\ssup\limits_{\gamma \inn [\ubar{\gamma},\bar{\gamma}]}\bigg\lvert\frac{\partial}{\partial \gamma}\DD^{-1}(u; \gamma)\bigg\rvert \leq  R(u;\ubar{\gamma}) \coloneqq |\tilde{D}^{-1}(u;\ubar{\gamma})|/\sqrt{4\ubar{\gamma}(\ubar{\gamma}-2)}.
\end{equation}
Thus, the assumption is satisfied for \(\dot{Q}_F(u): [0,1] \rightarrow \mathbb{R}\) given by \[\dot{Q}_F(u) \coloneqq  \sqrt{\mmax\{\bar{\gamma},1/(\ubar{\gamma}-2)\}}|\tilde{D}^{-1}(u,\ubar{\gamma})|.\] 

\begin{remark} \normalfont
Having verified Assumption \ref{asu-D3} for the \(t\)-distribution \(\DD(\gamma_0)\), \(\gamma_0 \in (2,\infty)\), one may use this result to verify the case of the skewed \(t\)-distribution; i.e., \(F\) is distributed according to Hansen's skewed \(t\)-distribution, denoted by \(\FF(\gamma_{0,1},\gamma_{0,2})\), say, with degrees of freedom parameter \(\gamma_{0,1}  \in (2,\infty)\) and asymmetry parameter \(\gamma_{0,2} \in (-1,1)\); i.e. \(\PP(F \leq x) = \FF(x; \gamma_{0,1},\gamma_{0,2})\) and
\[
\FF^{-1}(u; \gamma_1,\gamma_2) = \begin{cases}\displaystyle\frac{1}{b}\left((1-\gamma_2)\DD^{-1}(u/(1-\gamma_2),\gamma_1)-a\right) & u \in (0,(1-\gamma_2)/2) \\
\displaystyle\frac{1}{b}\left((1+\gamma_2)\DD^{-1}((u+\gamma_2)/(1+\gamma_2),\gamma_1)-a\right) & u \in [(1-\gamma_2)/2,1), 
\end{cases}
\]
where
\[
c \coloneqq \frac{\Gamma((\gamma_1+1)/2)}{\sqrt{\pi(\gamma_1-2)\Gamma(\gamma_1/2)}},\quad b \coloneqq \sqrt{1+3\gamma_2^2-a^2},\quad a \coloneqq 4\gamma_2c\frac{\gamma_1-2}{\gamma_1-1};
\]
see, e.g., \citet[Appendix A.2]{fanta08}.
\end{remark}

\end{document}